\documentclass[conference]{IEEEtran}
\IEEEoverridecommandlockouts
\makeatletter
\newif\if@restonecol
\makeatother

\usepackage[utf8]{inputenc}
\usepackage{adjustbox}
\usepackage{graphicx}
\usepackage{color}
\usepackage{caption}
\usepackage{ulem}
\usepackage{float}
\usepackage{CJK}
\usepackage{subfigure}
\usepackage[caption=false,font=footnotesize]{subfig}
\usepackage{placeins}
\usepackage{url}
\usepackage{listings}
\usepackage{algcompatible}
\usepackage{amsmath,amssymb,amsfonts}
\usepackage{algorithmicx}
\usepackage[lined,boxed,commentsnumbered, ruled,vlined,linesnumbered, frenchkw]{algorithm2e}
\usepackage{graphicx}
\usepackage{textcomp}
\usepackage{xcolor}
\DeclareUnicodeCharacter{3000}{}
\DeclareUnicodeCharacter{FF1A}{}
\usepackage[backend=biber]{biblatex}
\usepackage[english]{babel}

\def\BibTeX{{\rm B\kern-.05em{\sc i\kern-.025em b}\kern-.08em
    T\kern-.1667em\lower.7ex\hbox{E}\kern-.125emX}}
    
\begin{document}

\title{Task allocation for decentralized training in heterogeneous environment
}

\author{\IEEEauthorblockN{1\textsuperscript{st} Yongyue Chao}
\IEEEauthorblockA{\textit{Institute of Automation, Chinese Academy of Sciences } \\
Beijing, China \\
chaoyongyue2020@ia.ac.cn}
\and
\IEEEauthorblockN{2\textsuperscript{nd} Mingxue Liao}
\IEEEauthorblockA{\textit{Institute of Automation, Chinese Academy of Sciences} \\
Beijing, China \\
mingxue.liao@ia.ac.cn}
\and
\IEEEauthorblockN{3\textsuperscript{rd} Jiaxin Gao}
\IEEEauthorblockA{\textit{Institute of Automation, Chinese Academy of Sciences} \\
Beijing, China\\
jiaxin.gao@ia.ac.cn}
}

\maketitle

\begin{abstract}
The demand for large-scale deep learning is increasing, and distributed training is the current mainstream solution. Ring AllReduce is widely used as a data parallel decentralized algorithm. However, in a heterogeneous environment, each worker calculates the same amount of data, so that there is a lot of waiting time loss among different workers, which makes the algorithm unable to adapt well to heterogeneous clusters. Resources are not used as they should be. In this paper, we design an implementation of static allocation algorithm. The dataset is artificially allocated to each worker, and samples are drawn proportionally for training, thereby speeding up the training speed of the network in a heterogeneous environment. We verify the convergence and influence on training speed of the network model under this algorithm on one machine with multi-card and multi-machine with multi-card. On this basis of feasibility, we propose a self-adaptive allocation algorithm that allows each machine to find the data it needs to adapt to the current environment. The self-adaptive allocation algorithm can reduce the training time by nearly one-third to half compared to the same proportional allocation.In order to better show the applicability of the algorithm in heterogeneous clusters, We replace a poorly performing worker with a good performing worker or add a poorly performing worker to the heterogeneous cluster. Experimental results show that training time will decrease as the overall performance improves. Therefore, it means that resources are fully used. Further, this algorithm is not only suitable for straggler problems, but also for most heterogeneous situations. It can be used as a plug-in for AllReduce and its variant algorithms.   
\end{abstract}


\section{INTRODUCTION}
Artificial intelligence is widely applied in various fields as well as deep learning plays an important role in it.Establishing deep neural network model within big data is the  main method in deep learning. With the development of technology, more complex model and larger dataset appear but it is hard to train by single machine. To speed up, distributed training steps in large scaled deep learning. It is the idea that all data is distributed to nodes for calculating and then results are aggregated together to update parameters of model.

At present, small memory and long running time are two primary shortcomings of single machine. In distributed training, model parallelism\cite{2014Primitives} and data parallelism\cite{2018Demystifying} are proposed to solve the problems. Owing to high throughput, data parallelism is extensively used to train large scaled neural network. most of data parallelism methods are constructed based on gradient aggregation. In 2013, Li et al. achieved parameter server({\tt PS}) which is one of common data parallelism framework as figure 1(a) shows. It consists of two functional nodes called worker and server. Workers pull latest parameters from servers to compute and then push local gradients to servers. Server aggregates all local gradients to update parameters of network model. However, communication bottleneck\cite{1992Overcoming} exists in {\tt PS} framework because of centralized communication. In 2017, All reduce is applied in deep learning by Ring allreduce algorithm as figure 1(b) shows. It eliminated communication bottleneck by letting each node only communicate with adjacent nodes. All nodes in Ring Allreduce has same amount of communication. Local gradients is passed to each node through Allreduce, then all nodes update parameters of model individually. Ring Allreduce is a synchronization algorithm. Aggregating local gradients happens after that all nodes have finished computing. Therefore, in heterogeneous environment, due to barrel effect, training velocity relies on the slowest node. The later algorithms focused on improving Ring Allreduce as an asynchronous algorithm.They modified the method of aggregating gradients to reduce training time. 

\begin{figure}[ht]
\centering
\subfigure[Parameter Sever]{
\begin{minipage}[t]{0.5\linewidth}
\includegraphics[width=1.6in]{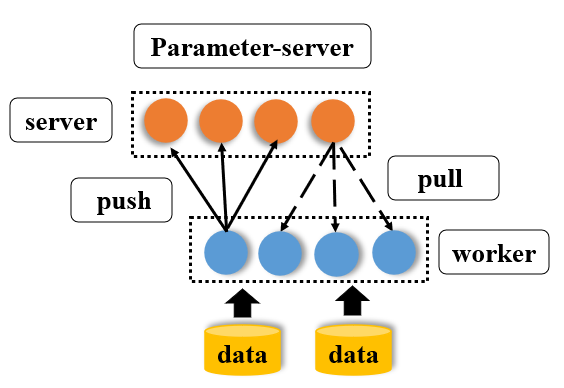}
\end{minipage}
}%
\subfigure[Ring AllReduce]{
\begin{minipage}[t]{0.5\linewidth}
\centering
\includegraphics[width=1.6in]{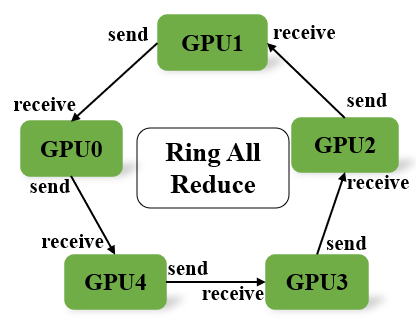}
\end{minipage}
}
\centering
\caption{the training time of models results in calculating time, the ratio of weights and training time from two machines with RTX1080ti and V100.}
\end{figure}

Using Ring Allreduce to complete the training is divided into two steps. First, separate dataset equally to each node and train. In this step, all nodes obtain the same number of data. There is no interference between nodes. Each node just calculate the gradient of parameters. After computing, node will arrive at barrier to wait for other nodes together. Second, renew parameters. Through Allreduce, each node gathers global gradients from others and uses back propagation\cite{1989Theory} to update model parameters. In homogeneous environment, due to equal performance and same amount data, all nodes reach barrier almost at the same time. Therefore, there is no time wasted by waiting. In heterogeneous environment, this problem can slow down training greatly. However, the second step received most attention from researchers. Most of existing methods focusing on Ring Allreduce make progress in acceleration by asynchronous algorithm. Just few researchers have considered how to realize the partition of dataset based on Ring Allreduce and how to separate data set appropriately to nodes with different performance. Theoretically, through assigning reasonable task to different nodes, synchronization waiting time can be shortened and thus total training time can be shortened too.

In this paper, we proposed an implement of assigning tasks to nodes with different performance. It is called static allocation Allreduce algorithm. It could ensure that the dataset is allocated to each node under the premise of neural network convergence. To prove it, we train a simple convolutional neural network\cite{2006Notes} model on MNIST dataset\cite{2012The}, some complex network such as ResNet18, ResNet50\cite{2016Inception}, VGG16 and VGG19\cite{2014Very} model on CIFAR10 dataset\cite{2009Learning}. Experimental results show that changing the ratio of dataset or tasks on different nodes has little effect on the number of network convergence epochs but can adjust the training time of each epoch. On the basis of this implement, we further proposed an adaptive algorithm to compute how to distribute data to different nodes without manufacturer information of nodes. We set a series of experiments to prove the improvement of training velocity. We use two nodes equipped with v100 and RTX 2080ti respectively to confirm the training velocity will be increased along with the ratio of dataset on nodes and epochs changed. When the ratio of dataset stabilized, training time per epoch could be reduced 20~40 percentage than the same ratio. When we replace one of weak GPUs with strong GPU or add one GPU
We also compared results. Under the same network bandwidth, the training time will be reduced along with the improvement of performance.   

\section{BACKGROUND AND MOTIVATION}
\subsection{Heterogeneity}
From the perspective of heterogeneity, in data parallelism , whether PS or Ring Allreduce, although the training process seems to be accelerated through the way that each node computes local gradients, the occurrence of synchronization and communication among different nodes slows down the training time after that. Backup and asynchronization explored solution to ignore or put off straggler\cite{2017Gradient} in synchronization. They fully applied time difference between data computation doing something to make up. These methods still allocate tasks evenly to different nodes. However, they don't work in some specific situations. For example, imagine that if there is one node faster than others, it must wait for aggregating gradients with another node at least. No one could complete with that fast node immediately. Actually, the effect of that faster node is equal with other nodes. It is obvious that computing resources are wasted. In a word, Only when the amount of data held by each node is different, can the maximum performance be exerted as much as possible.               
\subsection{Ring Allreduce}
When Ring Allreduce is utilized in distributed deep learning by Baidu, Pytorch\cite{2021PyTorch} and Horovod\cite{2018Horovod} successionly designed their own distributed training framework based on the algorithm. However, the human understanding of this algorithm is not yet complete. Before starting our algorithm, let's review Ring Allreduce in detail. Assuming there are $n$ workers waiting for training, they are distributed on a ring and the total dataset is divided into $n$ parts for $n$ workers. After all workers finished calculate local gradients of subset, each worker cuts local gradients into $n$ parts. There are two steps in communication. First, for the $k$th worker, this worker will send the $k$th data to the next worker, and at the same time receive the $k-1$th data from the previous worker. After looping $N$ times, each worker will contain a copy of the final integration result. In the second stage, each worker sends the integrated part to the next worker. After the worker receives the data, it can update the corresponding part of its own data. There is no communication bottleneck among workers through Ring Allreduce. From the process of Ring Allreduce, there is a barrier\cite{2010Inter} to synchronize all workers meanwhile distinguish two steps. Also, there are many synchronized operations during the second step. Everytime finishing changing data a round, the time consumed depends on the lowest worker. These synchronized operations are difficult to be reduced. Therefore, which we can extremely control is the first synchronization. We can make training time approach among workers to accelerate.

\section{METHODS}
In order to accelerate by balancing the number of tasks among different workers, we followed the idea of first experimental verification and then theoretical optimization. We proposed one implementation of static task allocation algorithm and one algorithm for adaptively adjusting the task volume. Static task allocation algorithm was achieved based on gradient accumulation to prove the feasibility of unequal tasks. After that, we proposed a self-adaptive algorithm to distribute tasks automatically without obtaining the external information of workers. We set $w_{1}, w_{2}, \cdots, w_{n}$ as how many samples each worker need to train in one gradient aggregation. 
\subsection{static allocation implement}
The purpose of static allocation implement is to verify that the ratio of $w_{i}$ among workers is one of significant velocity factors and has few influence on network convergence. First of all, we set the weight $w_{1}, w_{2}, \cdots, w_{n}$ to ensure that worker $i$ must wait for others after training $w_{i}$ samples in one gradient aggregation. Second, according to the ratio $\frac{w_{1}}{\sum_{i=1}^n w_{i}}, \frac{w_{2}}{\sum_{i=1}^n w_{i}}, \cdots, \frac{w_{n}}{\sum_{i=1}^n w_{i}}$, we assigned a corresponding proportion of training samples to each worker from the total dataset so that each worker holds unequal subdataset. Third, each worker computes independently. In one gradient aggregation, Worker $i$ draws $w_{i}$ samples from subdataset to calculate gradients. All workers need to accumulate gradients without back propagation. Specifically, $(1)$ After transferring one sample to the network and getting prediction results, calculate the loss value according to the prediction result and label. $(2)$ Use loss for back propagation and calculate parameter gradient. Accumulate the gradient instead of clearing up. $(3)$ Repeat from $(1)$ to $(2)$ steps until $w_{i}$ samples are transferred to the network. Finally, when all workers finished computing and arrived at the synchronization barrier, they will join in Ring Allreduce to aggregate gradients with local accumulation gradients. Inevitably, total minibatch size are expanded to $minibatch*(\sum_{i=1}^n w_{i})$. In figure 2, it is the flow chart showing the process of static allocation implement. 

\begin{figure*}[ht]
\centering
\includegraphics[scale=0.58]{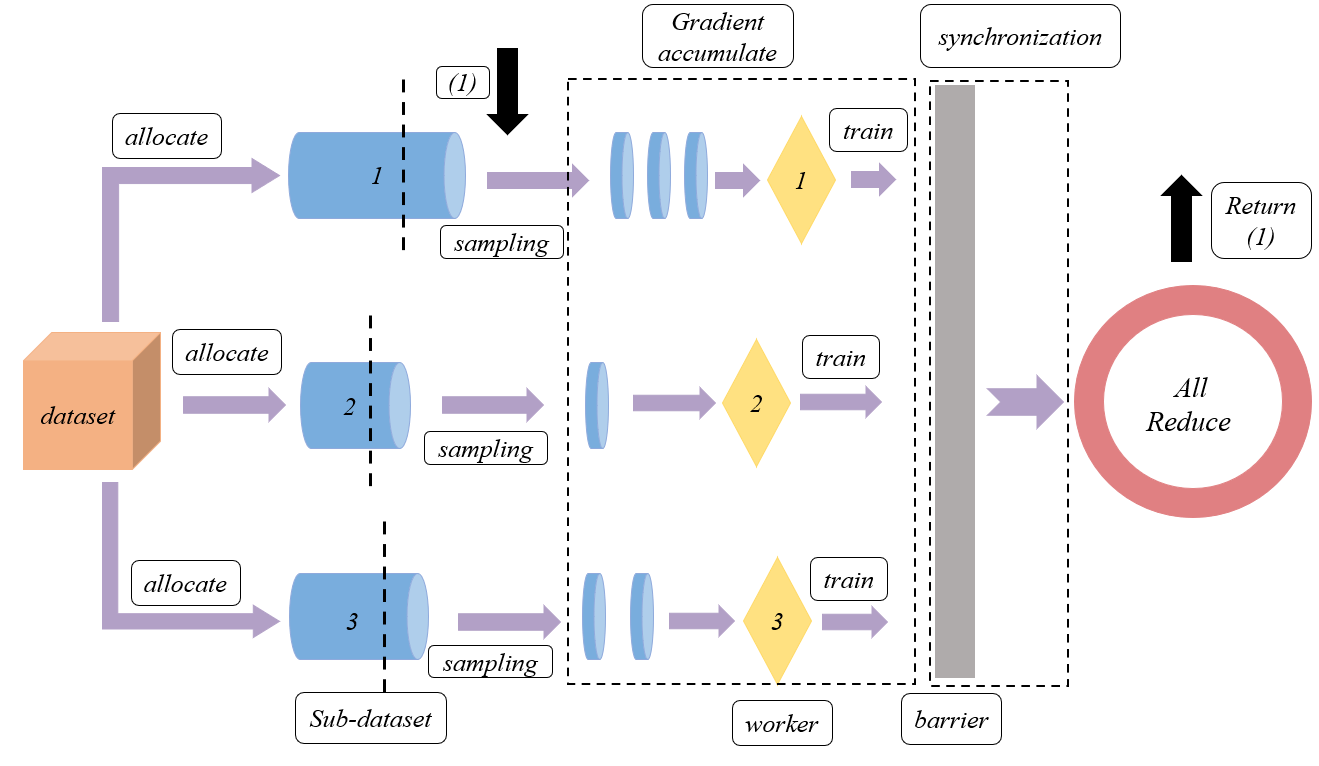}%
\caption{the procedure of static allocation algorithm with three GPUs or machines}
\label{figl}
\end{figure*}
\subsubsection{Acceleration Analysis}
Worker $i$ draws $w_{i}$ samples from subdataset to accumulate gradients can be able to ensure that there are no remaining samples without training after one epoch. Also, it's the most significant step to accelerate. Normally, whatever the worker is fast or slow, it must only train original $minibatch$ samples and then wait for other workers to synchronize. It means that faster workers are stuck at synchronization barrier without any computing. The average allocation consumes resources and wastes time. However, accumulating gradient fully made use of time gap. As figure 3 shows, it delayed the faster worker entering synchronization by expanding batchsize. In the same time interval, Fast workers take it for granted. Each worker tries the best to avoid waiting for others. Ideally, the straggler accumulated least times while other workers accumulated more times. To some extent, though this procedure don't break synchronization, it is approaching heterogeneous environment to homogeneous environment.
\begin{figure}[ht]
\centering
\includegraphics[scale=0.4]{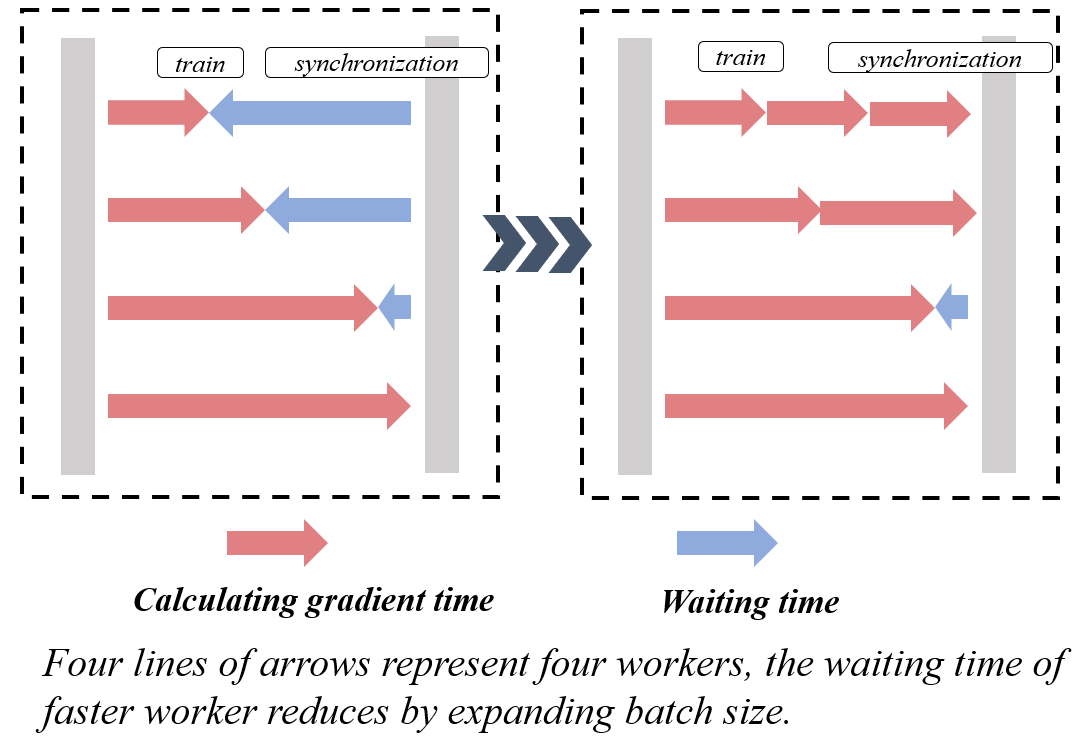}%
\caption{the procedure of static allocation algorithm with three GPUs or machines}
\label{figl}
\end{figure}
\subsubsection{Convergence Analysis}
Static allocation implement can guarantee model parameter dropping along the gradient direction and converge to stable loss. Figure 4 shows Ring AllReduce when changing allocating tasks among workers. Although the number of tasks was changed, the specific link of Ring AllReduce will not change. The only change of Ring AllReduce happens in back propagation. As formula (1) shows, 
\begin{equation} 
w_{k}=w_{k-1}-\eta \frac{1}{\text { N }} \sum_{i=1}^{N} \Delta f(w_{k-1}) 
\end{equation}
$N$ represents total batchsize of all minibatch at different workers, when task allocation occurred, $N$ depends on $\sum_{i=1}^n minibatch*w_{i}$. As long as total tasks maintain abiding, $N$ and gradient direction of descent will maintain too. Therefore, the final convergence depends on model parameters' initial weight and the whole process of training seems to be the same as equal task allocation. For some special situations, gradient aggregation SGD\cite{2017Accurate} must be modified on the basis of the importance of samples. Setting the weighted sum of local gradients in SGD to highlight difference. That's what we need to continue exploring later. In section 4, we set experiments to prove the convergence of static allocation.
\begin{figure}[ht]
\centering
\includegraphics[scale=0.5]{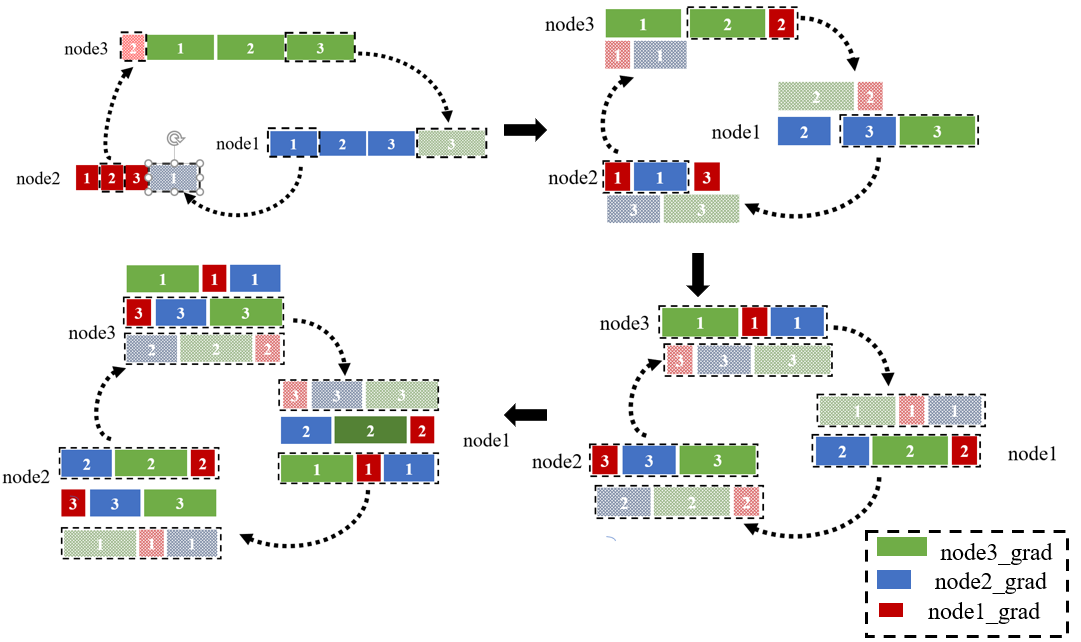}%
\caption{the procedure of static allocation algorithm with three GPUs or machines}
\label{figl}
\end{figure}
\subsection{Self-adaptive allocation algorithm}
Static allocation laid the foundation for task allocation, but specific allocated ratio of tasks is completely adjusted based on experience. It is full of uncertainty. Besides, manufacturer information of worker or GPU generally can't show accurate computing power. Load and network bandwidth during training will change slightly. They also influence the velocity of worker computing. Therefore, it's difficult to allocate tasks in accordance with existing information. In this part, we proposed a self-adaptive algorithm to solve this problem based on the training information.
\subsubsection{Notation}
\begin{itemize}
\item Gradient aggregation time: $t_{c}^{1}, t_{c}^{2}, \cdots, t_{c}^{n}$: In the $\mathrm{k}$th epoch, the time to perform allreduce and update the global model parameters after the local gradient of each node is accumulated. 
\item Gradient computing time: $t_{s}^{1}, t_{s}^{2}, \cdots, t_{s}^{n}$: In the $\mathrm{k}$th epoch, each node calculates the gradient of each sample before aggregation, including the sum of calculation and communication time.
\item Synchronization waiting time: $t_{w}^{1}, t_{w}^{2}, \cdots, t_{w}^{n}$: In the $\mathrm{k}$th epoch, the time that each node waits for other nodes to synchronize and aggregate.
\item Total training time: $T_{1}, T_{2}, \cdots, T_{n}$: The total time for each node training to complete one aggregation in the $\mathrm{k}$th epoch, and $T_{i}=t_{c}^{i}+t_{s}^{i}+t_{w}^{i}$.
\item Calculation speed: $v_{1}, v_{2}, \cdots, v_{n}$: the speed at which each node in the $\mathrm{k}$th epoch calculates the sample gradient before aggregation.Moreover, $v_{ i}=\frac{D_{i}}{t_{s}^{i}} $ where $D_{i}=D *\frac{w_{i}}{\sum_{i=1}^{ n} w_{i}}$, $ \quad \mathrm{D}$ represents the total number of samples.
\item In the $\mathrm{k}$th epoch, the samples computed by each node in one gradient aggregation: $w_{1}^{(k)}, w_{2}^{(k)}, \cdots, w_{n}^{(k)}$, the ratio of tasks obtained by each node: $\frac {w_{1}^{(k)}}{\sum_{i=1}^{n} w_{i}^{(k)}}, \frac {w_{2}^{(k)}}{\sum_{i=1}^{n} w_{i}^{(k)}}, \cdots, \frac {w_{n}^{(k)}}{\sum_{i=1}^{n} w_{i}^{(k)}}$.
\item Difference in synchronization waiting time: $\Delta t_{w}^{ij}=t_{w}^{i}-t_{w}^{j}$, in the $\mathrm{k}$th epoch, the time to wait when worker $\mathrm{i}$ synchronizes with worker $\mathrm{j}$. 
\item The amount of increased samples required for the new proportion of deployment in one gradient aggregation: $u_{1}, u_{2}, \cdots, u_{n}$
\end{itemize}
\subsubsection{Hypothesis}
In order to describe the algorithm better, we made the following assumptions based on the real situation.
\begin{itemize}
\item Due to synchronizing before gradient aggregation and sending global model parameter gradients to all workers after aggregation, as well as there are many synchronization operations during AllReduce, so it can be approximated that all workers started and ended at the same time in the process of AllReduce. Therefore, gradient aggregation time of all workers is equal. We set:\begin{equation}
    t_{c}^{1}=t_{c}^{2}=\cdots=t_{c}^{n}
\end{equation}
\item Total training procedure includes three steps: computing, synchronization and AllReduce(update). All workers attain the first minibatch at the same time. After computing, they will be blocked at barrier.Integrating the first assumption, it can be approximated that the total time for each node training to complete one aggregation is equal. We set:
\begin{equation}
    T_{1}=T_{2}=\cdots=T_{n}
\end{equation}
\item To avoid modifying learning rate along with the ratio of task allocation changing, the total batchsize should be stable. Therefore, the sum of samples computed by each node in one gradient aggregation is set  as a constant $C$. We set: 
\begin{equation}
w_{1}^{(k)}+ w_{2}^{(k)}+\cdots+w_{n}^{(k)}=C
\end{equation}
\begin{equation}
u_{1}+u_{2}+\cdots+u_{n}=0
\end{equation}
\end{itemize}
\subsubsection{Derivation}
The self-adaptive allocation algorithm mainly relies on adjusting the calculation time of different workers, shortening the waiting time of fast workers, thereby saving overall training time. Before the start of each epoch, the system will recalculate $w$ for all workers based on the information in the previous epoch, and allocate the corresponding task amount. When calculating the gradient, the new $w$ is also used for accumulation. After 4-5 epochs of training, The ratio of allocated tasks is basically stable, that is, the amount of redistributed tasks is stopped, and the amount of tasks for subsequent training is fixed. The self-adaptation is reflected in the amount of tasks that can be assigned by itself in each epoch. The specific idea is to use the available information in the previous epoch to get the amount of assigned tasks in the next epoch. 

\begin{algorithm}
\caption[]{self-adaptive allocation algorithm}
\KwIn{Randomly specify the sample distribution ratio of each node $w_{1}^{(k)}, w_{2}^{(k)}, \cdots, w_{n}^{(k)}$, and the calculation sample gradient time on each worker $t_{s}^{1}, t_{s}^{2}, \cdots, t_{s}^{n}$ is set to 0 }

\For{epoch}{
\textbf{step 1:}
　　\eIf{node==$i$}{
　　The worker broadcasts its own statistics of the last round of calculation gradient time.\\ }{
　　The worker accepts and updates the calculated gradient time broadcast by other workers.\\ 
　　}
\textbf{step 2:} Calculate the new sample distribution ratio of all workers:

$w_{i}^{(k+1)}=u_{i}+w_{i}^{(k)}=\frac{w_{i}^{(k)} / t_{s}^{i}}{\sum_{i=1}^{n} w_{i}^{(k)} / t_{s}^{i}} \sum_{i=1}^{n} w_{i}$

\textbf{step 3:}  Redistribute the subdataset of each worker according to the sample ratio.

(Step 2 and step 3 could be cancelled when the ratio is not fluctuating.)

\While{All data is unused}{
\textbf{step 4:} Proportionally draw samples from the sub-data set for training and accumulate gradients.

\textbf{step 5:} Record the calculation time and enter synchronization to wait for other workers.

\textbf{step 6:} Update network model parameters by AllReduce.}}

\end{algorithm}

To explore which parameters are available information to accelerate,  we analyzed from destination to what we chose to utilize. Our objective functions are reflected as following：
\begin{equation}
\underset{D}{\min} \sum_{i=1}^{n} \sum_{j !=i}^{n} \Delta t_{w}^{i j} \quad s.t. (2)(3)(4)
\end{equation}
\begin{equation}
\quad T_{1}, T_{2}, \cdots, T_{n} \quad \downarrow 
\end{equation}
Formula (6) and (7) represent that our destination is minimizing time of synchronization and training.According to notations and hypothesises, difference of synchronization waiting time are expected equivalent to
$\Delta t_{w}^{ij}=t_{w}^{i}-t_{w}^{j}=t_{s}^{j}-t_{s}^{i}=\frac{D_{j}}{v_{j}}-\frac{D_{i}}{v_{i}}=D \frac{w_{j}}{\sum_{k=1}^{n} w_{k}} \cdot \frac{1}{v_{j}}-D \frac{w_{i}}{\sum_{k=1}^{n} w_{k}} \cdot \frac{1}{v_{i}}=0$    
. Due to $\sum_{i=1}^{n} w_{i}=C$, we got the following formula:
\begin{equation}
 \frac{D w_{j}}{C v_{j}}-\frac{D w_{i}}{C v_{i}}=0   
\end{equation}
Simply, we can discover that the ratio of samples in one gradient aggregation is reciprocal to the velocity of calculating gradients between any two workers. In appendix A, we computed the solution of increased samples in one gradient aggregation. $u=[u_{1}, u_{2}, \cdots  u_{3}]$, where $u_{i}$ is: 
\begin{equation}
\begin{gathered}
u_{i}=\frac{v_{i}}{\sum_{i=1}^{n} v_{i}} \sum_{i=1}^{n} w_{i}^{(k)}-w_{i}^{(k)}
\end{gathered}
\end{equation}
Coincidentally, the equation (8) reflected the final sulotion of $w_{i}^{(k+1)}$, because of $w_{i}^{(k+1)}=w_{i}^{(k)}+u_{i}$. Eventually, we can obtain the amount of change $u$ in $k+1$th epoch as the following equation according to the fifth notation:
\begin{equation}
w_{i}^{(k+1)}=u_{i}+w_{i}^{(k)}=\frac{w_{i}^{(k)}/t_{s}^{i}}{\sum_{i=1}^{n} w_{i}^{(k)}/t_{s}^{i}}\sum_{i=1}^{n} w_{i}^{(k)}
\end{equation}
After derivation, it is found that the available information is the task amount $w_{i}$ and gradient calculation time $t_{s}^{(i)}$ of the previous epoch. In other words, only the task amount $w_{i}$ and gradient calculation time $t_{s}^{(i)}$ of the previous epoch need to be obtained to get the task amount of the next epoch. As figure 5 shows, we proposed an algorithm to achieve self-adaptive task allocation. The algorithm is improved from static allocation. The difference between them is that self-adaptive allocation algorithm computes $w_{i}$ by $u_{i}$ from previous epoch in next epoch and redistributes tasks by the ratio of $w_{i}/\sum_{k=1}^n w_{k}$ until $w_{i}$ stays still. The whole procedure is displayed in Algorithm 1, the most obvious modification is collecting gradient calculation time $t_{s}^{(i)}$. Each worker only get its own $t_{s}$, so they need to broadcast their own $t_{s}^{(i)}$ to other workers and receive others' $t_{s}$. After collecting, worker $i$ calculates the ratio $w_{(k+1)}$ and updates all subsequent operations. The reason why rounding decimals of $u_{i}$ is that $w^{(k+1)}$ is integer. 
\begin{figure*}[ht]
 
\centering
\includegraphics[scale=0.9]{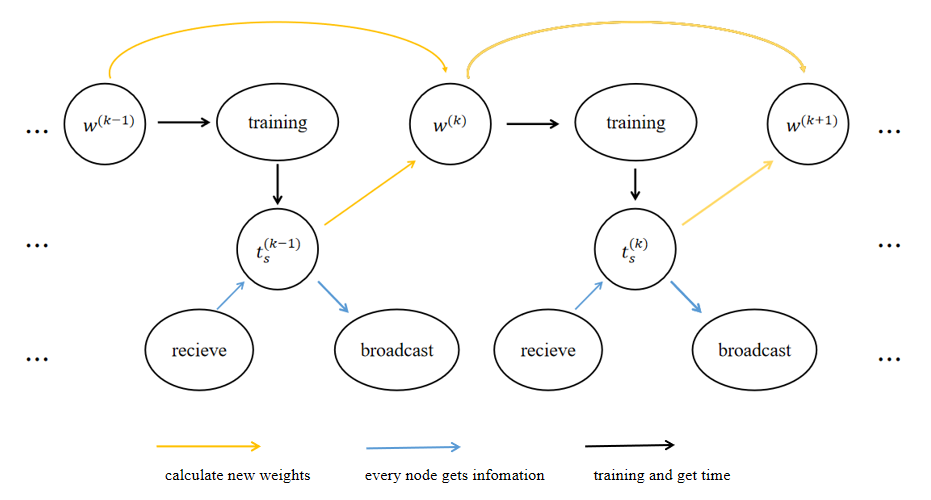}
\caption{the process of self-adaptive allocation algorithm}
\label{figl}
\end{figure*}

The algorithm searched the most suitable ratio of sample distribution through numerical solution. Though repeating transferring samples, it will be steady in few epochs. Therefore, redistributing will stop in later epochs and worker computes based on specific $w_{i}$ in almost epochs. The algorithm will revert to static allocation. This algorithm is flexibly suitable for Ring Allreduce and its variants. The only thing should be taken care of depends on parameters in loss function. Most variants are designed on partial Allreduce or partial gradient aggregation. Therefore, changing tasks can't influence the whole procedure.

\section{EXPERIMENTS}
In this section, we developed a series of experiments on static allocation and self-adaptive allocation to prove that the idea on task allocation will accelerate network training. We first tested the influence of static allocation on convergence to ensure whatever the ratio of samples has been changed, models of network are unaffected in terms of convergence. It guarantees the feasibility of task allocation. Then, we manually adjusted the ratio of samples to prove training time changes with the ratio. It reflected ratio of samples is the factor of training speed indeed. Third, we demonstrated self-adaptive allocation can search the ratio to speed network training up. It shows that equal ratio is not the best ratio of samples on training speed. Existing the better ratio accelerate the convergence of models. Fourth, we added one node in cluster or changed one to display self-adaptive allocation can eliminate part of the impact to tolerant heterogeneity on synchronization. Finally, we compare self-adaption results to other algorithms in special situations. Experimental and analysis results show self-adaption is suitable for complex heterogeneous environment. 
\subsection{experiment setup}
Due to the limitation of hardware resources, we used 2-3 machines in different experiments to illustrate the situation. We uses two machines configured with RTX2080ti and Tesla v100 respectively as well as one machine configured with RTX2080ti and GTX1080ti to do static allocation experiments. We uses three machines configured with 2*RTX2080ti and Tesla v100 respectively to get results of self-adaptive algorithm. In table 1, models, network and GPUs in these experiments are displayed. 
\begin{table}[hbpt]
\centering
\subtable[group 1]{ 
\begin{tabular}{|p{2cm}|p{5cm}|}
\hline
Processor & Intel(R) Xeon(R) Gold 5117 CPU @ 2.00GHz \\
\hline
GPUs & Tesla V100 \\
\hline
Network & Broadcom Inc. and subsidiaries NetXtreme BCM5720 Gigabit Ethernet PCIe\\
\hline
\end{tabular}
}
\subtable[group 2(we have two same machines)]{ 
\begin{tabular}{|p{2cm}|p{5cm}|}
\hline
Processor & Intel(R) Xeon(R) Gold 5117 CPU @ 2.00GHz \\
\hline
GPUs & RTX2080ti \\
\hline
Network & Broadcom Inc. and subsidiaries NetXtreme BCM5720 Gigabit Ethernet PCIe\\
\hline
\end{tabular}
}
\subtable[group 3]{ 
\begin{tabular}{|p{2cm}|p{5cm}|}
\hline
Processor & Intel(R) Xeon(R) Bronze 3104 CPU @ 1.70 GHz \\
\hline
GPUs & GTX 1080ti and RTX2080ti \\
\hline
Network & Intel Corperation Ethernet Connection (3) I219-LM\\
\hline
\end{tabular}
}
\caption{experiment setup}
\end{table}

\subsection{Dataset And network}
Experiments are conducted on some classical datasets and network models. We selected MNIST and CIFAR10 dataset to train network models. We train three-layer convolutional neural network called ConvNet on MNIST handwritten digit recognition dataset. The model contains two convolutional layers , two maxpooling layers and one fully connected layer. We train VGG11, VGG16, VGG19, ResNet18 and Resnet50 models on CIFAR10 to observe allocation algorithm. The configuration of the models is equal basically. The learning rate is taken to the negative 2 order of 10 and $weight \uline{~}decay=10^{-4}$. Total batchsize is controled from $2000$ to $3000$. Minibatchsize depends on the ratio of samples. For $\sum_{i=1}^n w_{i}=C$, $C*minibatch =total\ batchsize$. To precisely control the ratio of samples $w$, the value range of $C$ can be from $20$ to $30$. In the later experiment, we set different value of $w$. We will explain details at that time.   
\subsection{Effects of static ratio on convergence and velocity}
For the purpose of testing different tasks allocation on different workers, we designed experiments to confirm smaller impacts on convergence and larger impacts on training speed. We did experiments with one machine with multiple cards and multiple machines with multiple cards. 

Above all, we train ConvNet on MNIST dataset, ResNet18, ResNet50 and VGG11 on CIFAR10 dataset with one machine with multiple cards. The ratio of samples $w_{i}$ was set to four groups including equal and unequal tasks. Except the ratio, other variables are the same. Minibatchsize is equivalent to $100$, total batchsize is equivalent to $1000$, learning rate is equivalent to $10^{-2}$, and weight decay is equivalent to $10^{-4}$. As figure 6 shows, the same network models with different ratio converge to same points approximately. No matter it is accuracy or training epochs, there will be no big ups and downs. Therefore, it is credible that convergence will never change due to the ratio changing.
\begin{figure}
\centering
\subfigure[ResNet18-epoch]{
\begin{minipage}[t]{0.45\linewidth}
\includegraphics[width=1.5in]{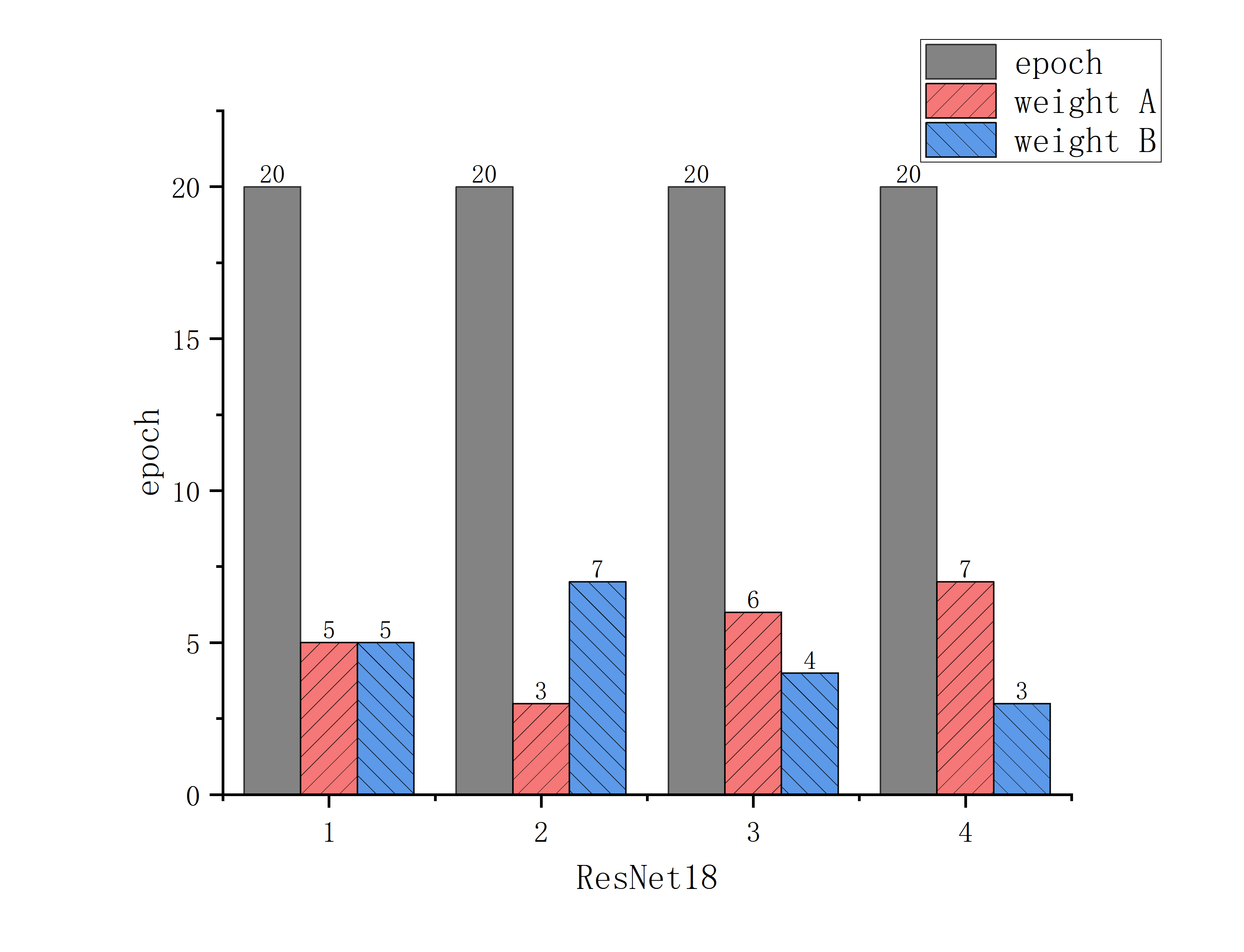}
\end{minipage}
}
\subfigure[ResNet18-accuracy]{
\begin{minipage}[t]{0.45\linewidth}
\centering
\includegraphics[width=1.5in]{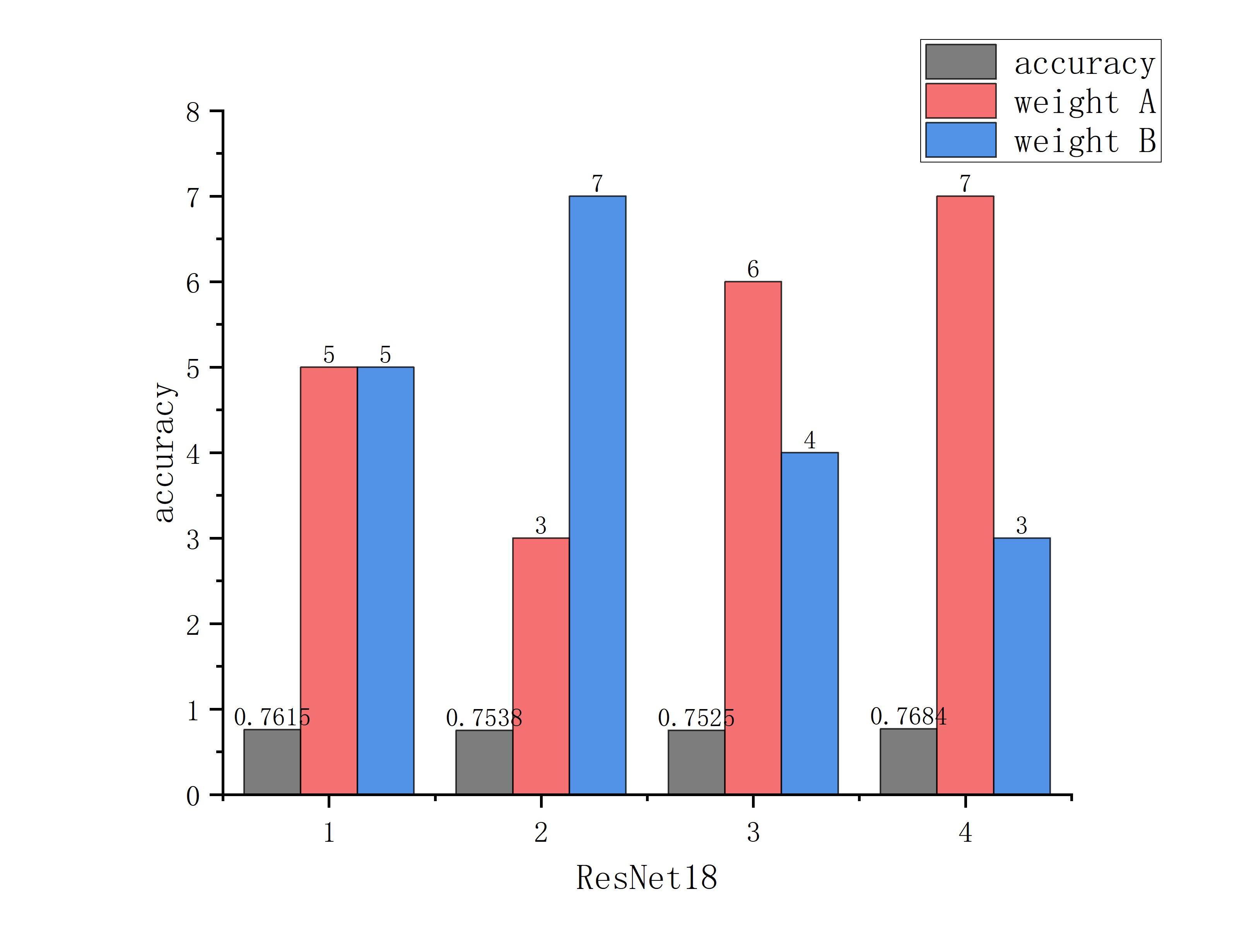}
\end{minipage}
}

\subfigure[ResNet50-epoch]{
\begin{minipage}[t]{0.45\linewidth}
\includegraphics[width=1.5in]{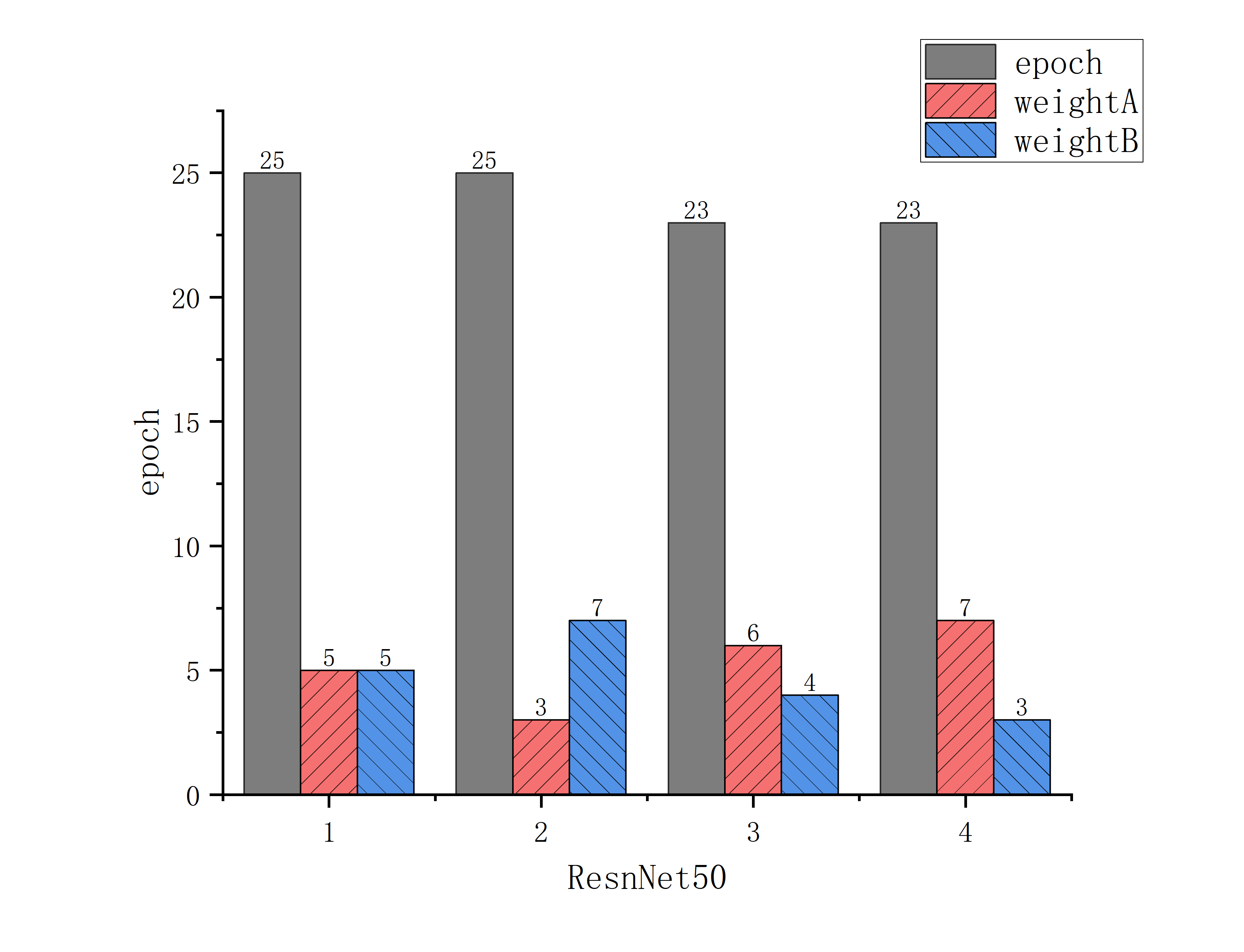}
\end{minipage}
}
\subfigure[ResNet50-accuracy]{
\begin{minipage}[t]{0.45\linewidth}
\centering
\includegraphics[width=1.5in]{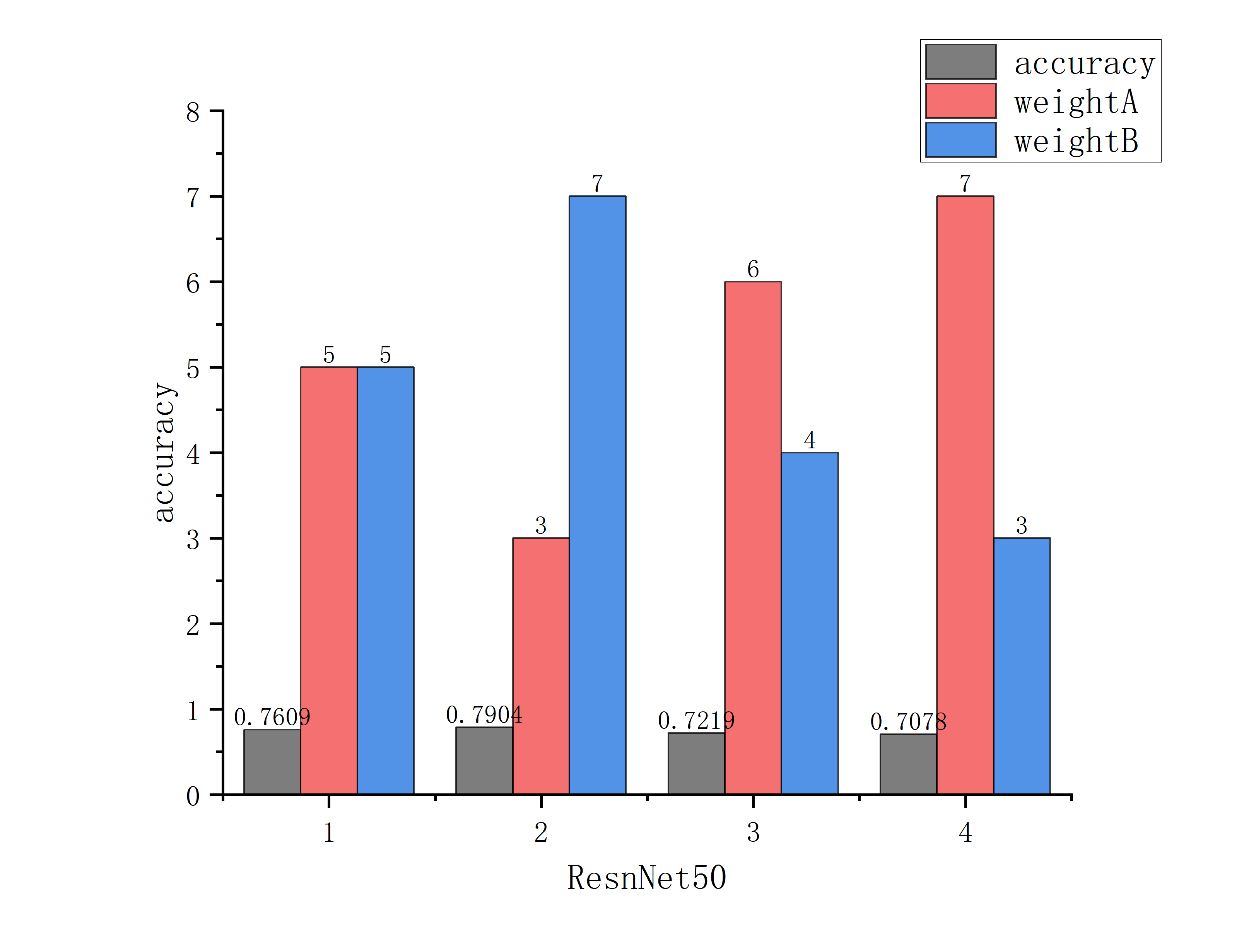}
\end{minipage}
}

\subfigure[ConvNet-epoch]{
\begin{minipage}[t]{0.45\linewidth}
\includegraphics[width=1.5in]{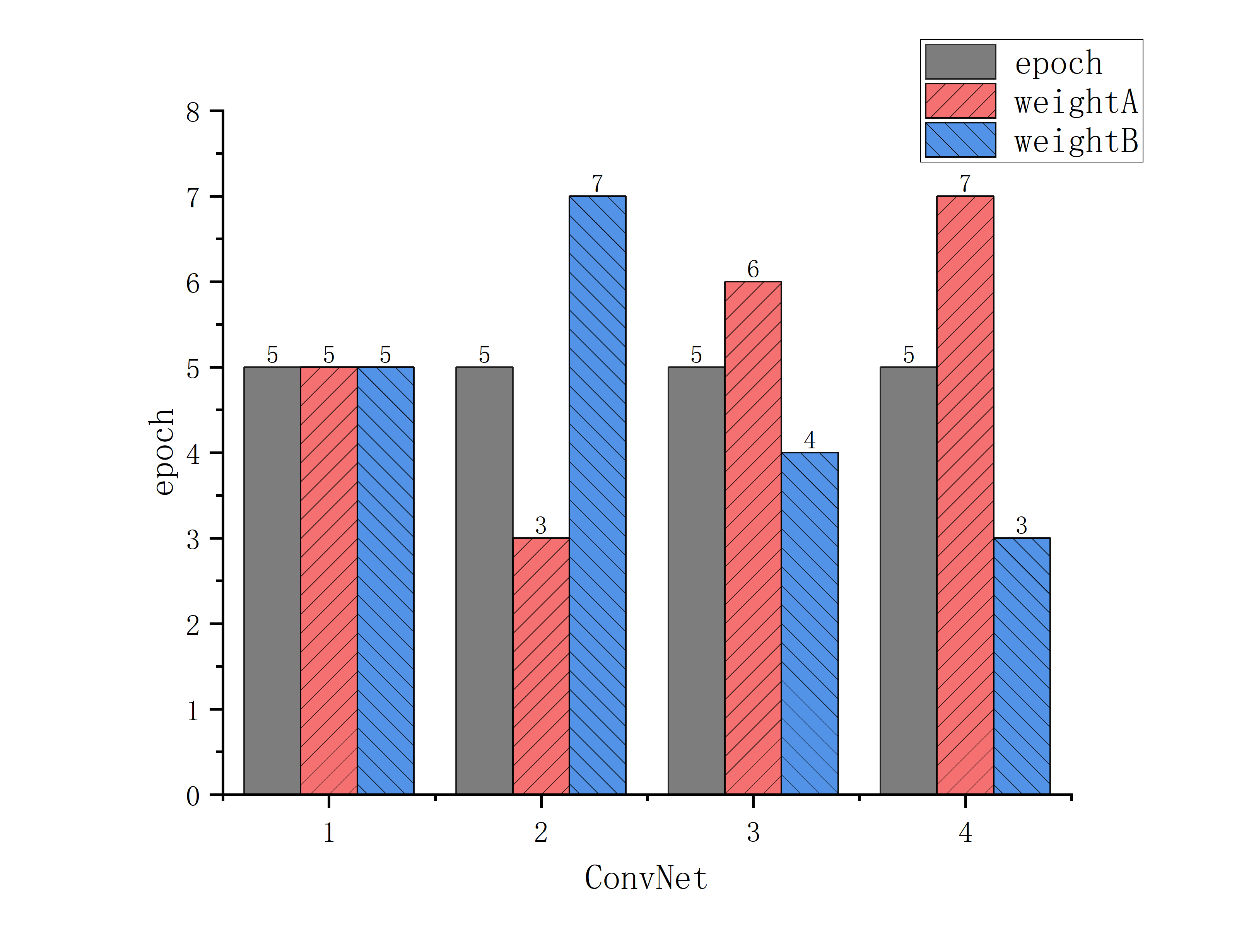}
\end{minipage}
}
\subfigure[ConvNet-accuracy]{
\begin{minipage}[t]{0.45\linewidth}
\centering
\includegraphics[width=1.5in]{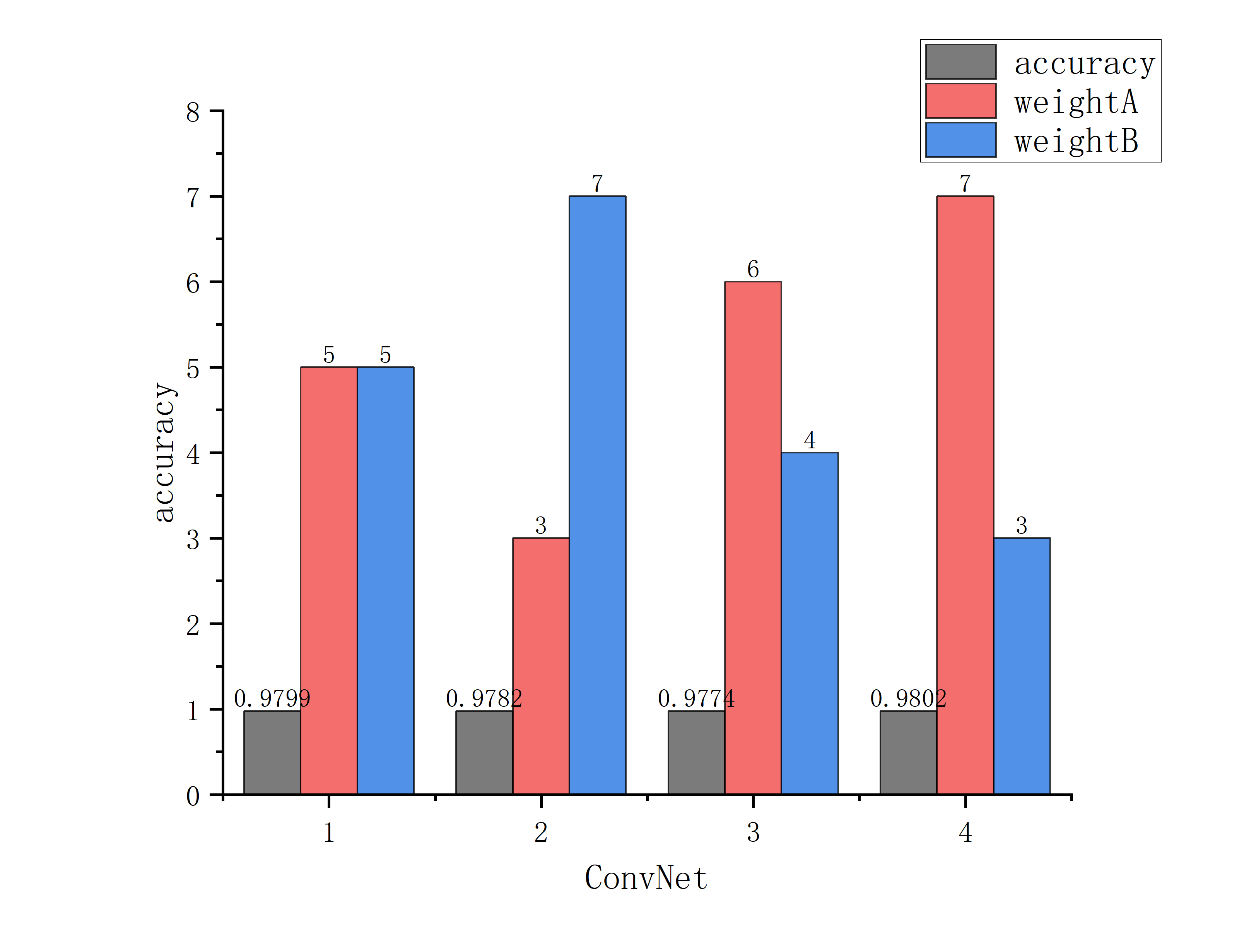}
\end{minipage}
}

\subfigure[VGG11-epoch]{
\begin{minipage}[t]{0.45\linewidth}
\includegraphics[width=1.5in]{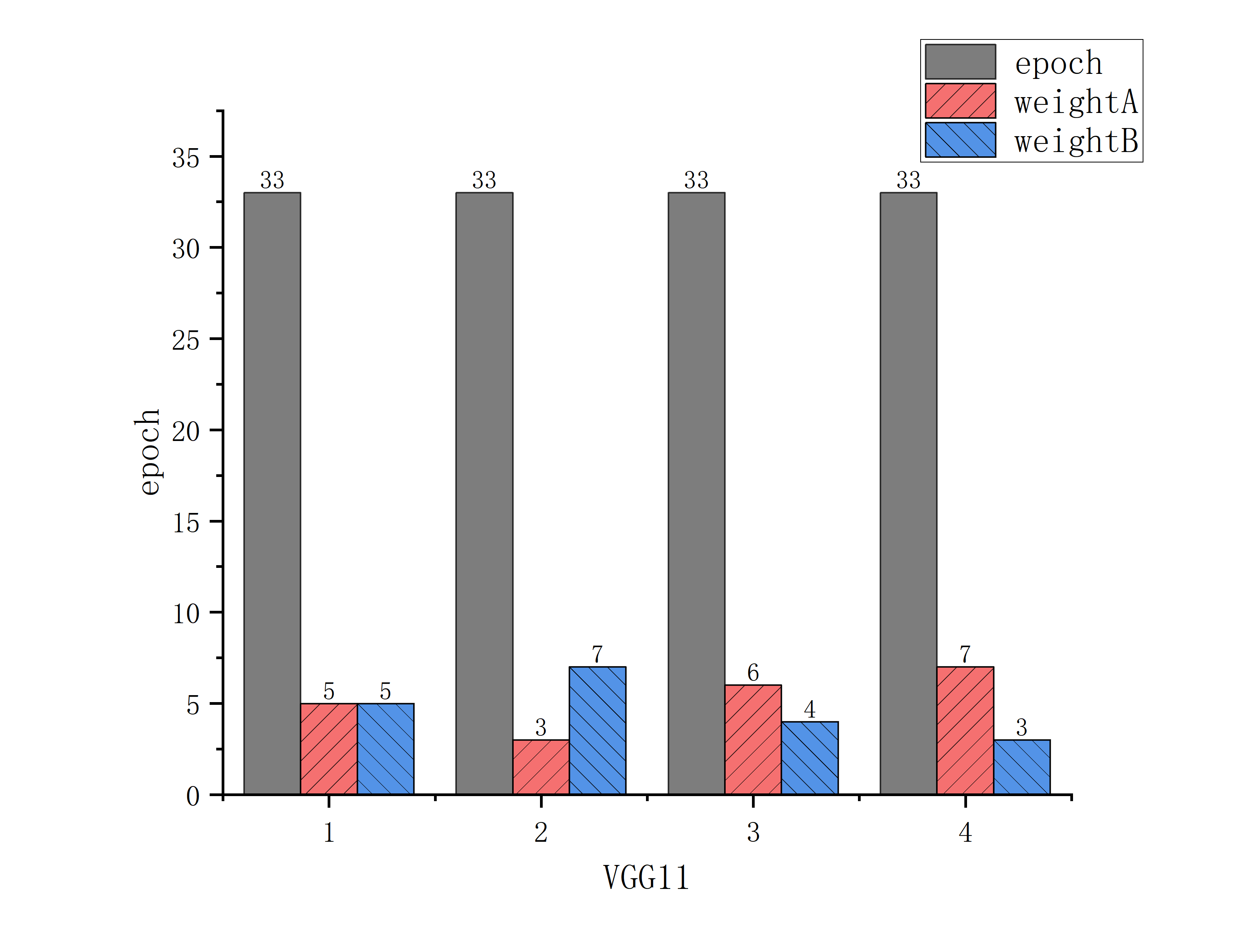}
\end{minipage}
}
\subfigure[VGG11-accuracy]{
\begin{minipage}[t]{0.45\linewidth}
\centering
\includegraphics[width=1.5in]{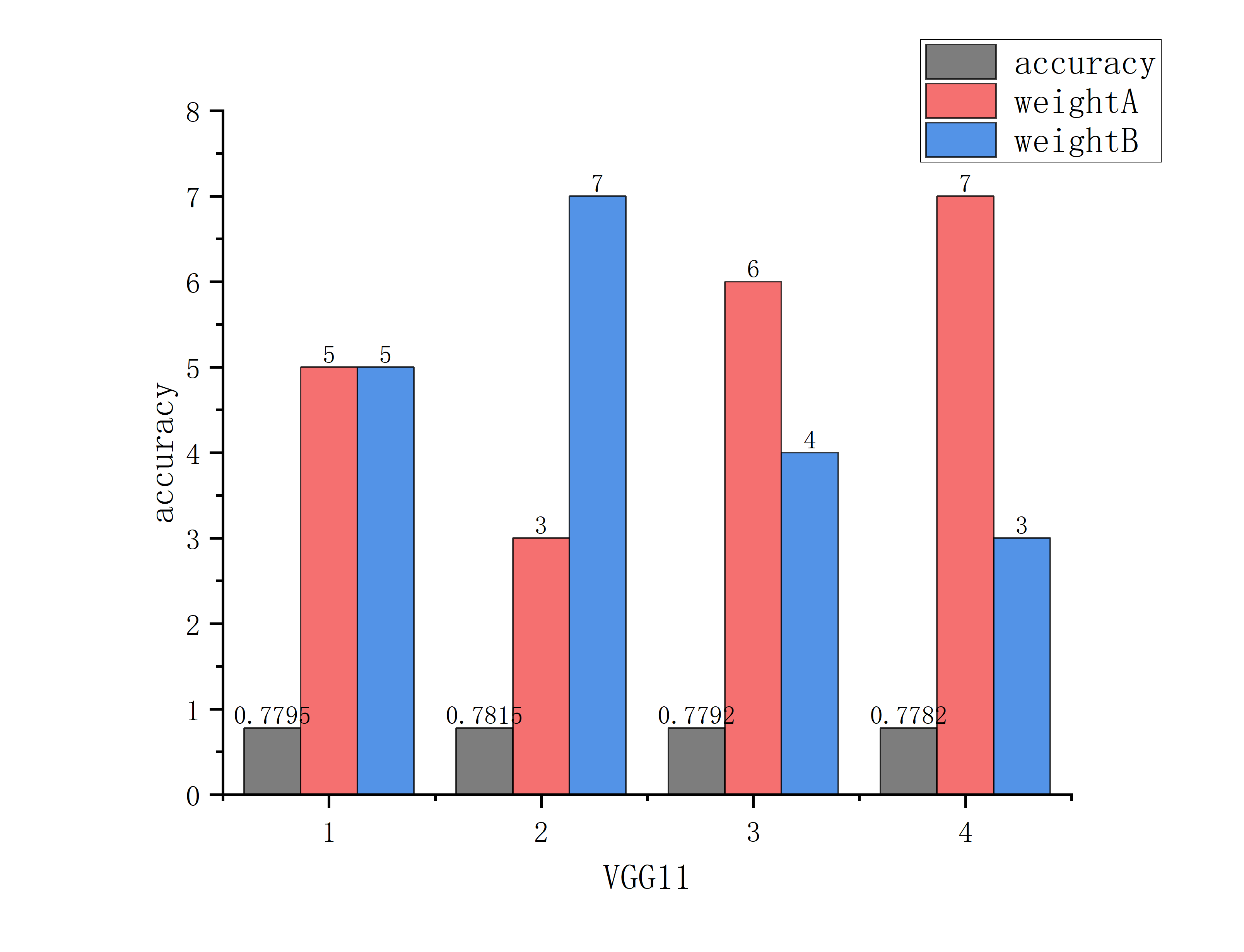}
\end{minipage}
}

\centering
\caption{the convergence data of models results in 4 groups $5:5,6:4,3:7,7:3$ from one machine with GTX1080ti and RTX2080ti. The gray pillars in pictures are epochs and accuracy.}
\end{figure}

\begin{figure}
\centering
\subfigure[ResNet18-time]{
\begin{minipage}[t]{0.45\linewidth}
\includegraphics[width=1.5in]{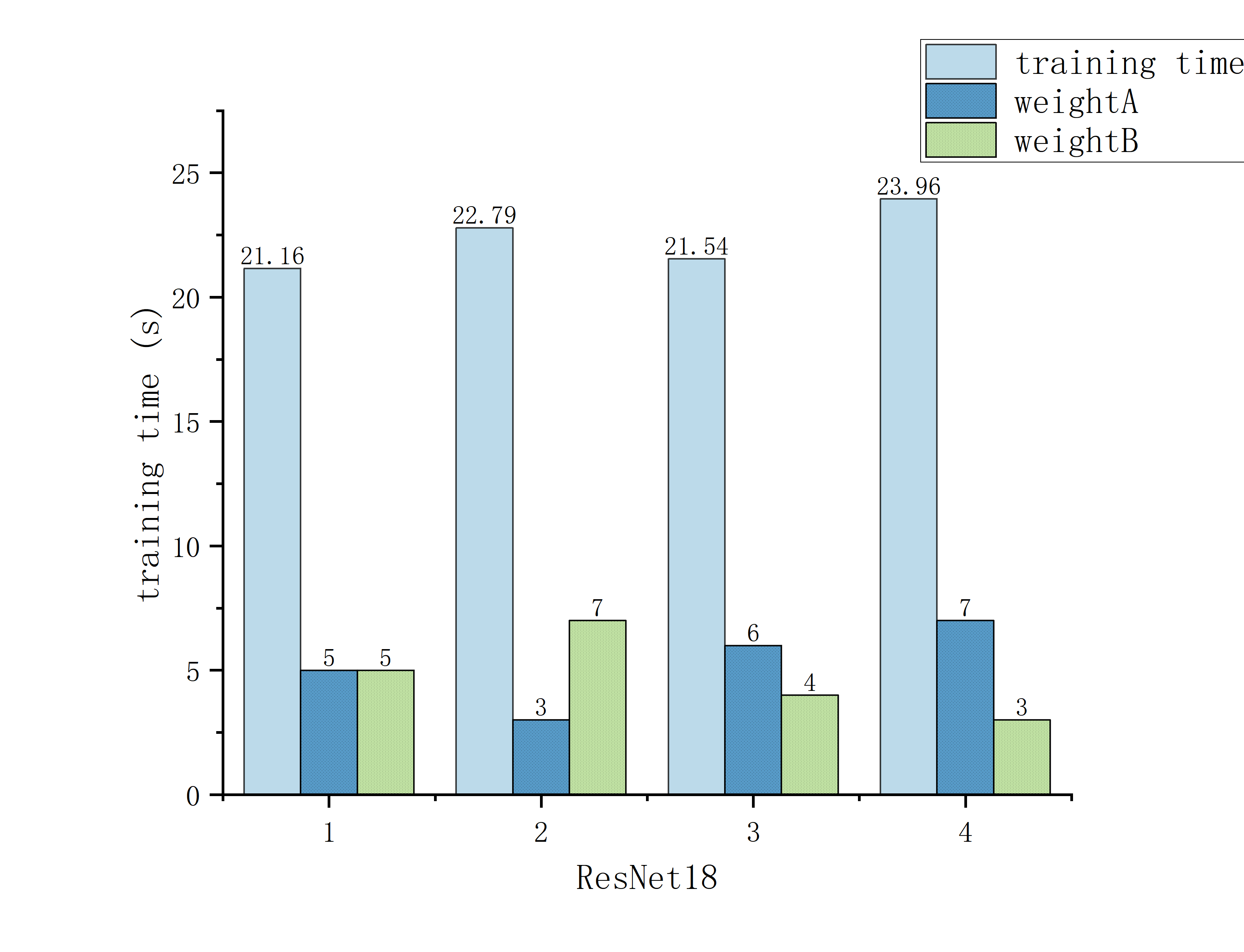}
\end{minipage}
}
\subfigure[ResNet50-time]{
\begin{minipage}[t]{0.45\linewidth}
\centering
\includegraphics[width=1.5in]{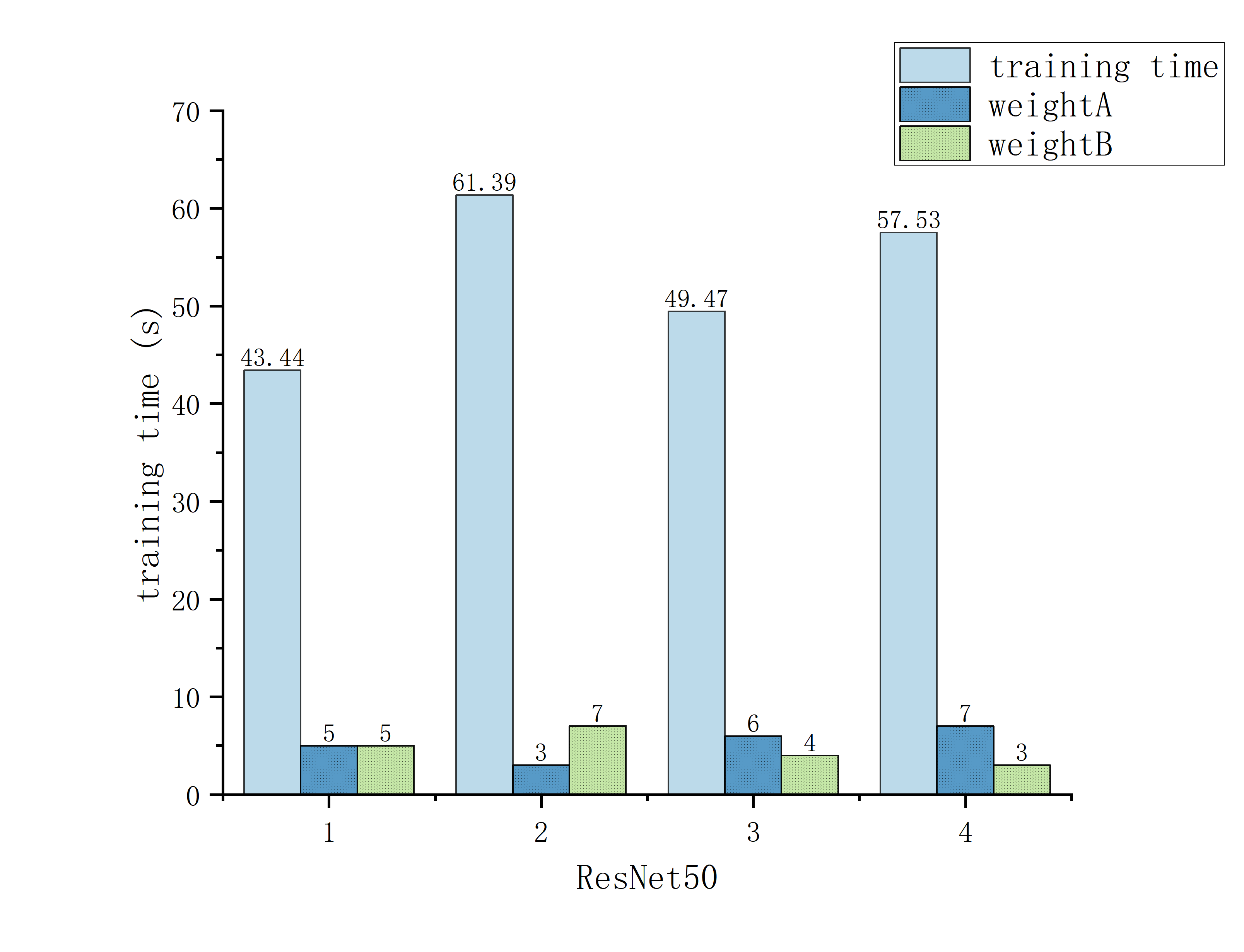}
\end{minipage}
}

\subfigure[ConvNet-time]{
\begin{minipage}[t]{0.45\linewidth}
\includegraphics[width=1.5in]{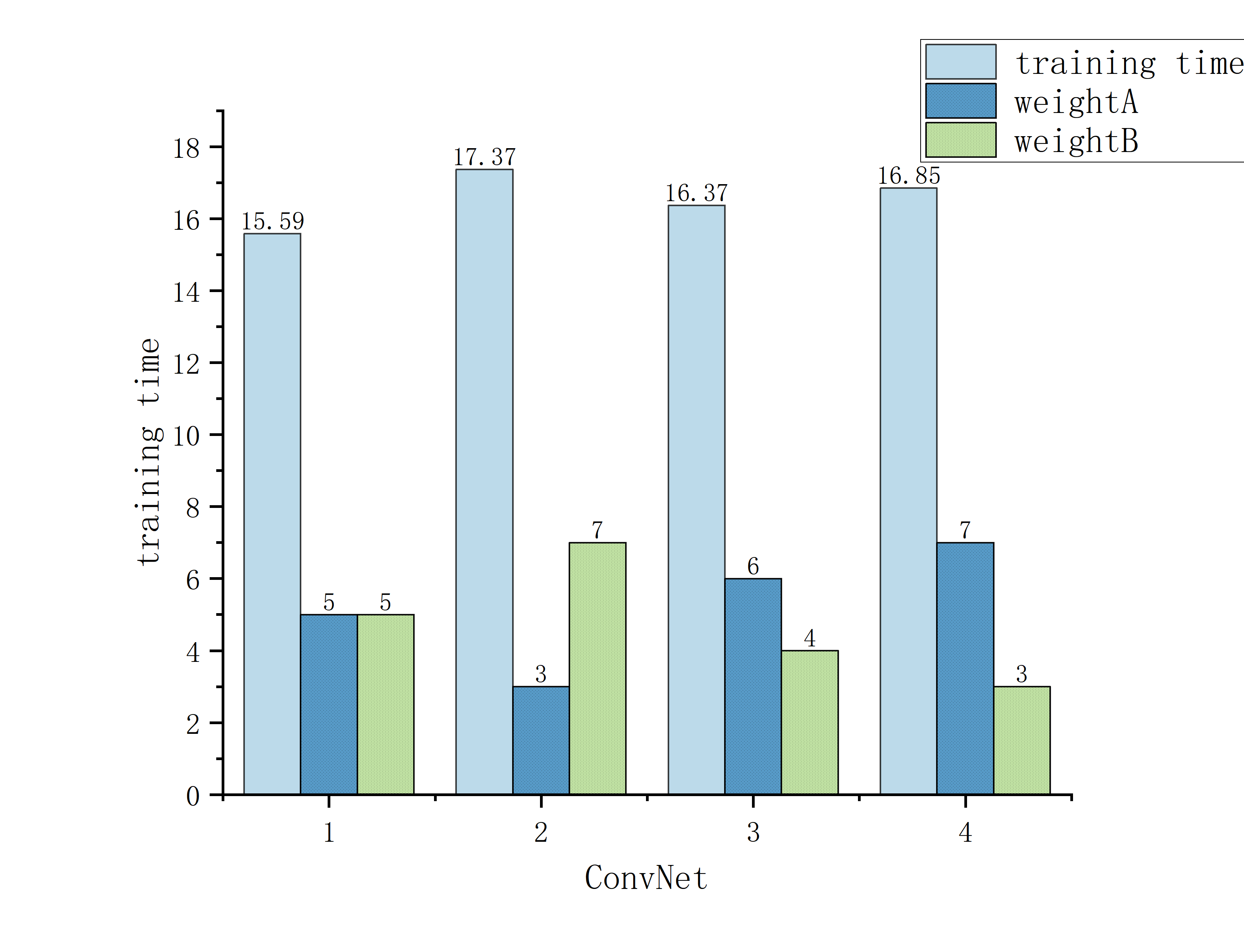}
\end{minipage}
}
\subfigure[VGG11-time]{
\begin{minipage}[t]{0.45\linewidth}
\centering
\includegraphics[width=1.5in]{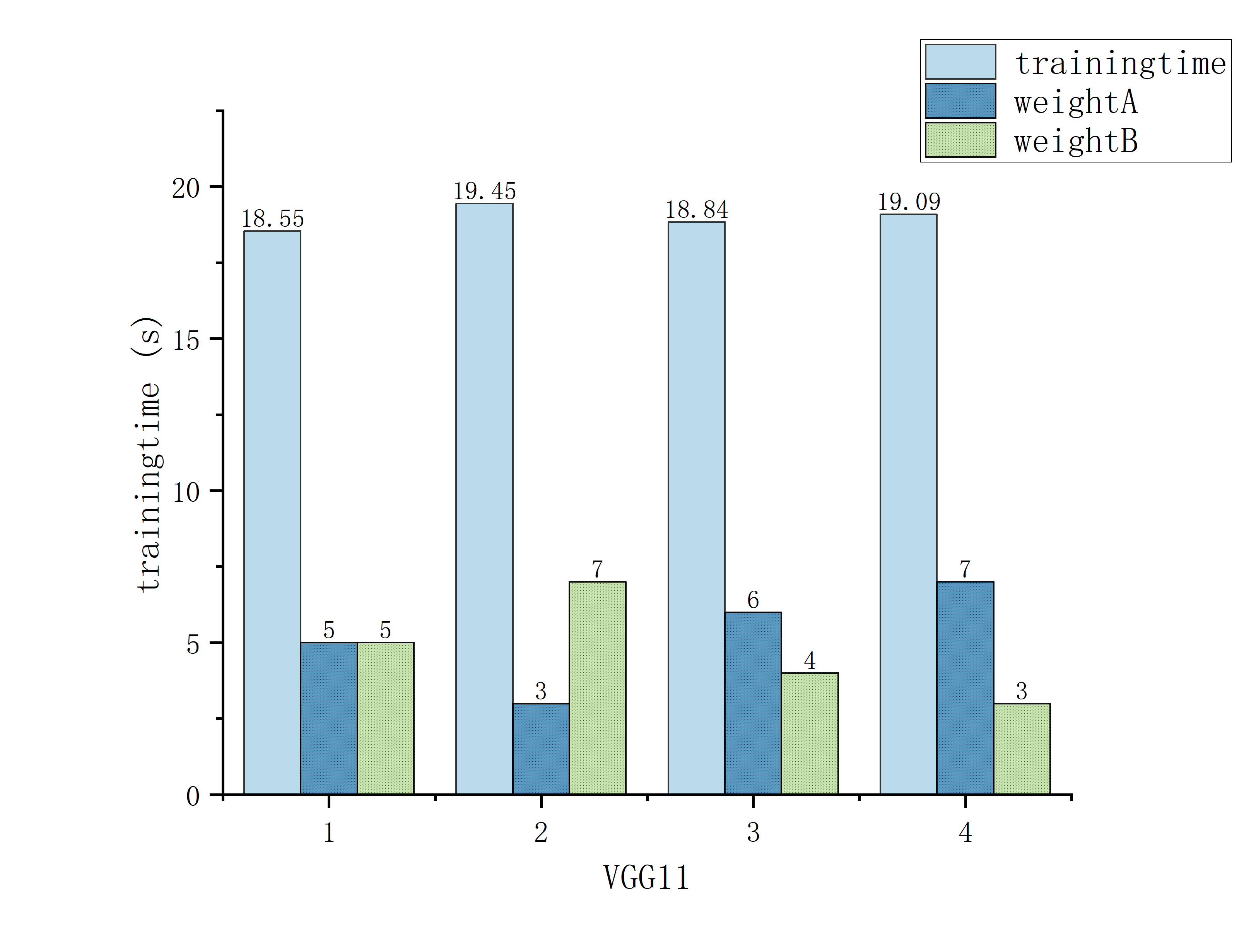}
\end{minipage}
}

\centering
\caption{the training time of models results in 4 groups $5:5,6:4,3:7,7:3$ from one machine with GTX1080ti and RTX2080ti.}
\end{figure}

\begin{figure}
\centering
\subfigure[ResNet18-time]{
\begin{minipage}[t]{0.45\linewidth}
\includegraphics[width=1.5in]{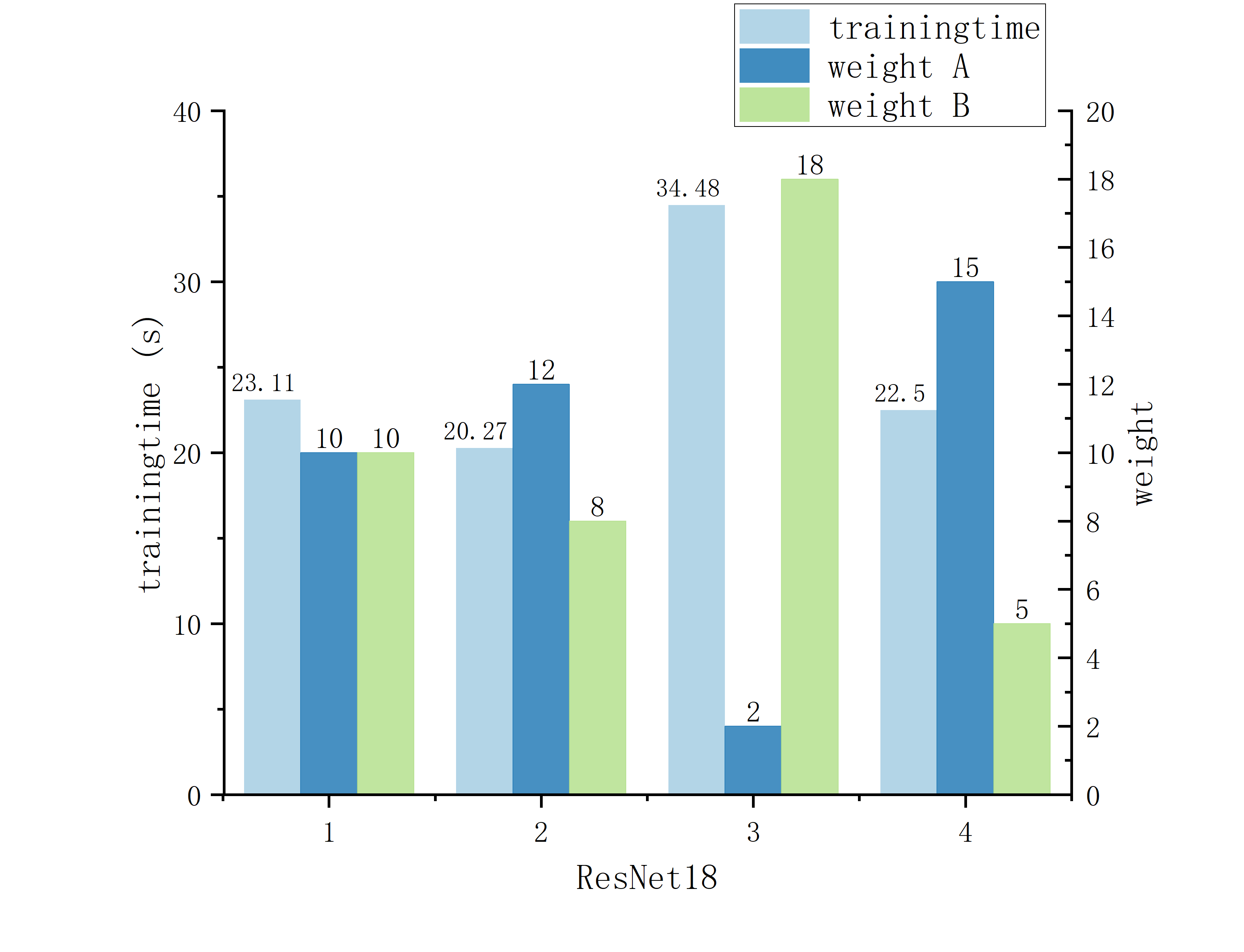}
\end{minipage}
}
\subfigure[ResNet50-time]{
\begin{minipage}[t]{0.45\linewidth}
\centering
\includegraphics[width=1.5in]{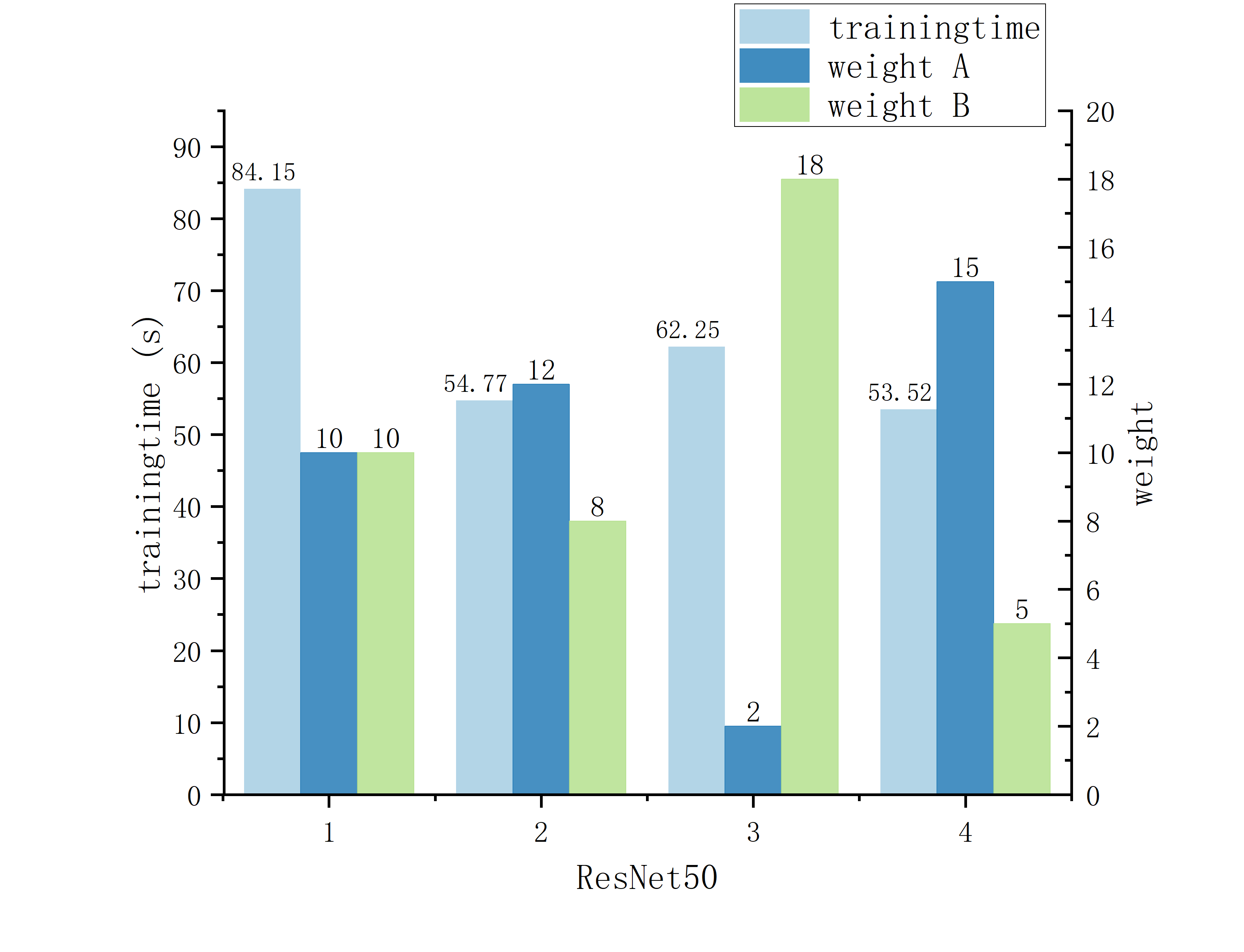}
\end{minipage}
}

\subfigure[ConvNet-time]{
\begin{minipage}[t]{0.45\linewidth}
\includegraphics[width=1.5in]{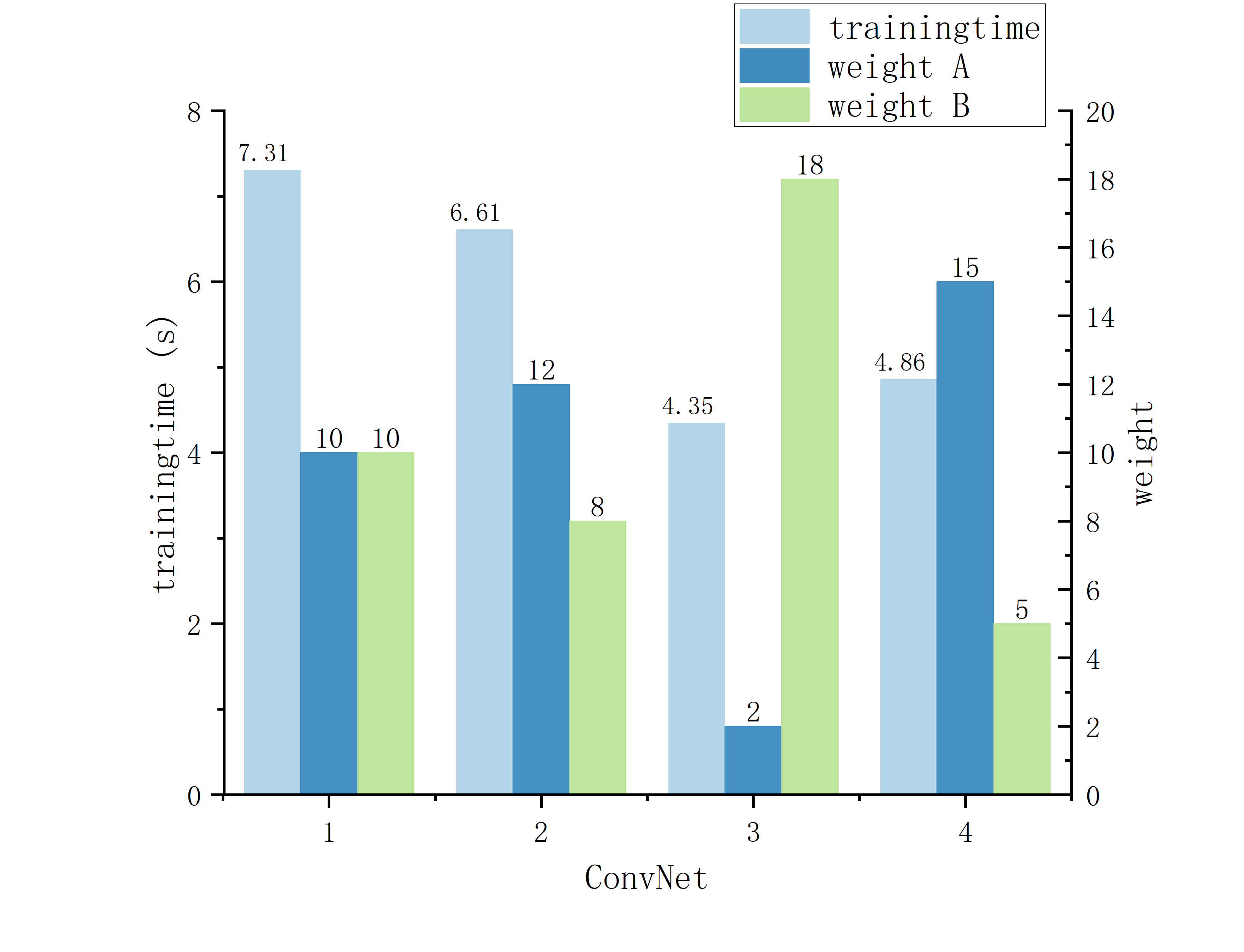}
\end{minipage}
}
\subfigure[VGG16-time]{
\begin{minipage}[t]{0.45\linewidth}
\centering
\includegraphics[width=1.5in]{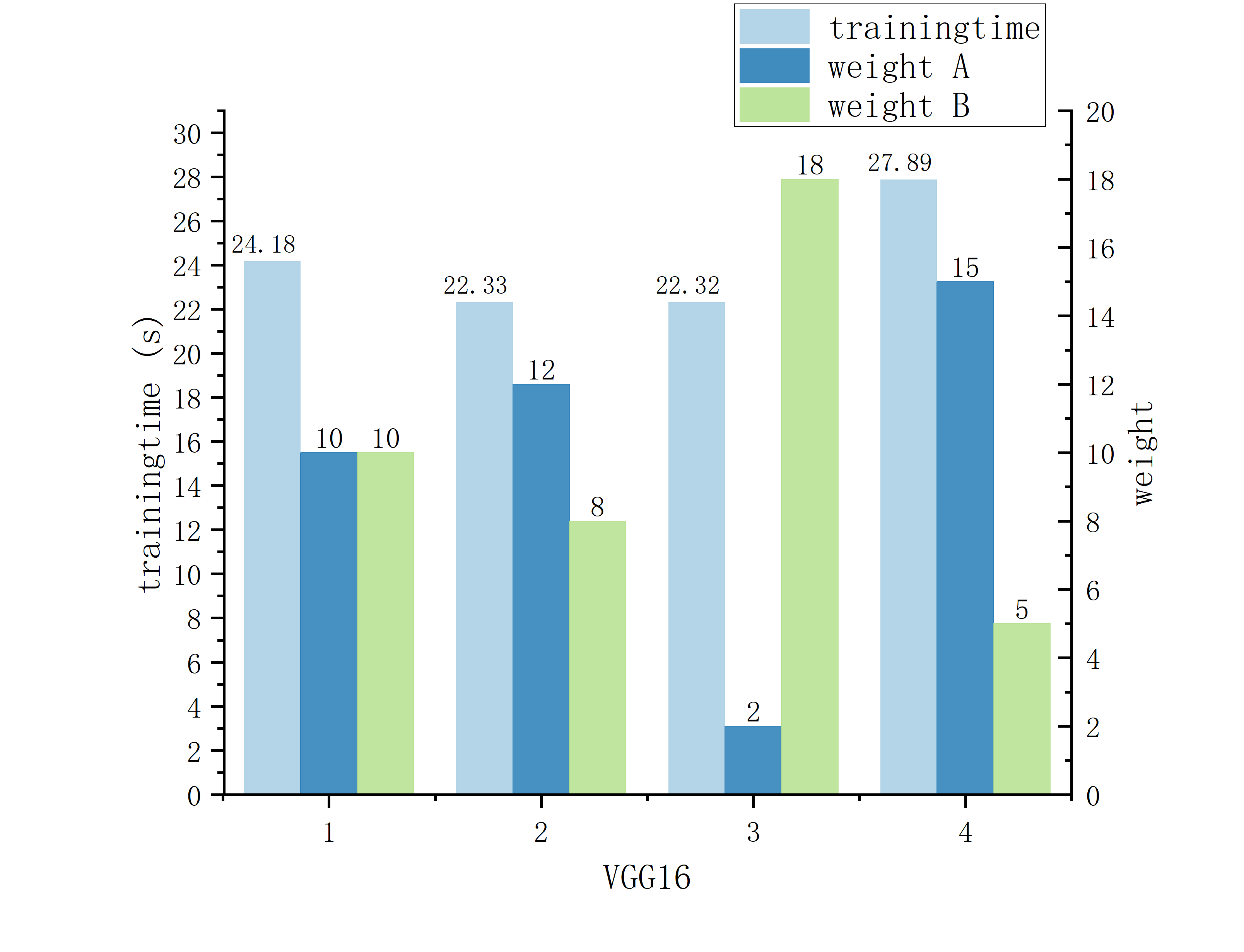}
\end{minipage}
}

\centering
\caption{the training time of models results in 4 groups $10:10,12:8,2:18,15:5$ from two machine with V100 and RTX2080ti.}
\end{figure}
Besides, we expand the sum of ratio and reduce the minibatchsize to show the variety of training speed. We do two sets of experiments with one machine with multiple cards and multiple machines with multiple cards respectively. The first set of experiment is implemented on Intel(R) Xeon(R) Bronze 3104 CPU @ 1.70 GHz with RTX2080ti and GTX1080ti. The second set of experiment is implemented on two Intel(R) Xeon(R) Gold 5117 CPU @ 2.00GHz with Tesla V100 and RTX1080ti. The ratio was set to four groups too. As figure 7 and 8 shows, when the performance of GPUs and network is similar, the ratio approaches same. The larger difference of performance between faster worker and slower worker is, the larger ratio of samples is. The results show there are appropriate ratio existing in heterogeneous environment, but it is difficult to search. Therefore, it is necessary to self-adaption.

\subsection{Speedup on self-adaptive allocation}
Further, we do experiments on multiple machines with multiple cards to examine results of self-adaptive allocation algorithm. Multiple machines with multiple cards are obvious in computing and communication performance. We train ResNet18, ResNet50, VGG16, VGG19 and ConvNet models on two nodes with different cards respectively. We record the ratio of samples $w_{1}, w_{2}, \cdots, w_{n}$, Gradient computing time: $t_{s}^{1}, t_{s}^{2}, \cdots, t_{s}^{n}$ and training time $T_{i}$in each epoch. 
In figure 9, training time of models on V100 and RTX2080ti is reduced along with the increasing of epoch. The gap between gradient computing time of two workers becomes smaller too. After 4 epochs, the ratio of samples becomes steady and the algorithm should be stopped. We make two sets of initial ratio of samples to confirm the convergence of the ratio to same point. 
In figure 10, we increased one machine with one card RTX2080ti to research more workers. We found it also became steady after several epochs. The speed of training increased along with epochs.
From the above results, the calculating time of workers approaches to same points, the ratio of weights approaches stability and the training time approaches declining.
\begin{figure}
\centering
\subfigure[ResNet50-$Ts$]{
\begin{minipage}[t]{0.3\linewidth}
\includegraphics[width=1.1in]{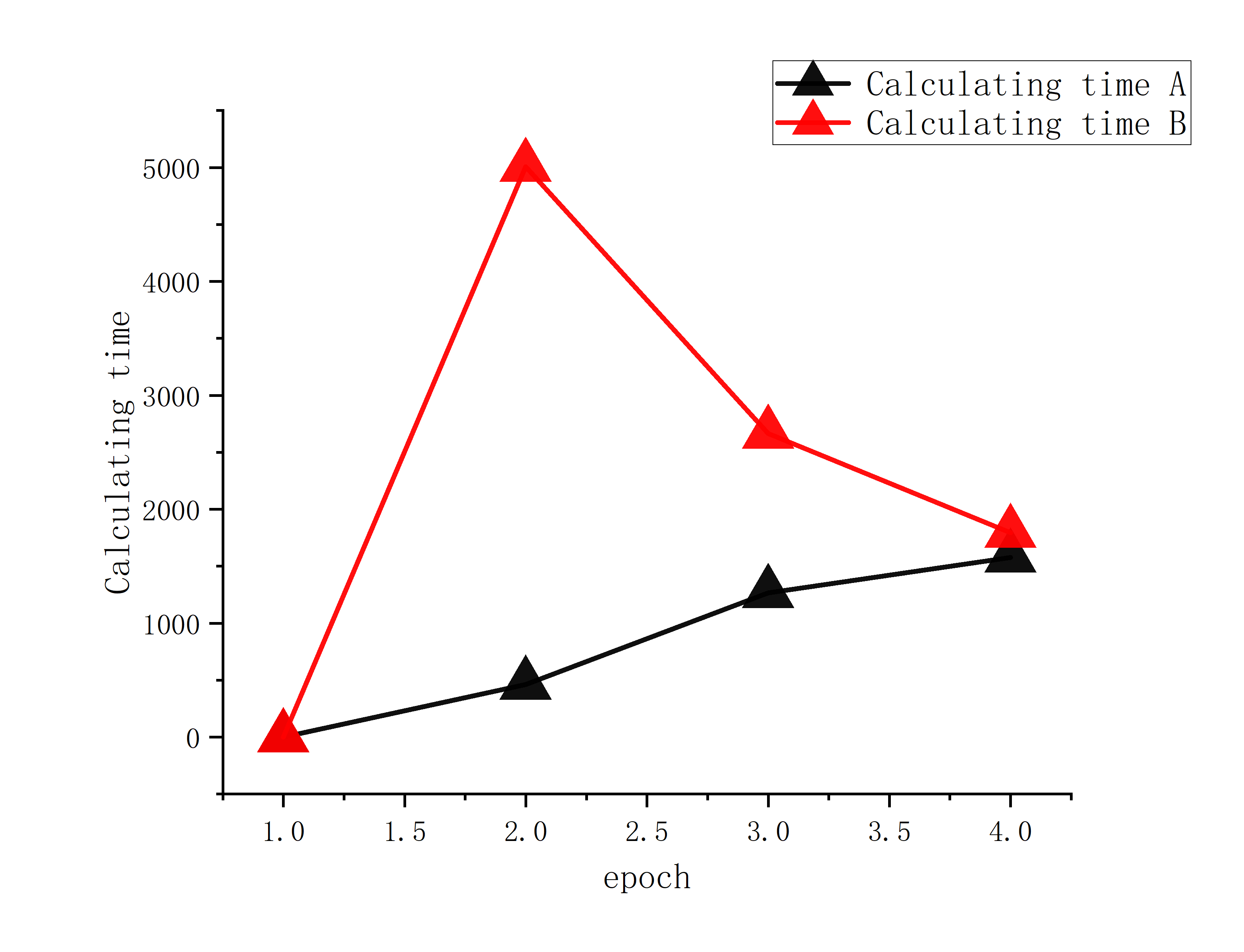}
\end{minipage}
}%
\subfigure[ResNet50-$W$]{
\begin{minipage}[t]{0.3\linewidth}
\centering
\includegraphics[width=1.1in]{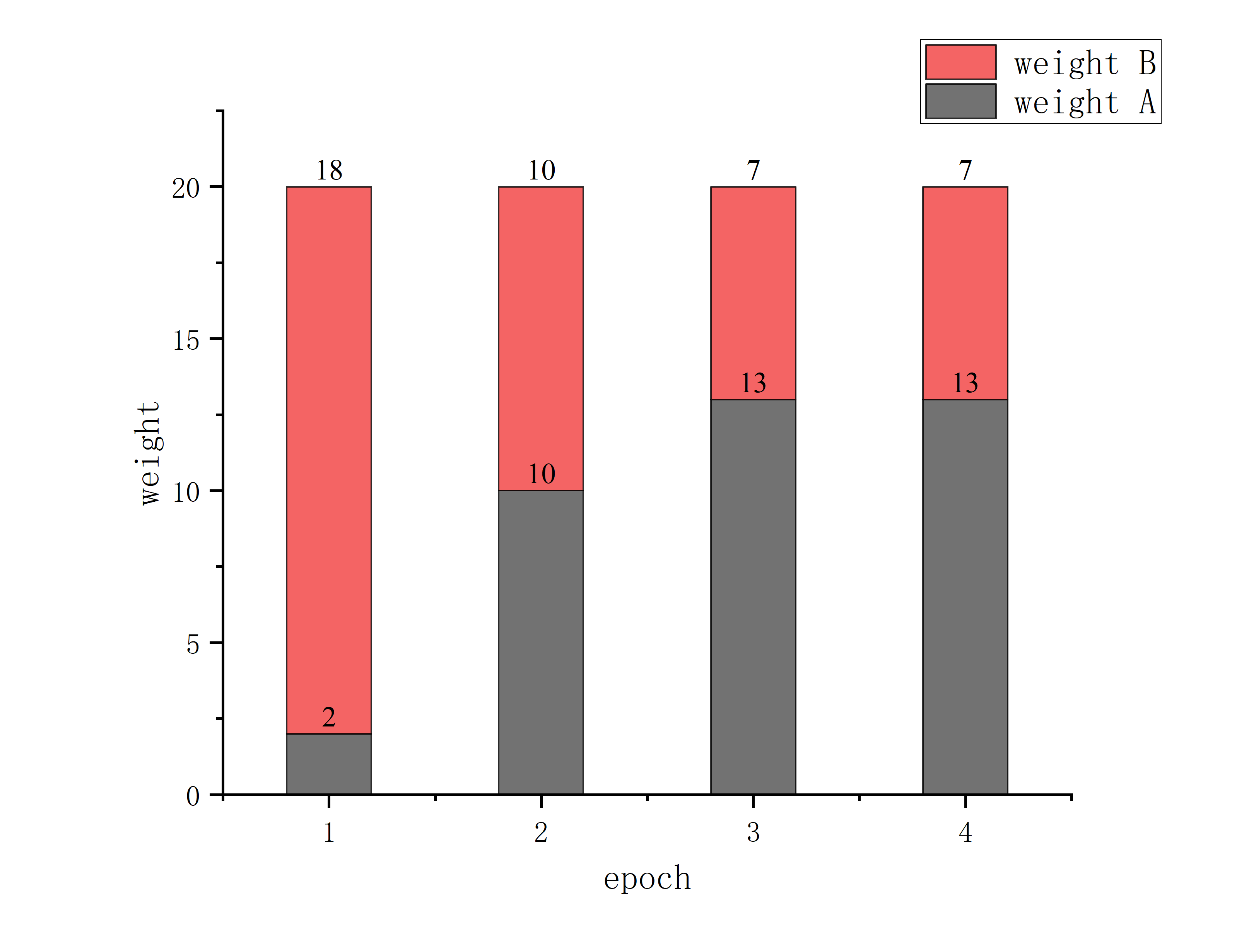}
\end{minipage}
}%
\subfigure[ResNet50-$t$]{
\begin{minipage}[t]{0.3\linewidth}
\includegraphics[width=1.1in]{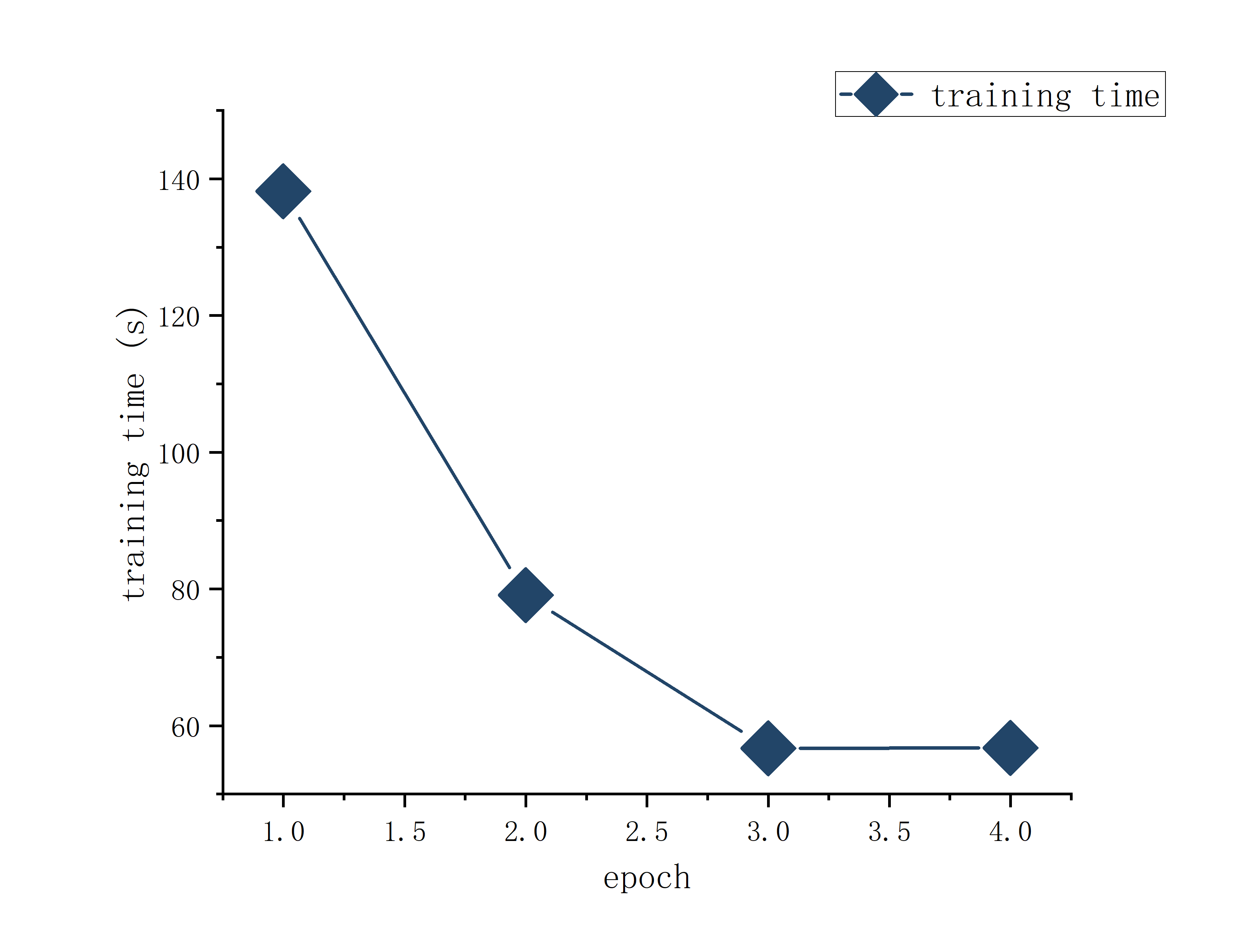}
\end{minipage}
}
\subfigure[ResNet50-$Ts$-$10$]{
\begin{minipage}[t]{0.3\linewidth}
\includegraphics[width=1.1in]{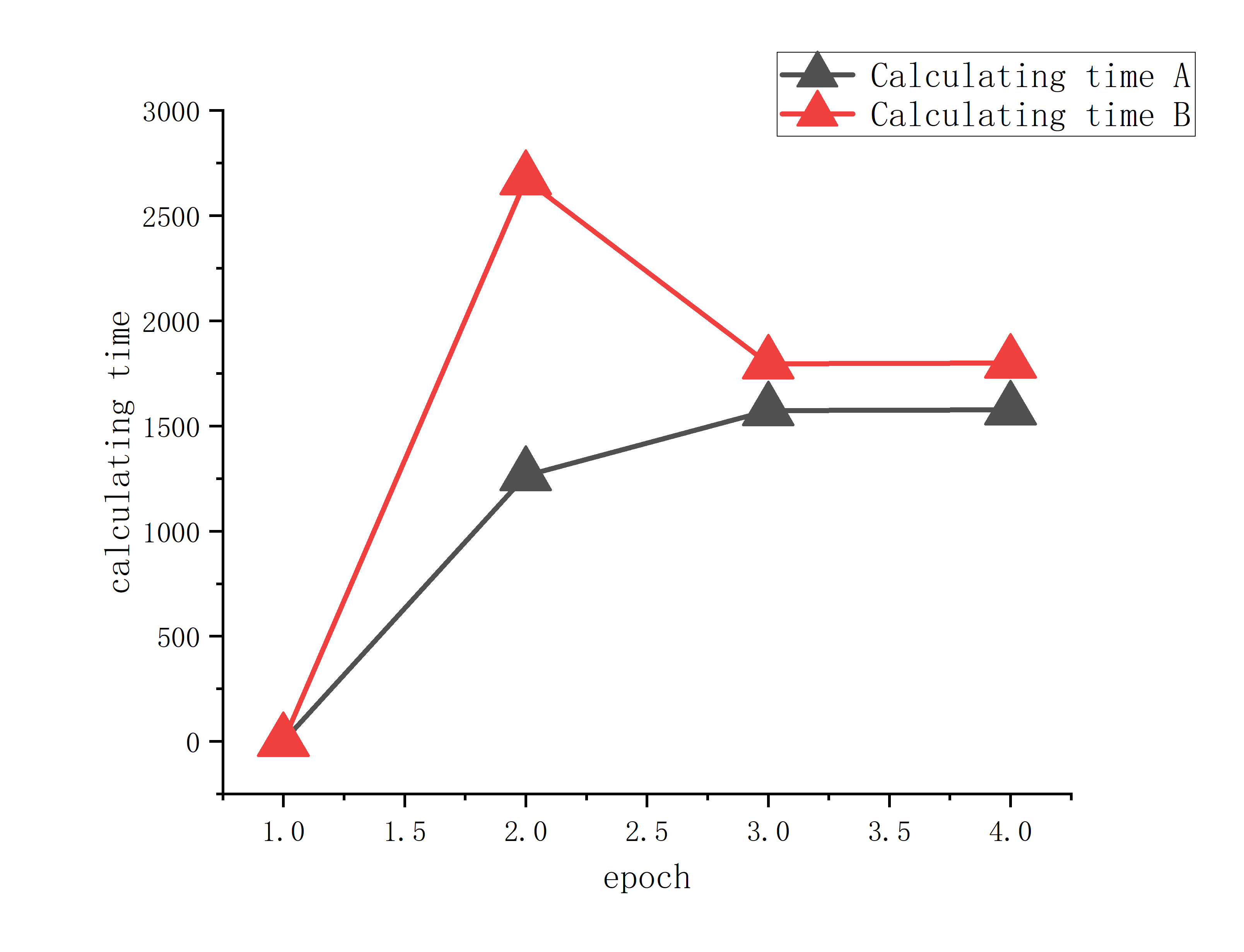}
\end{minipage}
}%
\subfigure[ResNet50-$W$-$10$]{
\begin{minipage}[t]{0.3\linewidth}
\centering
\includegraphics[width=1.1in]{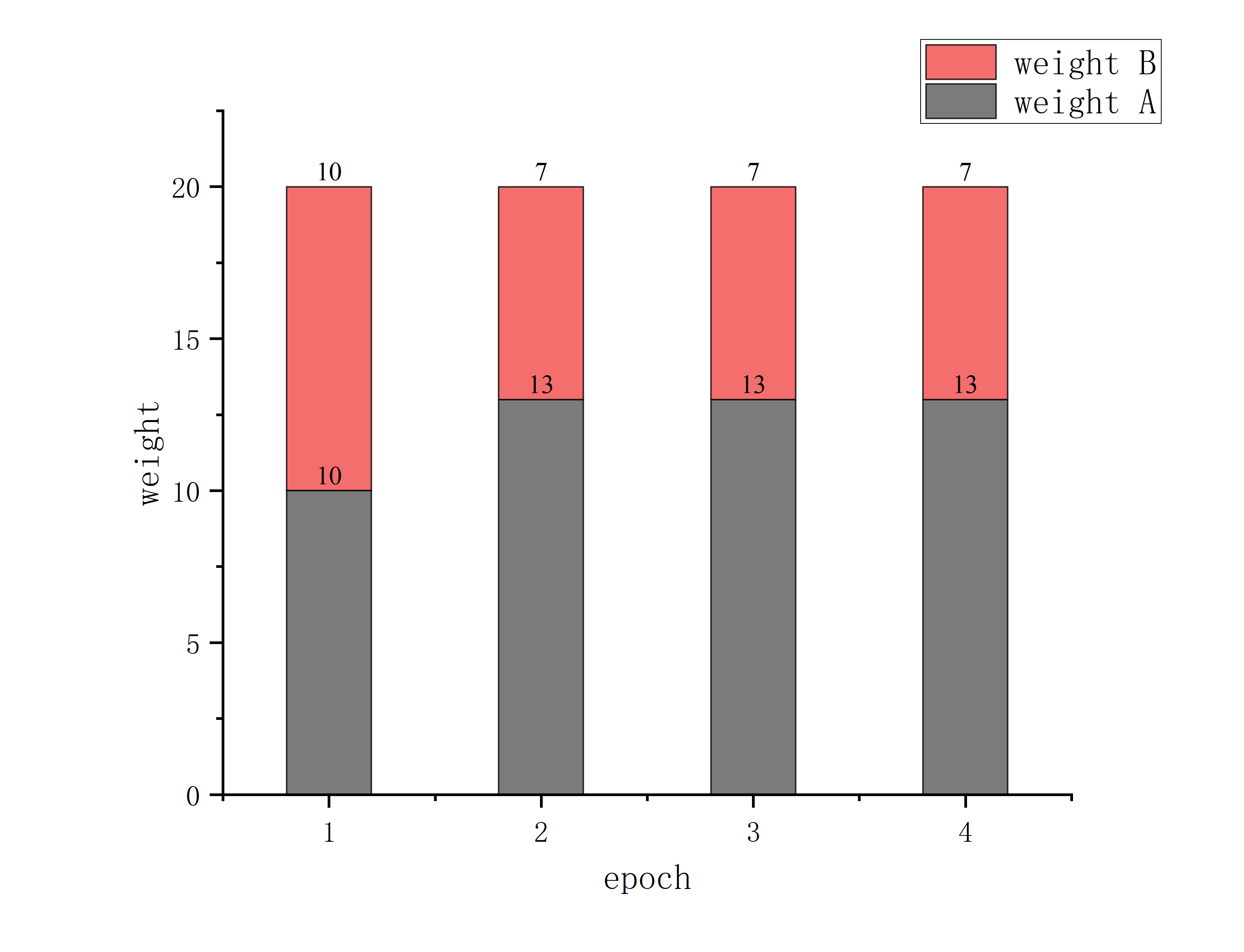}
\end{minipage}
}%
\subfigure[ResNet50-$t$-$10$]{
\begin{minipage}[t]{0.3\linewidth}
\includegraphics[width=1.1in]{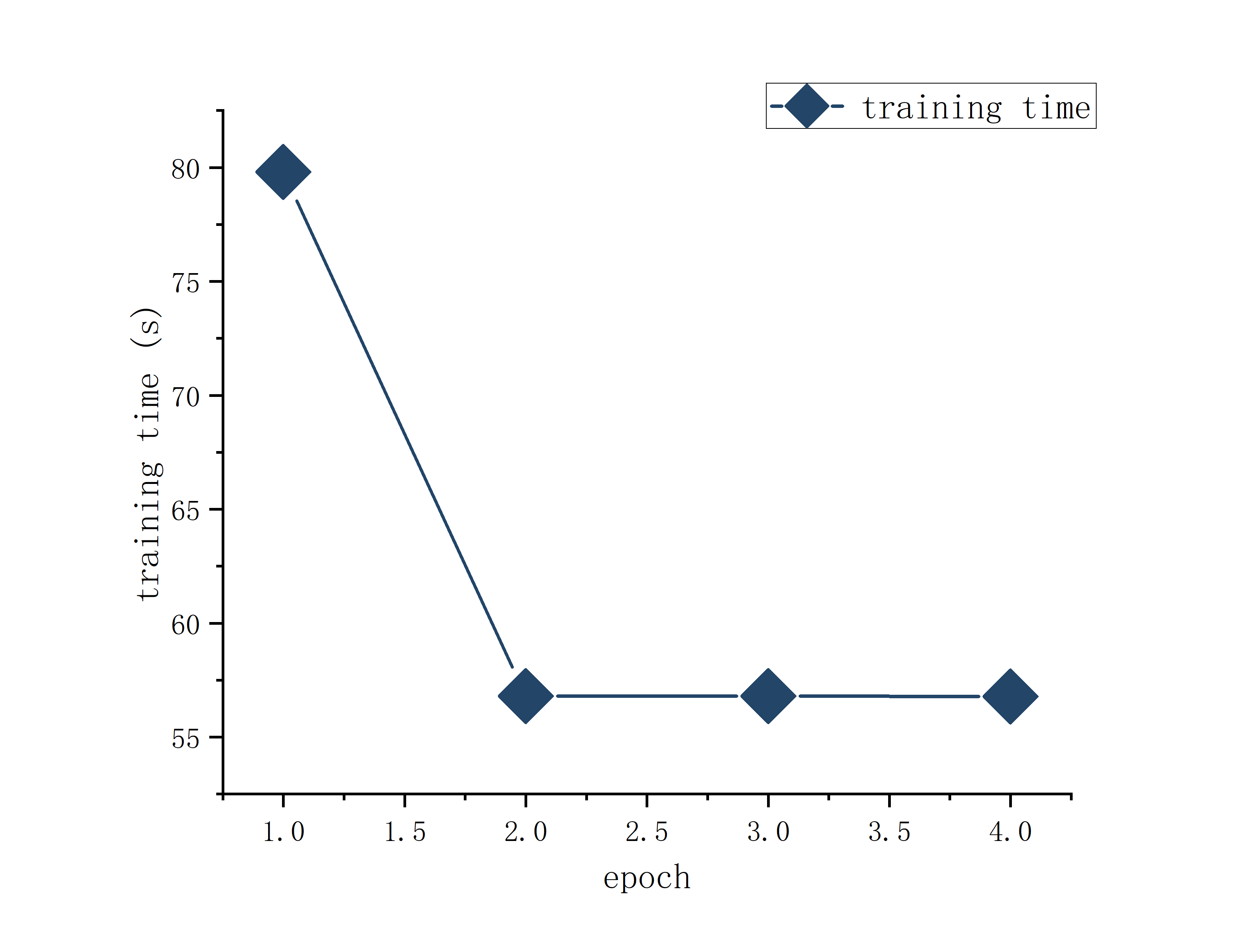}
\end{minipage}
}
\subfigure[VGG16-$Ts$-$10$]{
\begin{minipage}[t]{0.3\linewidth}
\includegraphics[width=1.1in]{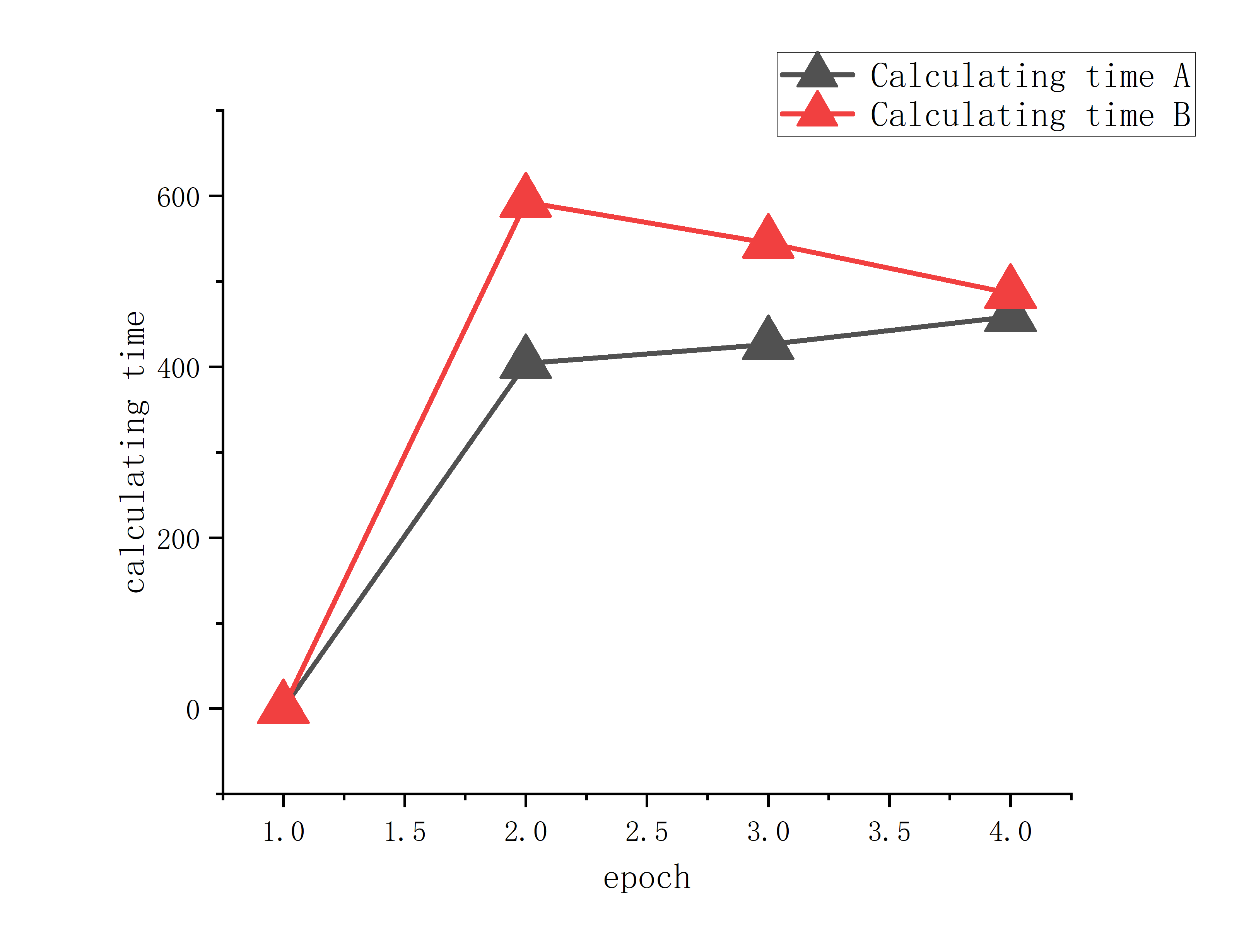}
\end{minipage}
}%
\subfigure[VGG16-$W$-$10$]{
\begin{minipage}[t]{0.3\linewidth}
\centering
\includegraphics[width=1.1in]{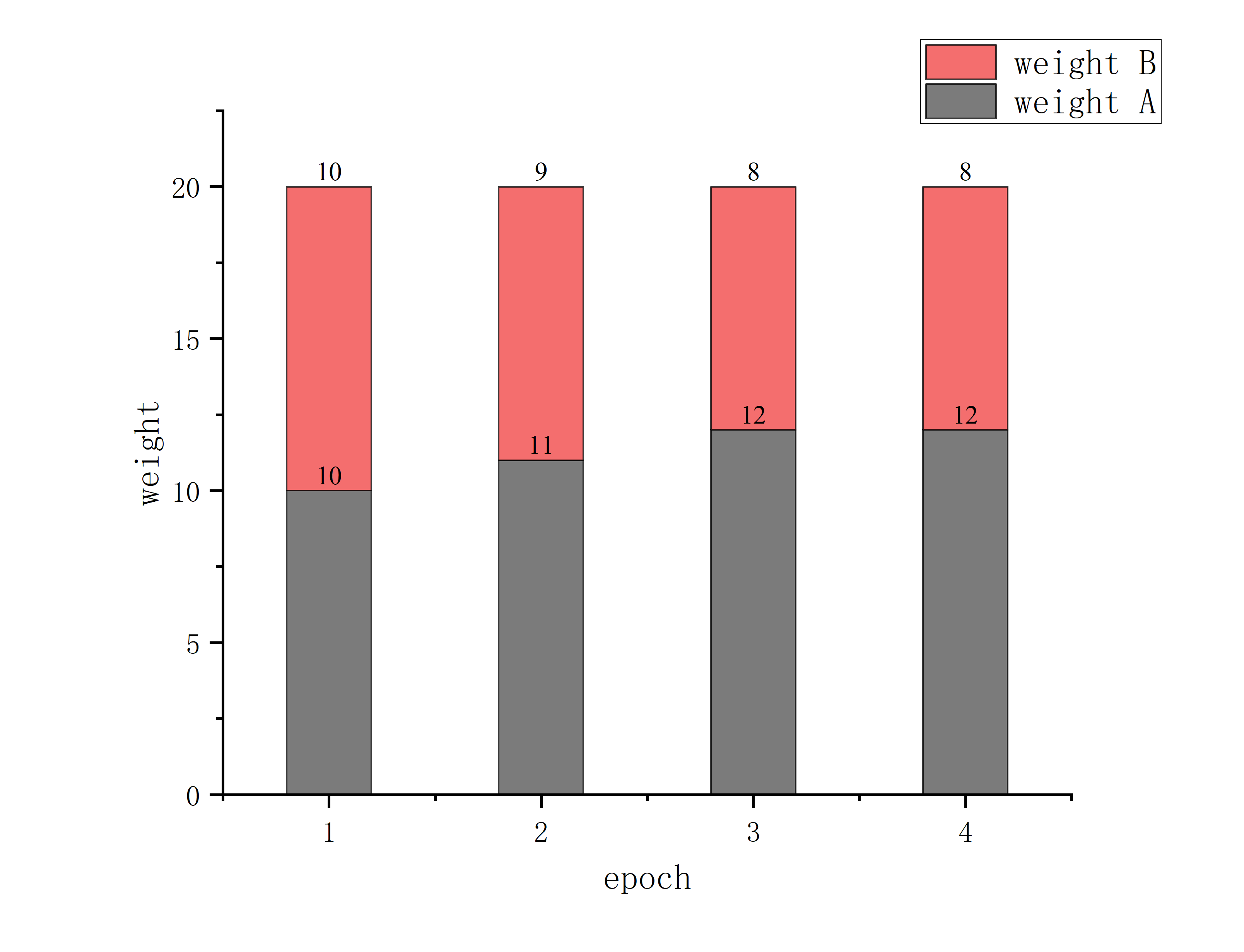}
\end{minipage}
}%
\subfigure[VGG16-$t$-$10$]{
\begin{minipage}[t]{0.3\linewidth}
\includegraphics[width=1.1in]{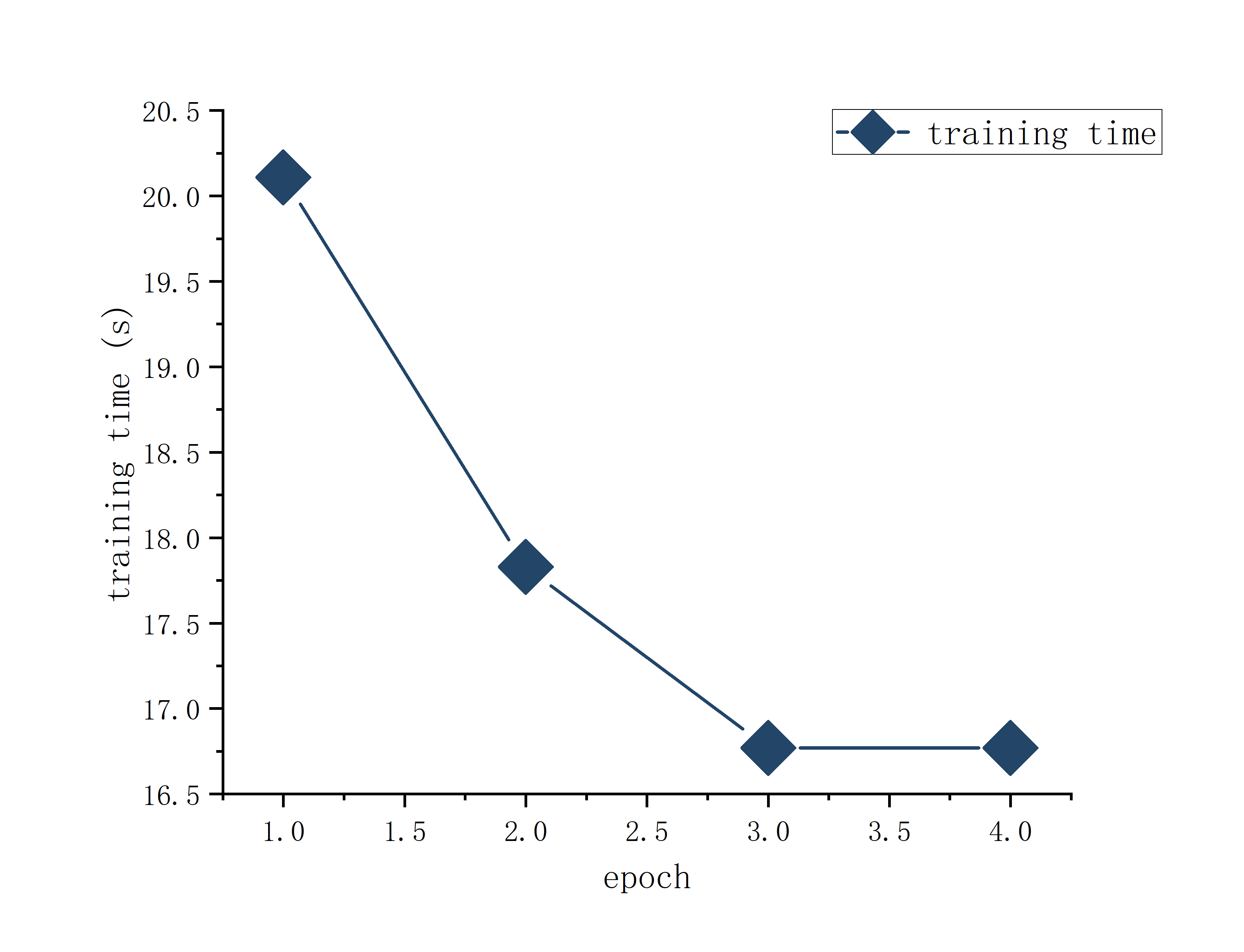}
\end{minipage}
}

\subfigure[VGG16-$Ts$]{
\begin{minipage}[t]{0.3\linewidth}
\includegraphics[width=1.1in]{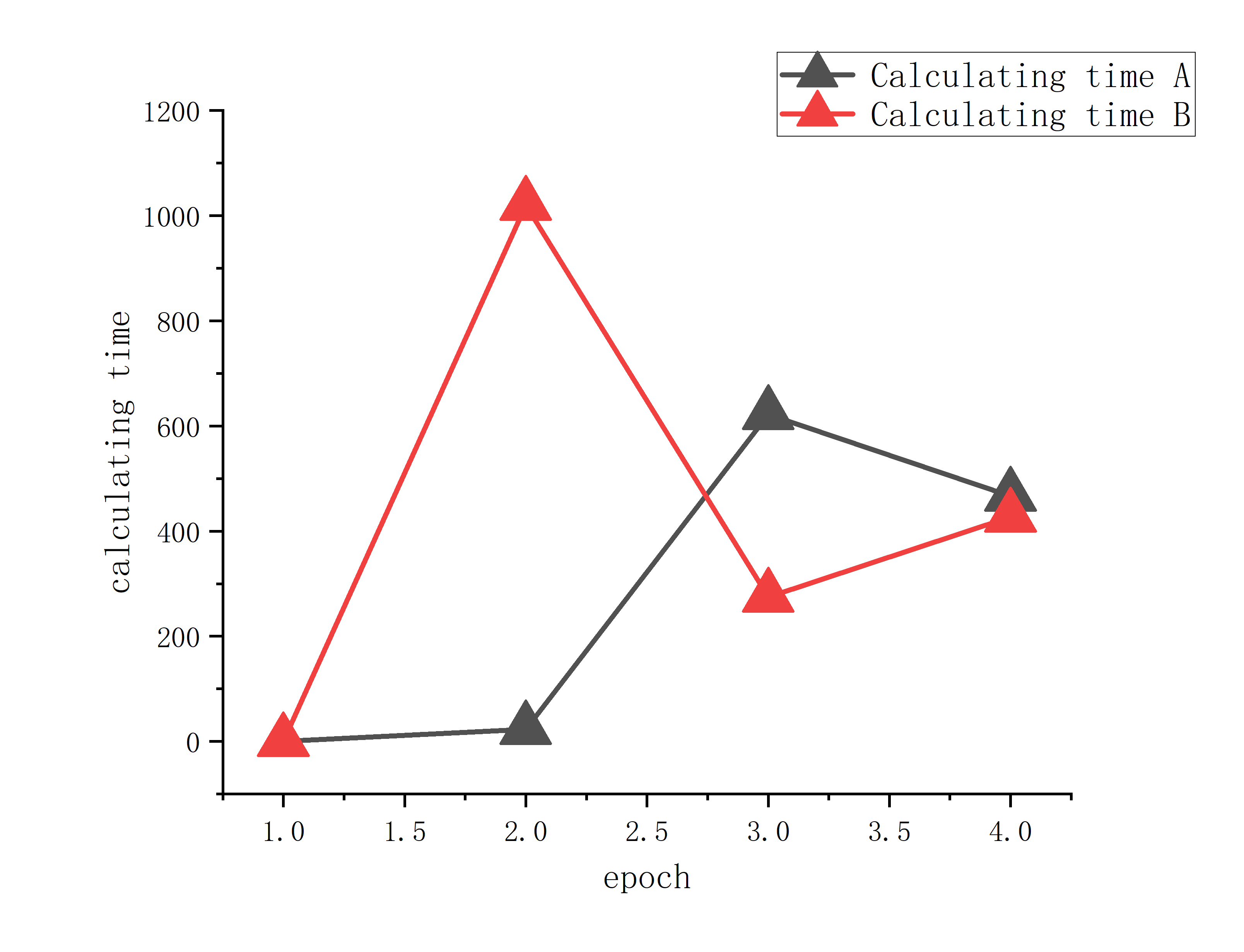}
\end{minipage}
}%
\subfigure[VGG16-$W$]{
\begin{minipage}[t]{0.3\linewidth}
\centering
\includegraphics[width=1.1in]{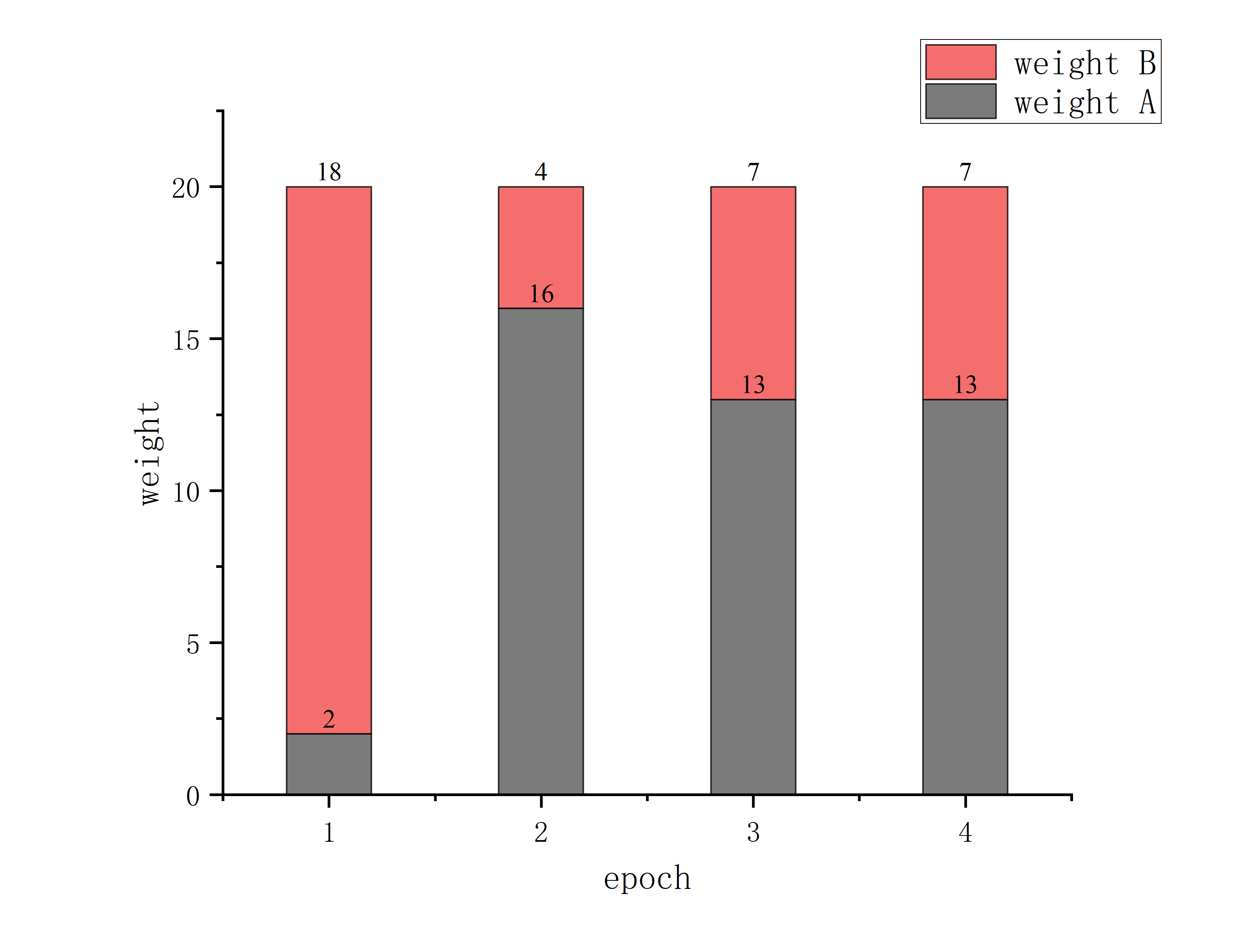}
\end{minipage}
}%
\subfigure[VGG16-$t$]{
\begin{minipage}[t]{0.3\linewidth}
\includegraphics[width=1.1in]{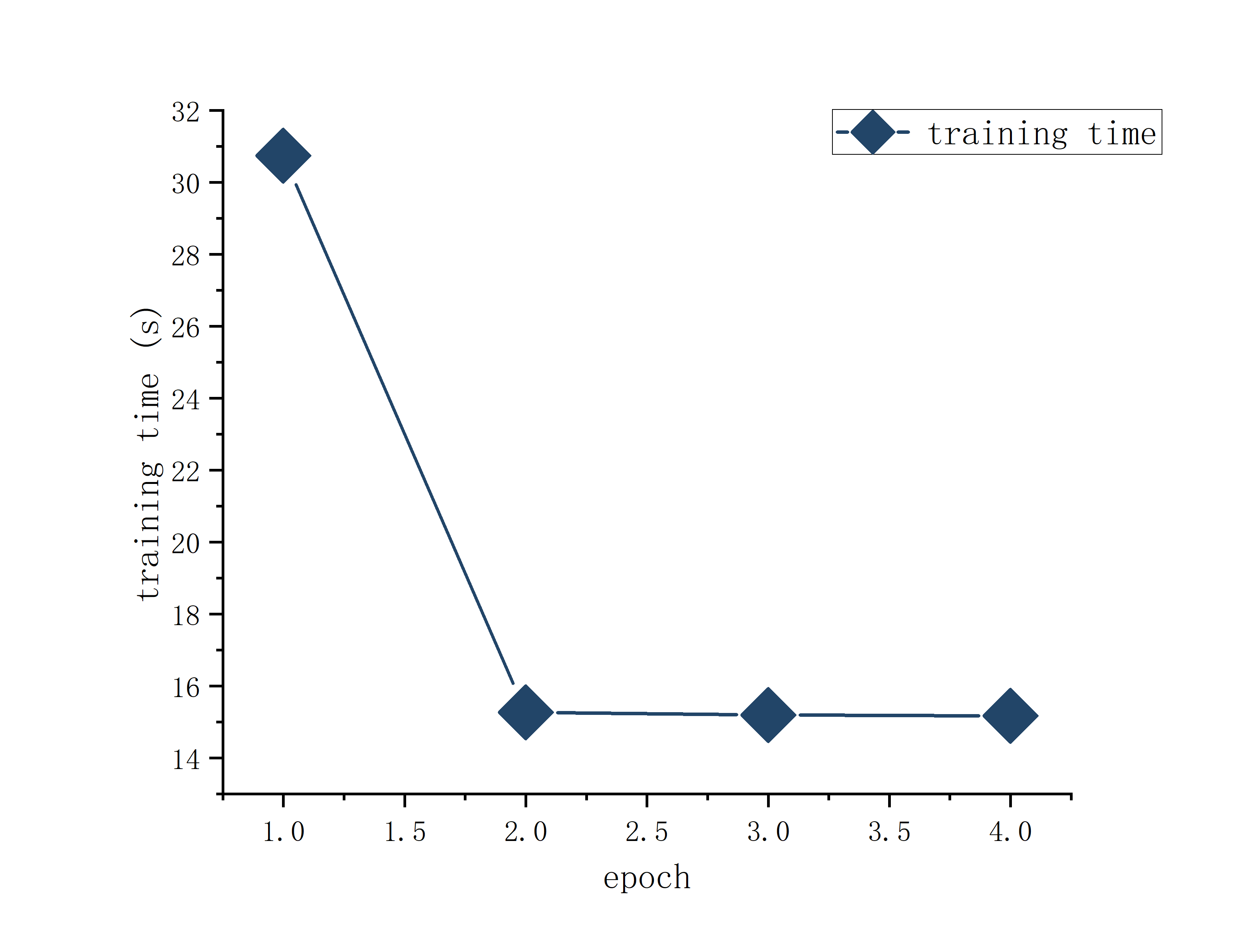}
\end{minipage}
}%
\centering
\caption{the training time of models results in calculating time, the ratio of weights and training time from two machines with RTX1080ti and V100.}
\end{figure}

\begin{figure}
\centering
\subfigure[ResNet50-$Ts$-3]{
\begin{minipage}[t]{0.3\linewidth}
\includegraphics[width=1.1in]{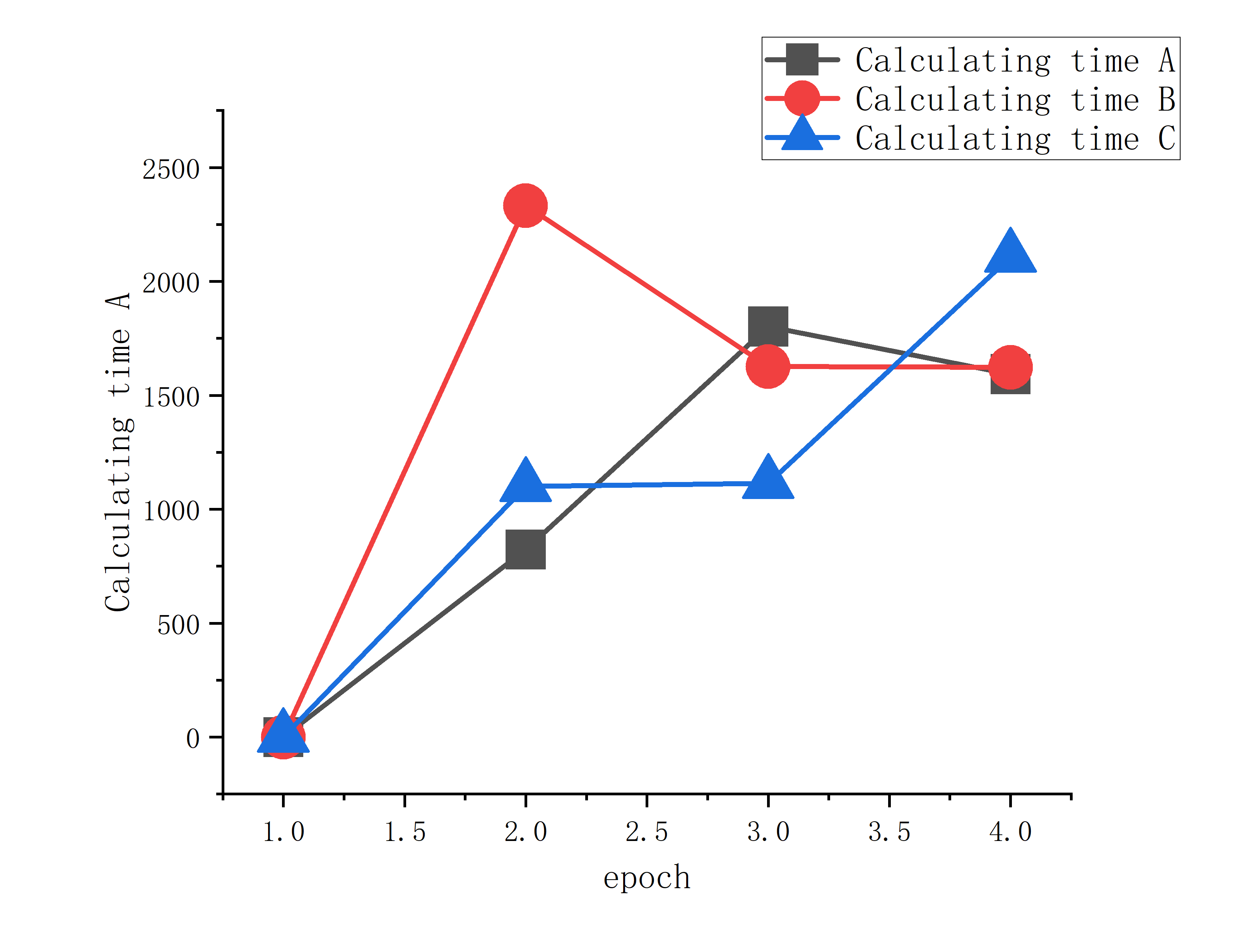}
\end{minipage}
}%
\subfigure[ResNet50-$W$-3]{
\begin{minipage}[t]{0.3\linewidth}
\centering
\includegraphics[width=1.1in]{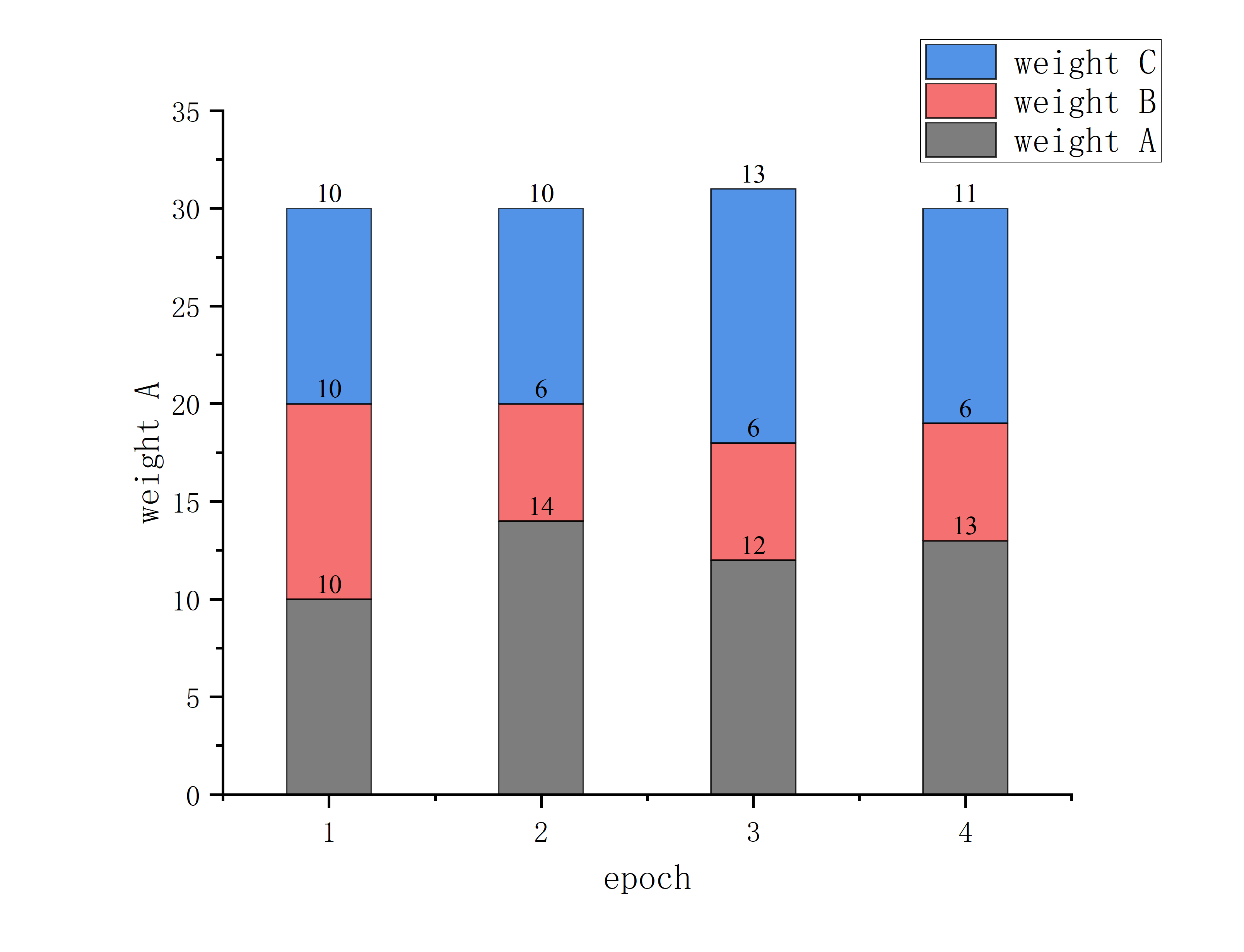}
\end{minipage}
}%
\subfigure[ResNet50-$t$-3]{
\begin{minipage}[t]{0.3\linewidth}
\includegraphics[width=1.1in]{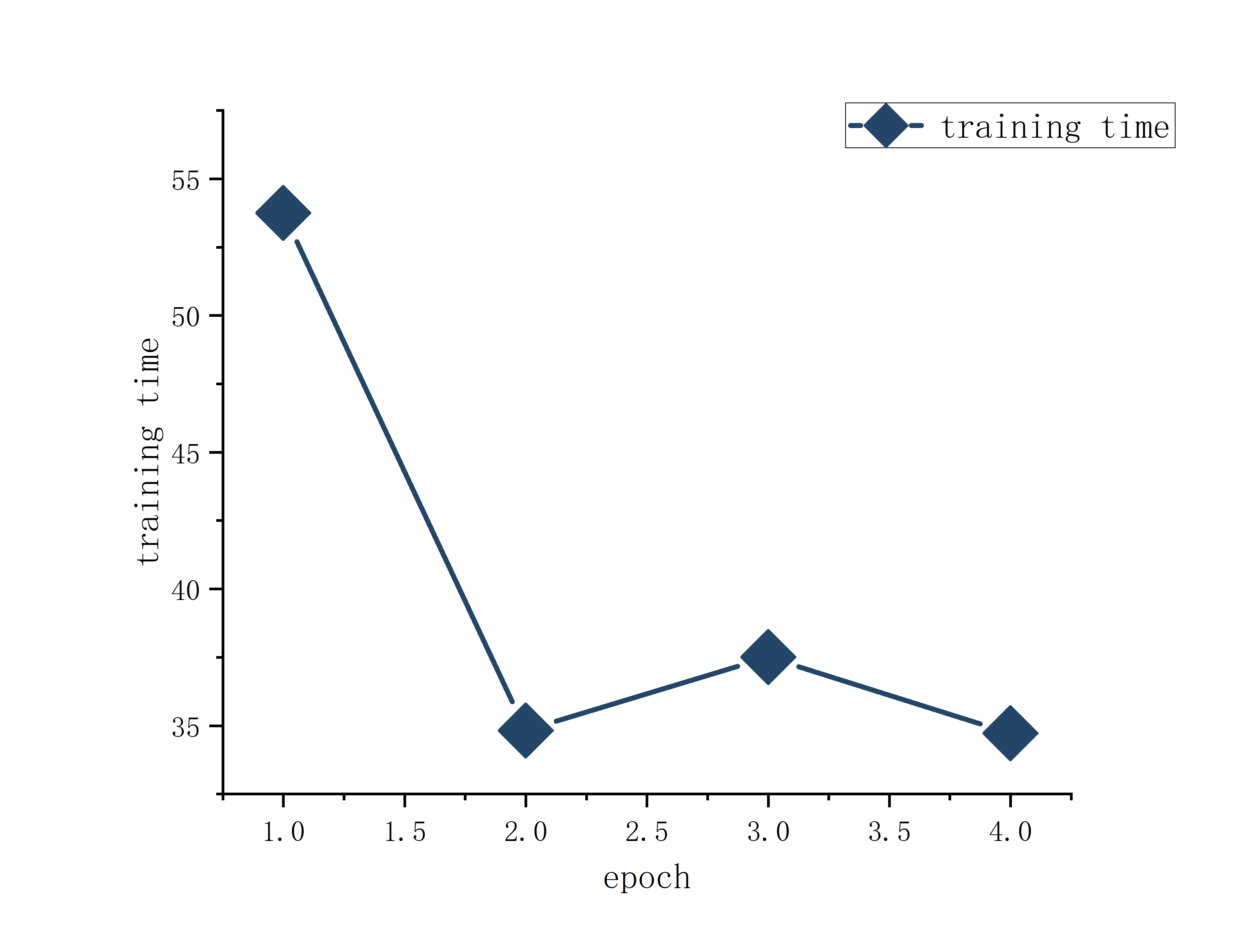}
\end{minipage}
}
\subfigure[ResNet18-$t$-$10$-3]{
\begin{minipage}[t]{0.3\linewidth}
\includegraphics[width=1.1in]{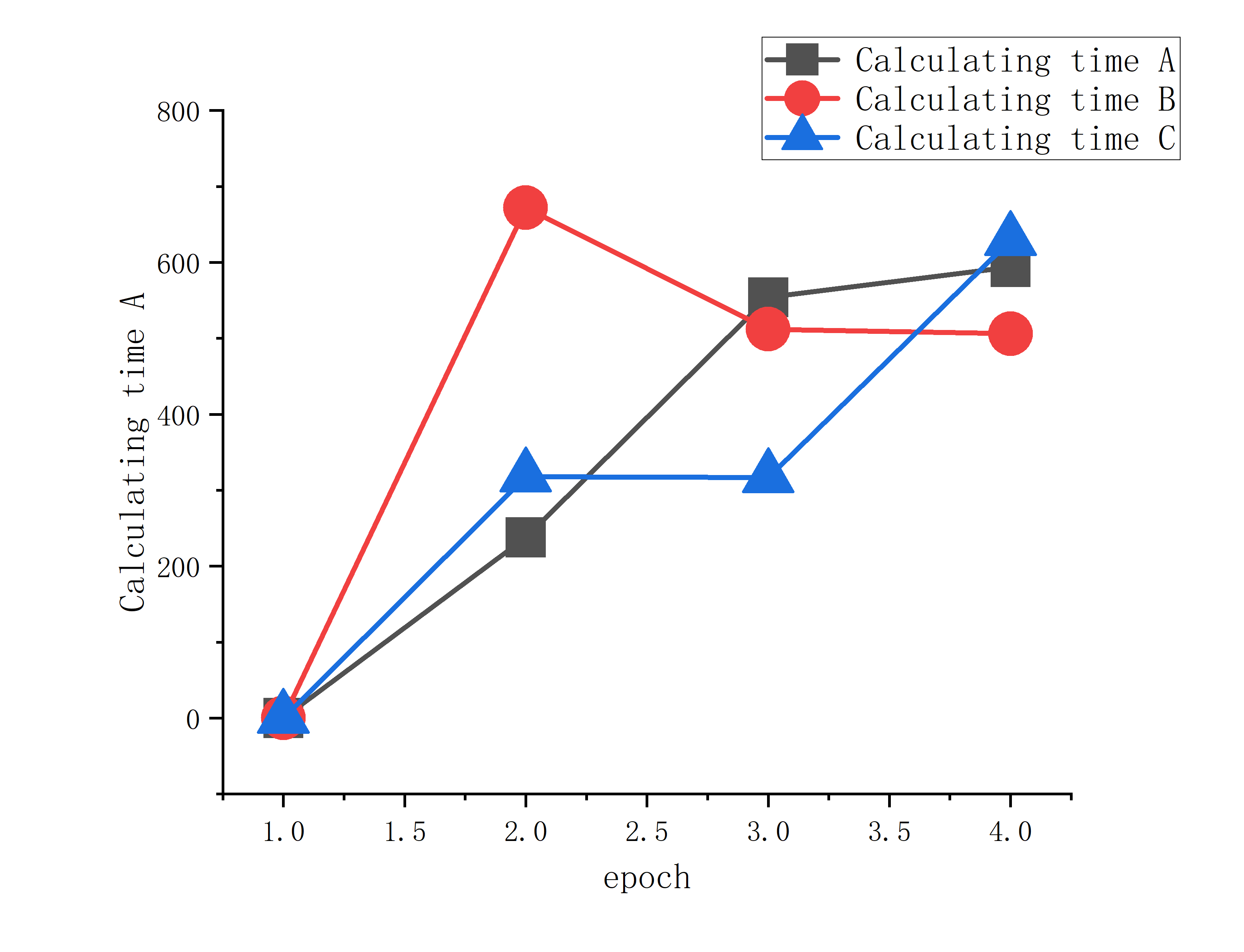}
\end{minipage}
}%
\subfigure[ResNet18-$Ts$-$10$-3]{
\begin{minipage}[t]{0.3\linewidth}
\centering
\includegraphics[width=1.1in]{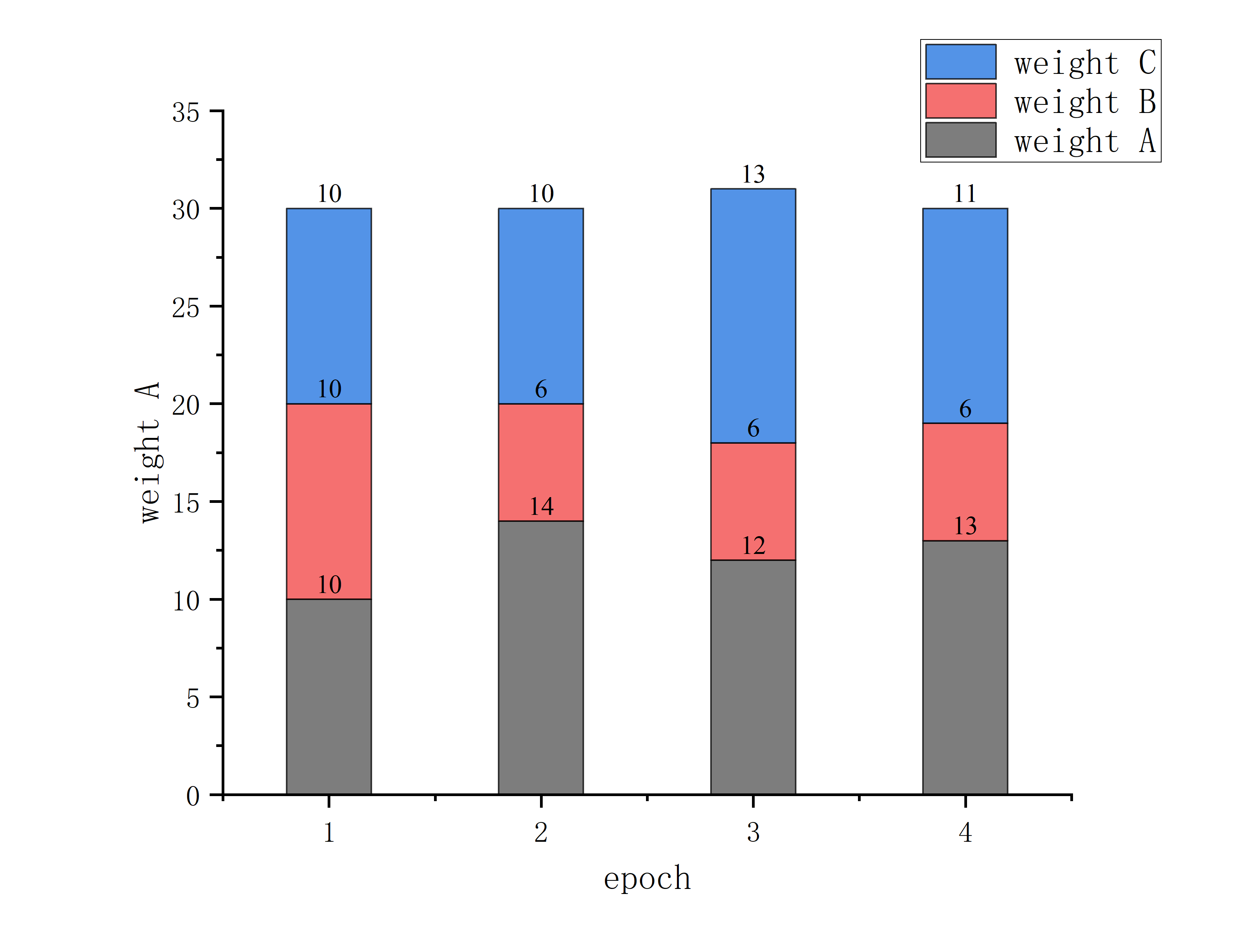}
\end{minipage}
}%
\subfigure[ResNet18-$W$-$10$-3]{
\begin{minipage}[t]{0.3\linewidth}
\includegraphics[width=1.1in]{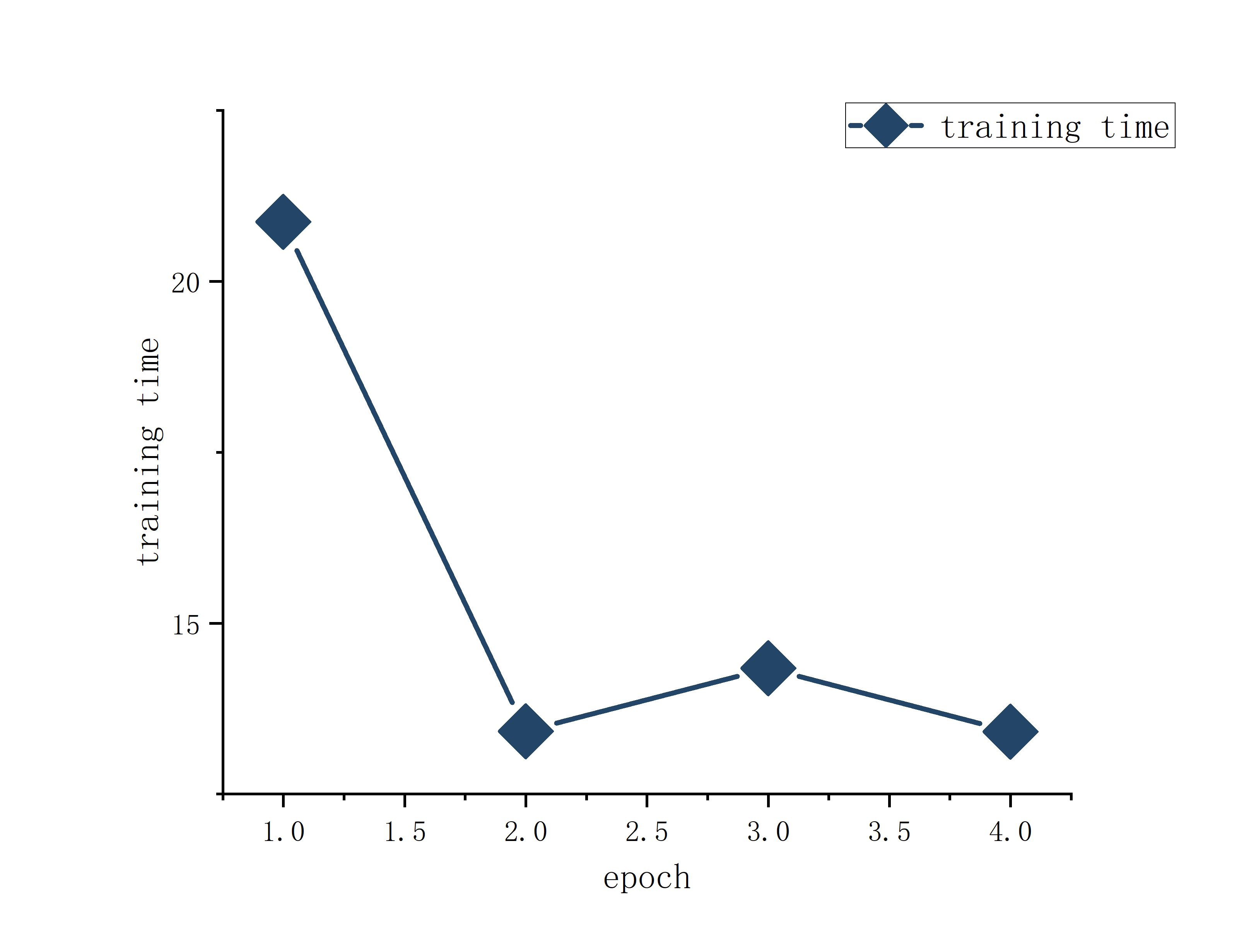}
\end{minipage}
}
\subfigure[VGG16-$Ts$-$10$-3]{
\begin{minipage}[t]{0.3\linewidth}
\includegraphics[width=1.1in]{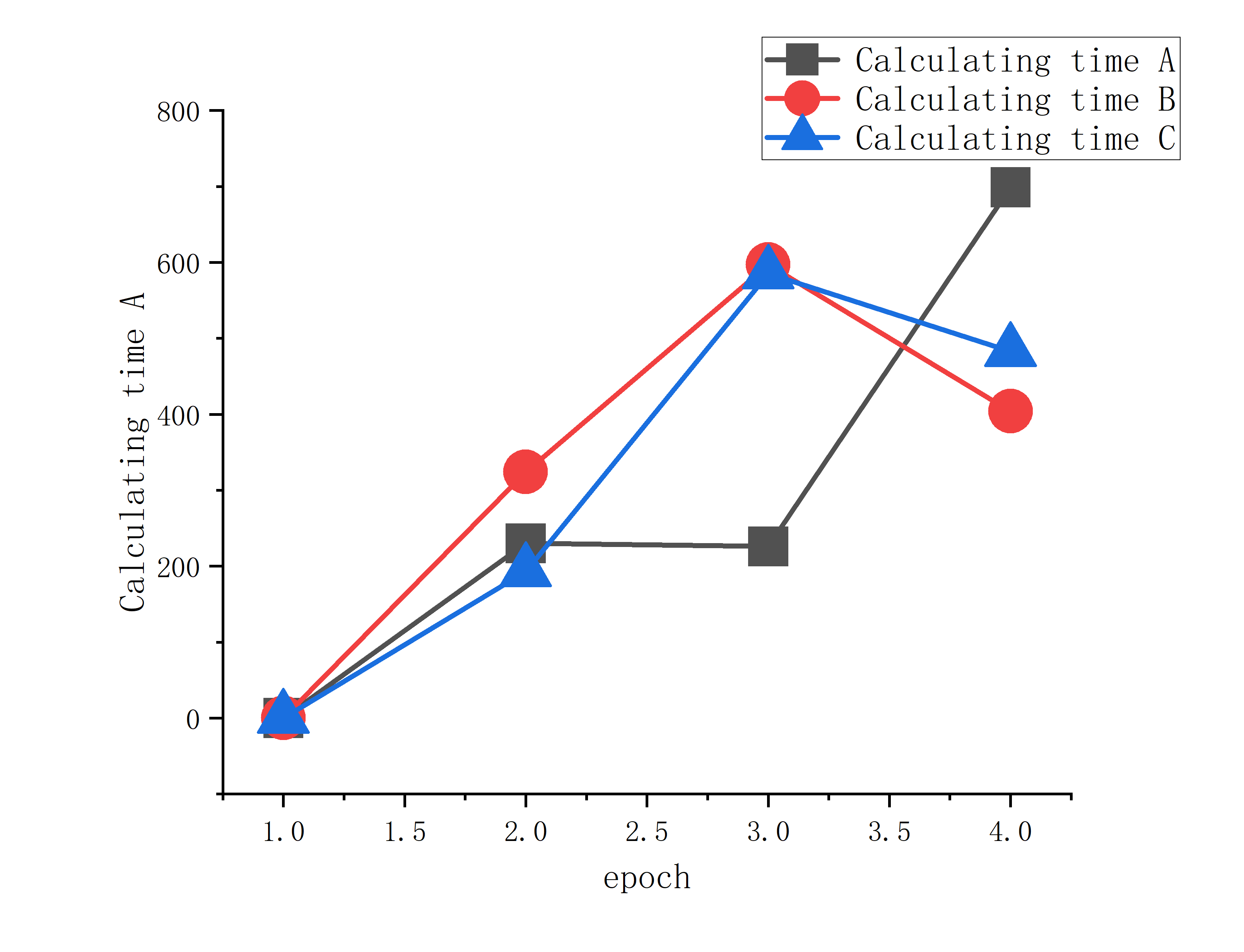}
\end{minipage}
}%
\subfigure[VGG16-$W$-$10$-3]{
\begin{minipage}[t]{0.3\linewidth}
\centering
\includegraphics[width=1.1in]{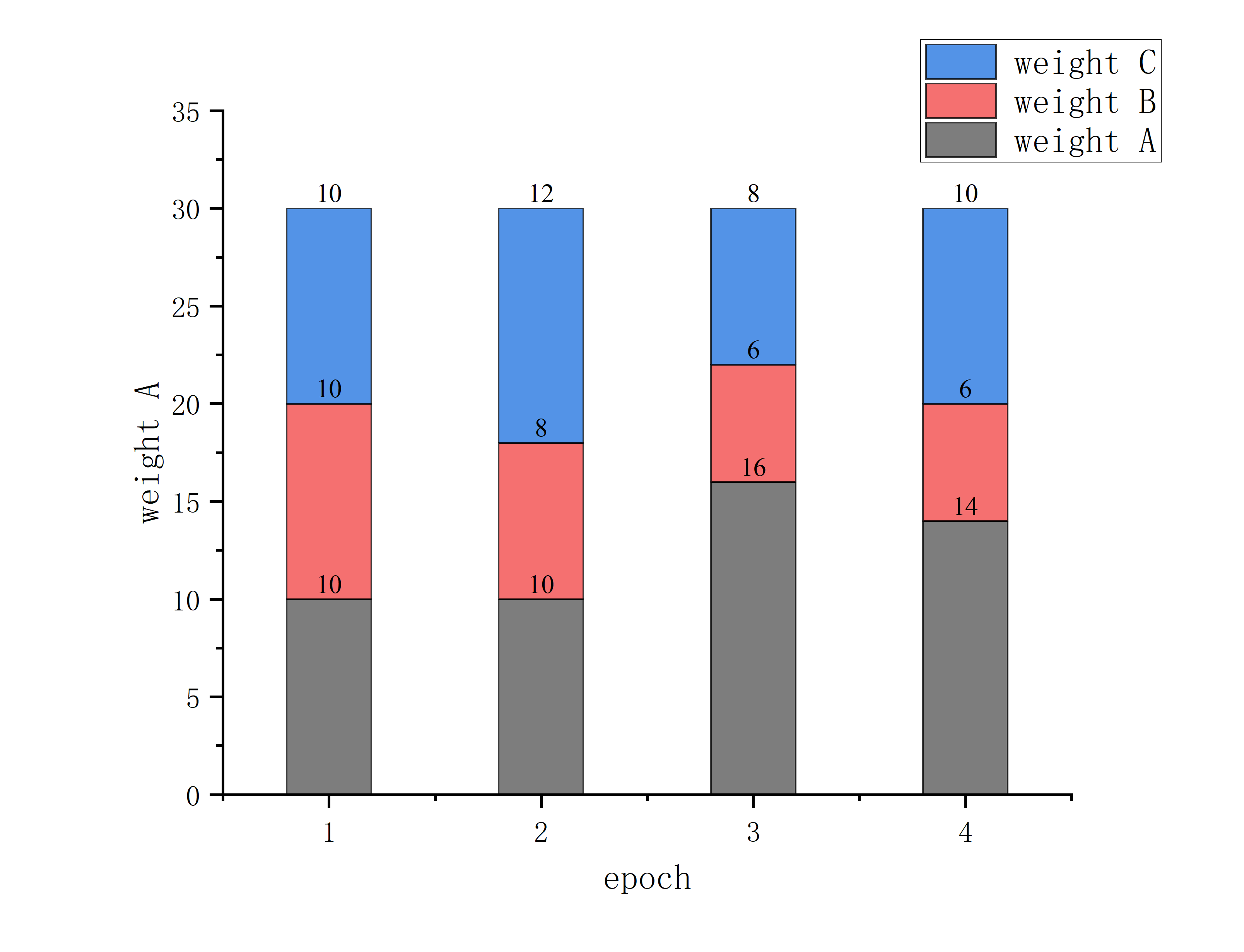}
\end{minipage}
}%
\subfigure[VGG16-$t$-$10$-3]{
\begin{minipage}[t]{0.3\linewidth}
\includegraphics[width=1.1in]{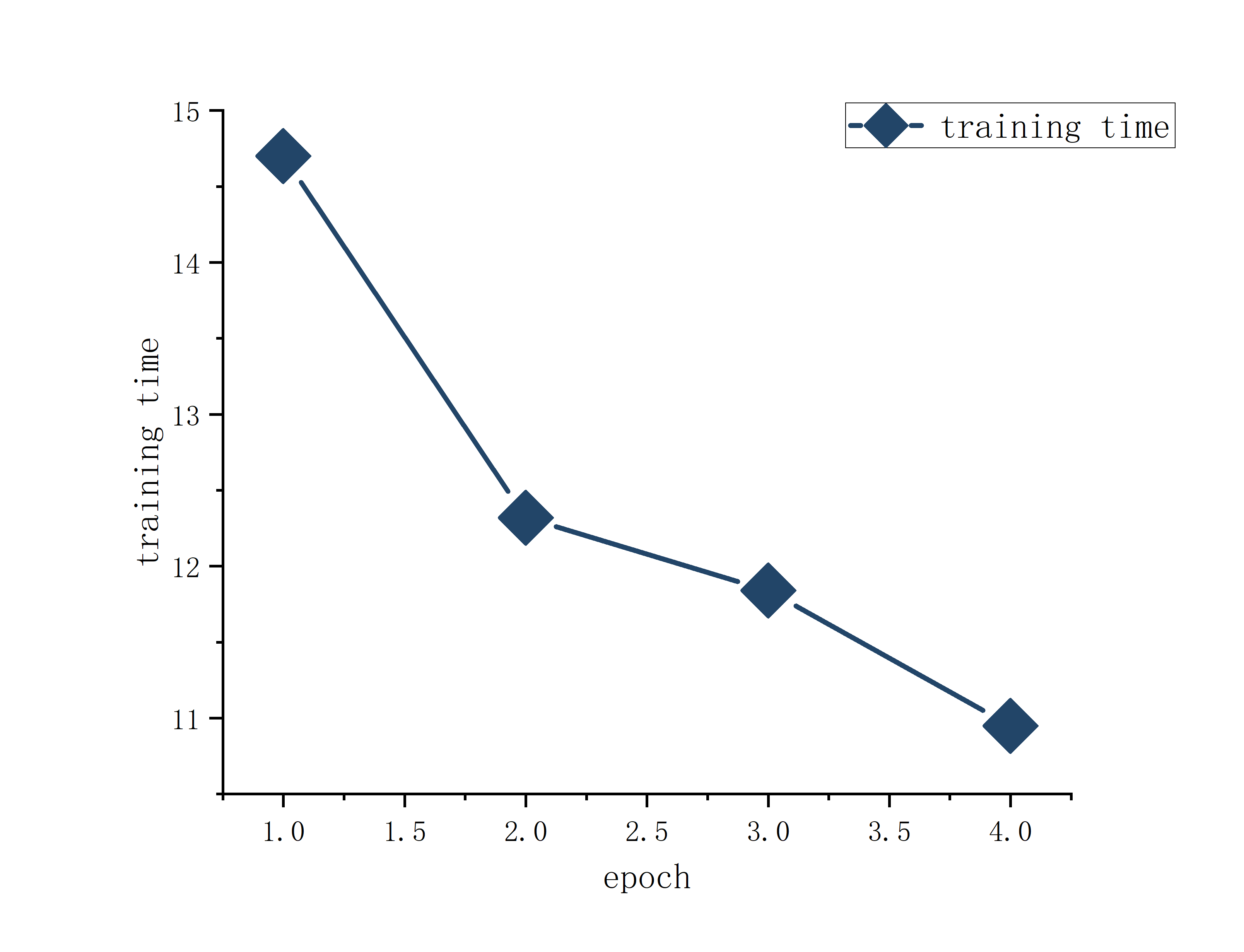}
\end{minipage}
}

\subfigure[VGG19-$Ts$-$10$-3]{
\begin{minipage}[t]{0.3\linewidth}
\includegraphics[width=1.1in]{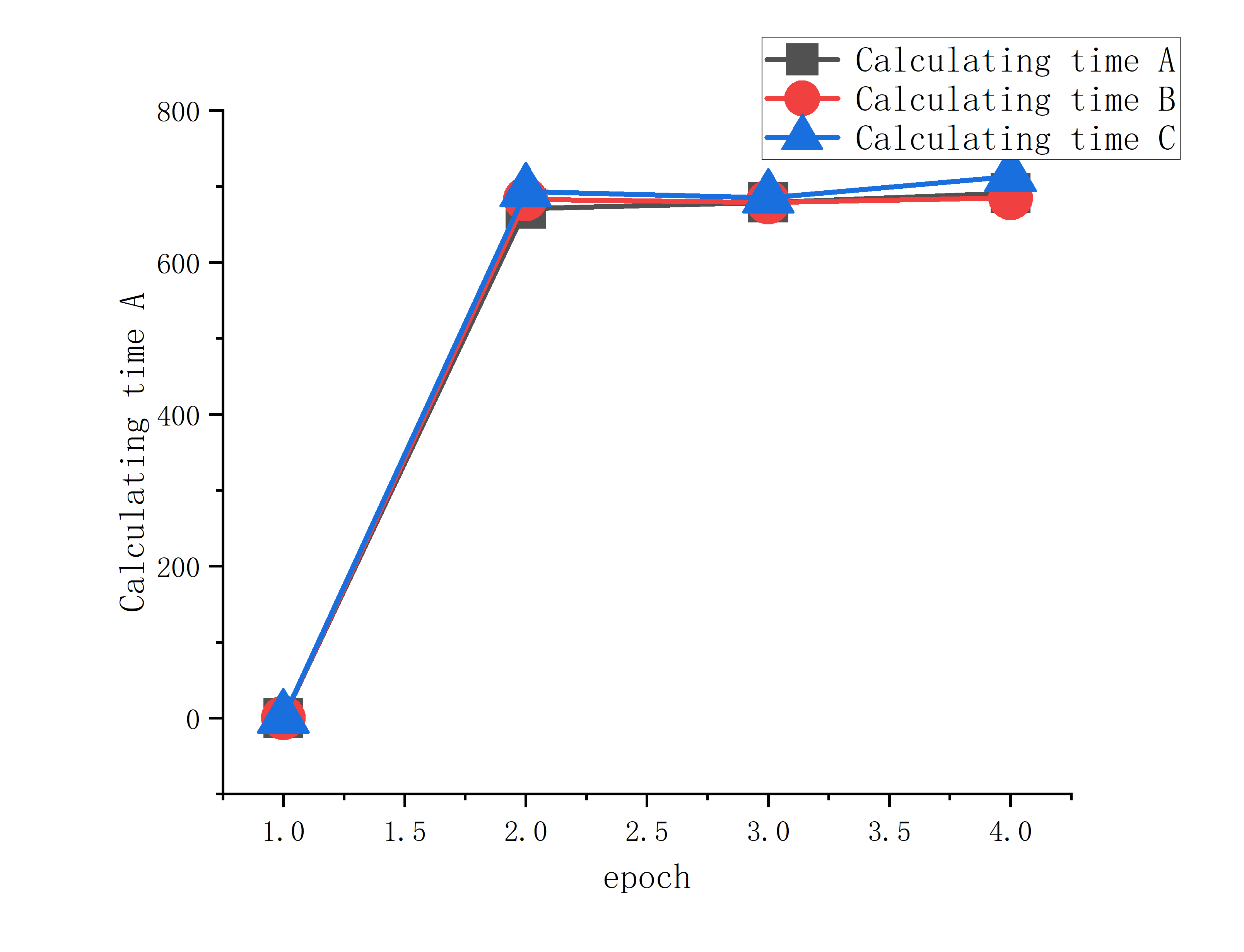}
\end{minipage}
}%
\subfigure[VGG19-$W$-$10$-3]{
\begin{minipage}[t]{0.3\linewidth}
\centering
\includegraphics[width=1.1in]{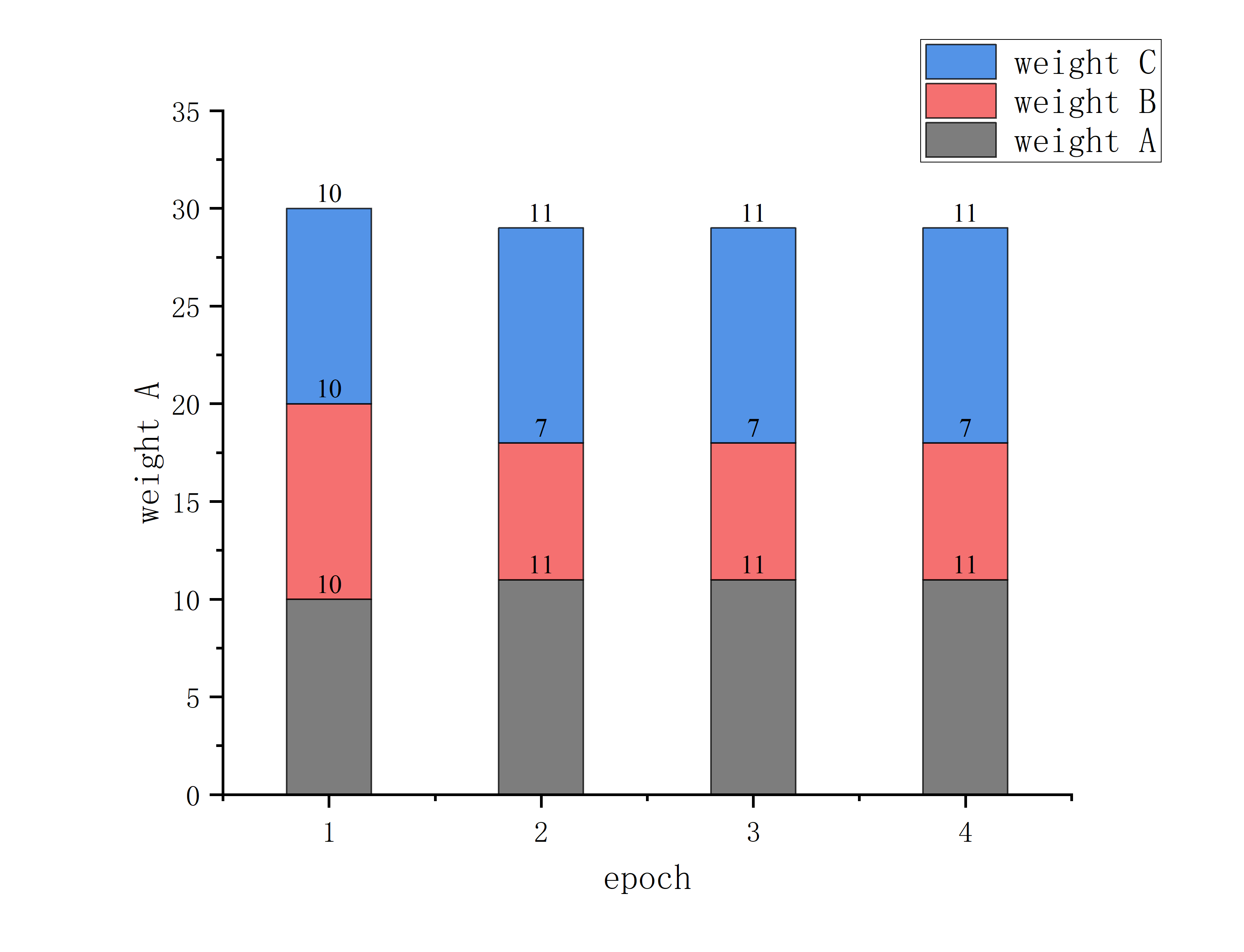}
\end{minipage}
}%
\subfigure[VGG19-$t$-$10$-3]{
\begin{minipage}[t]{0.3\linewidth}
\includegraphics[width=1.1in]{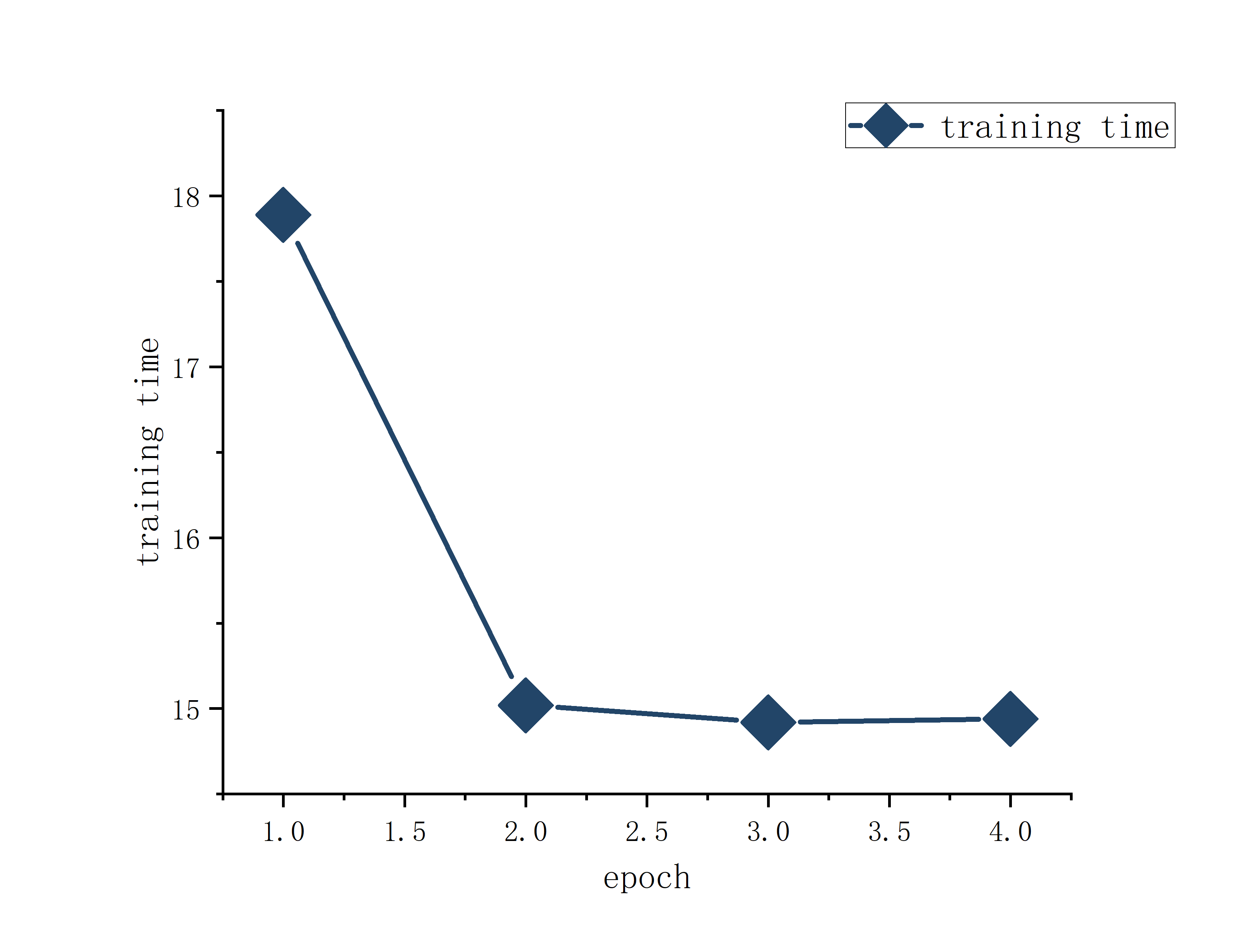}
\end{minipage}
}%
\centering
\caption{the training time of models results in calculating time, the ratio of weights and training time from three machines with two RTX1080ti and V100.}
\end{figure}
\subsection{performance on heterogeneity}
It is not enough to illustrate the importance of ratio $w_{i}$. We do the following experiments. We compared different group machines with little variety. For example, we compared V100+RTX2080ti with 2*RTX2080ti and V100+RTX2080ti with V100+2*RTX2080ti. In figure 11, under the same total batchsize, we can observe the velocity of training is increasing when adding a new card or replacing the weak card with strong card. It means that the performance of card is demonstrated to some extent.
\begin{figure}[ht]
\centering
\includegraphics[scale=0.2]{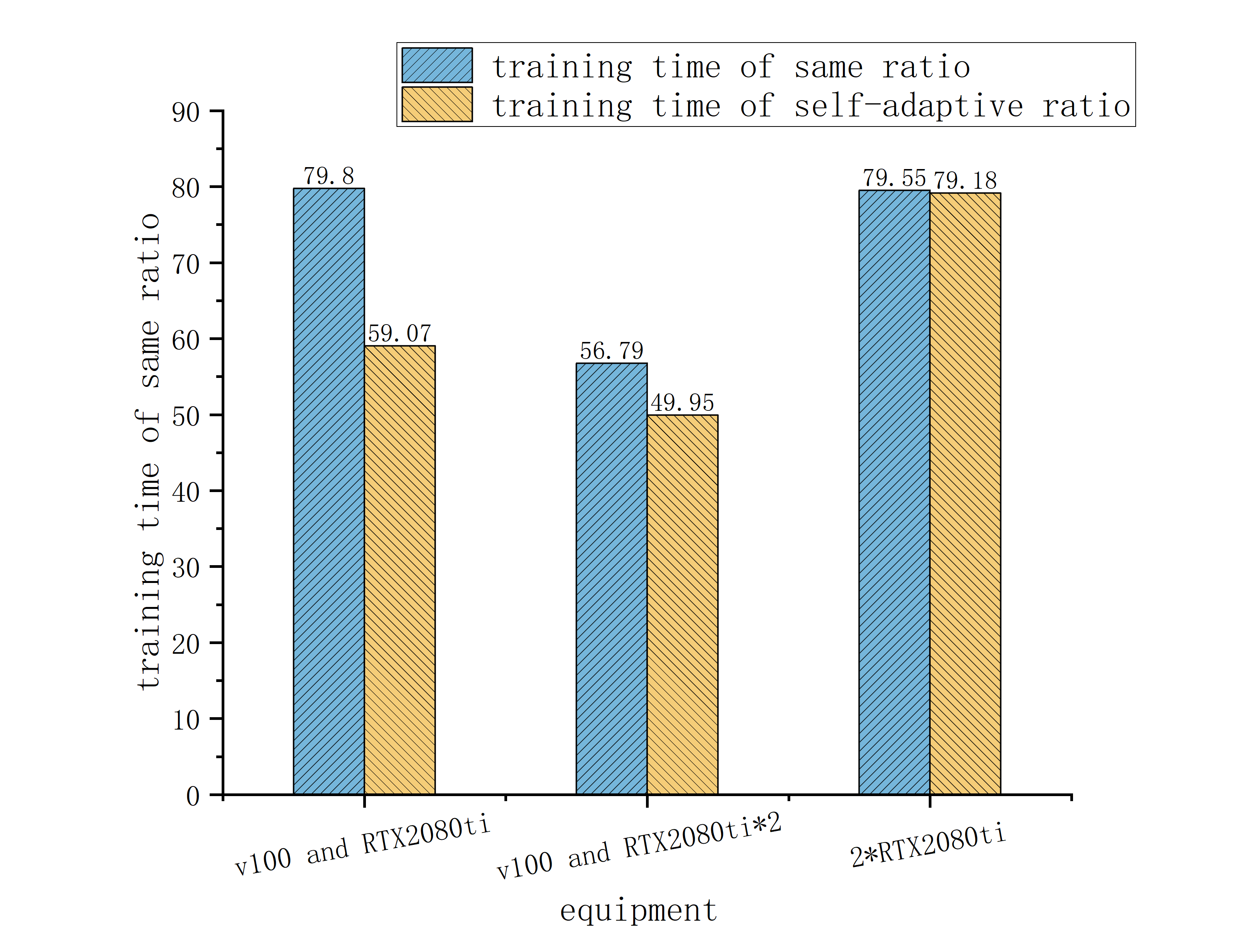}
\caption{the training time of models compared three groups, two machines with RTX1080ti and V100, three machines with two RTX1080ti and V100 as well as two machines with two RTX2080ti.}
\label{figl}
\end{figure}
\subsection{Universality in complex heterogeneous environment}
Through the combination of experiment and theory, we found that the existing AD-PSGD, AllReduce algorithm, etc., cannot guarantee the acceleration function in some heterogeneous environments. For example, when there are only two workers, AD -PSGD training speed is almost the same as AllReduce. As shown in Figure 12, we have drawn the convergence curve of each algorithm on GTX1080ti and RTX2080ti dual cards. It can be found that the allocation algorithm will have a significant acceleration effect. Another example, when a worker has a fast computing speed and other workers are relatively slow, the current asynchronous SGD algorithm cannot make full use of resources, which is equivalent to n slow nodes, but it can also be accelerated by allocating data. In addition, there is another possibility that the allocation algorithm can be used as a plug-in of AllReduce and combined with other algorithms.
\begin{figure}[ht]
\centering
\subfigure[total training loss]{
\begin{minipage}[t]{0.3\linewidth}
\includegraphics[width=1.3in]{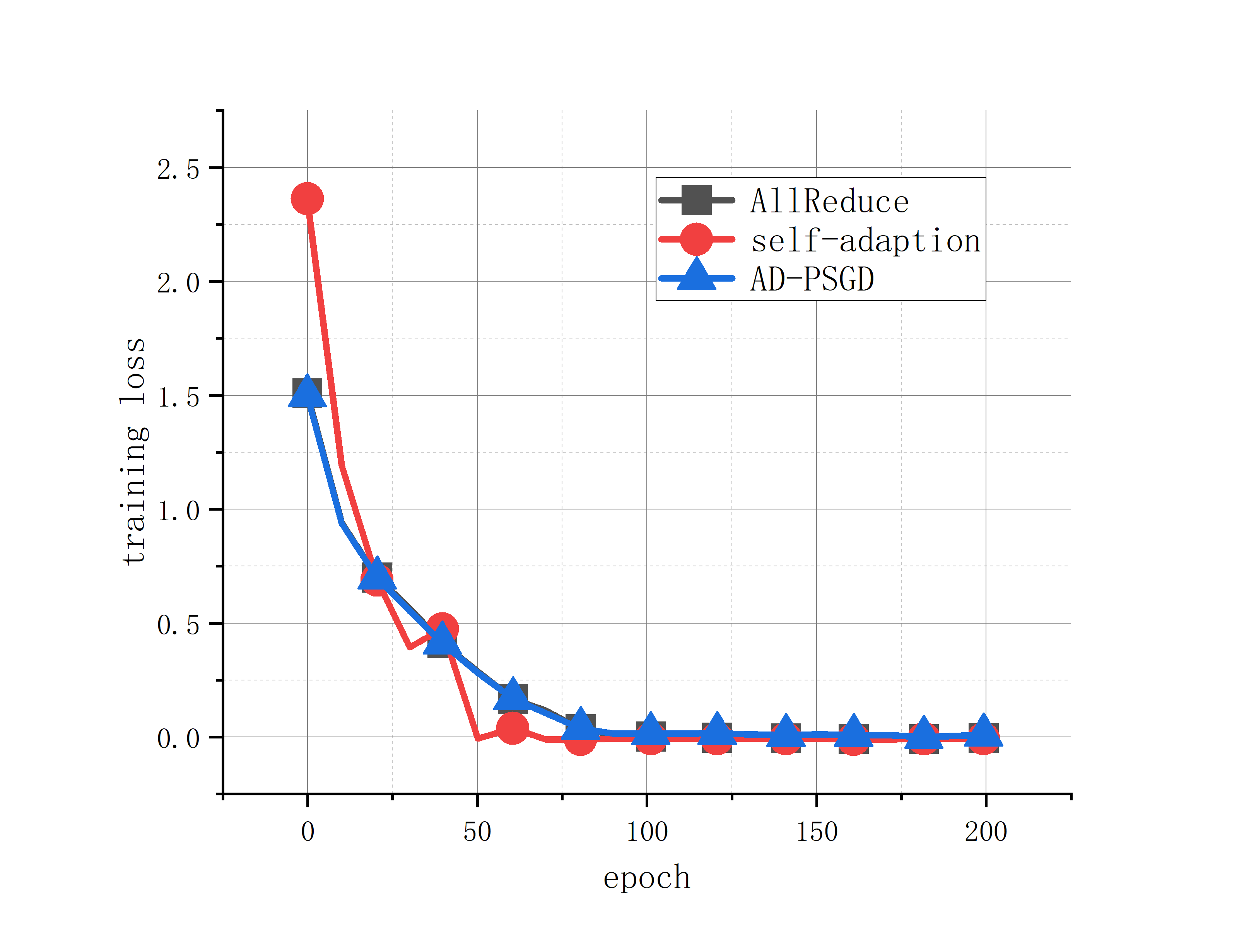}
\end{minipage}
}%
\subfigure[training loss of 2X slowdown]{
\begin{minipage}[t]{0.3\linewidth}
\centering
\includegraphics[width=1.3in]{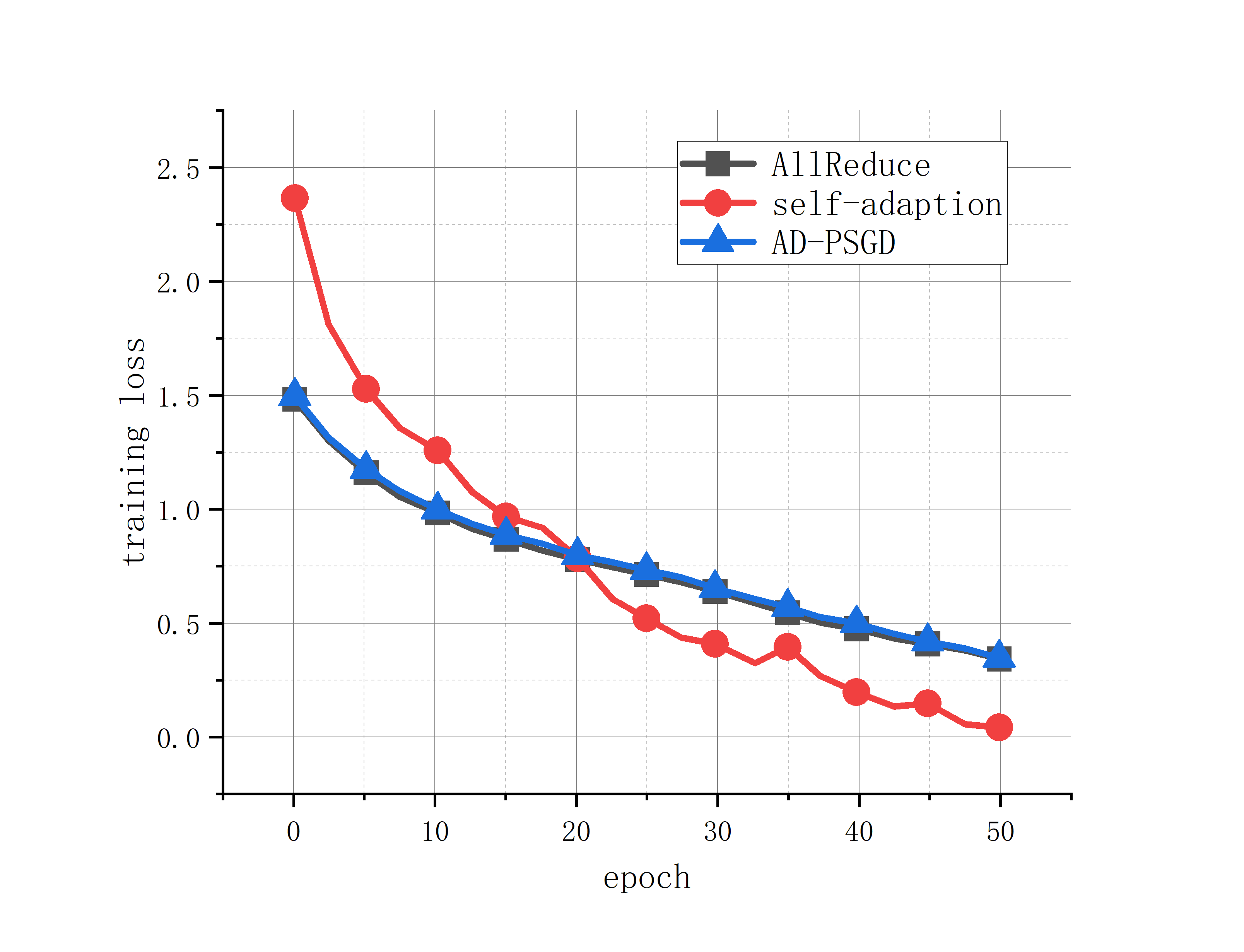}
\end{minipage}
}%
\subfigure[training loss of 10X slowdown]{
\begin{minipage}[t]{0.3\linewidth}
\centering
\includegraphics[width=1.3in]{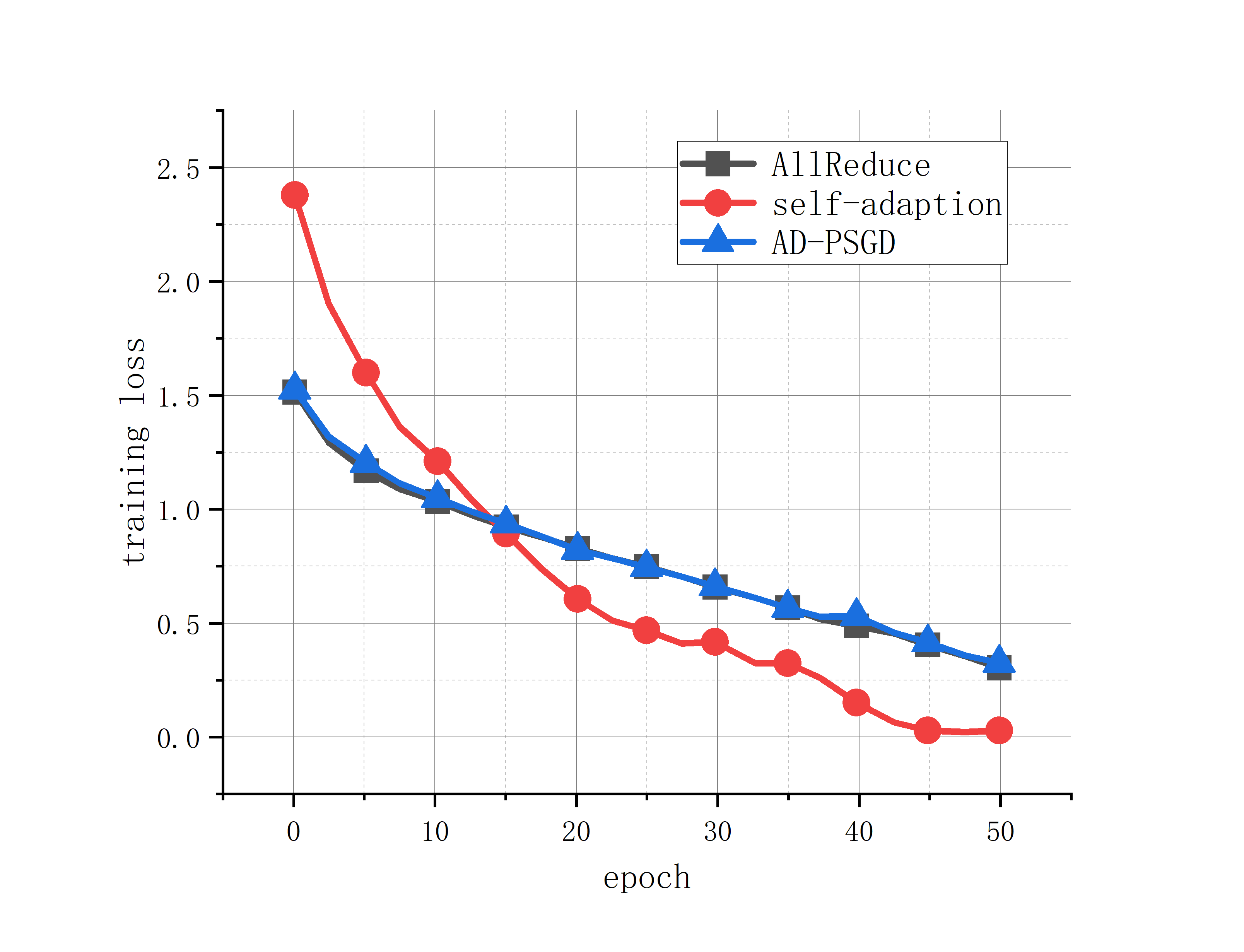}
\end{minipage}
}
\centering
\caption{the training loss of models results in one machines with RTX1080ti and GTX1080ti.}
\end{figure}

Figure 13 shows the approximate speedup ratios of several algorithms. Because the allocation algorithm itself has the behavior of expanding batchsize and reducing gradient aggregation, the speedup ratio of this algorithm is based on the parameter server, when the straggler iteration time is twice that of other cards It can reach about 5.36X, and when the straggler iteration time is 5 times that of other cards, it can reach 2.75X, which is not very exaggerated, and its speedup is basically about 3.3X that of AllReduce. 

\begin{figure}[ht]
\centering
\subfigure[speedup in 2X slowdown]{
\begin{minipage}[t]{0.5\linewidth}
\includegraphics[width=1.6in]{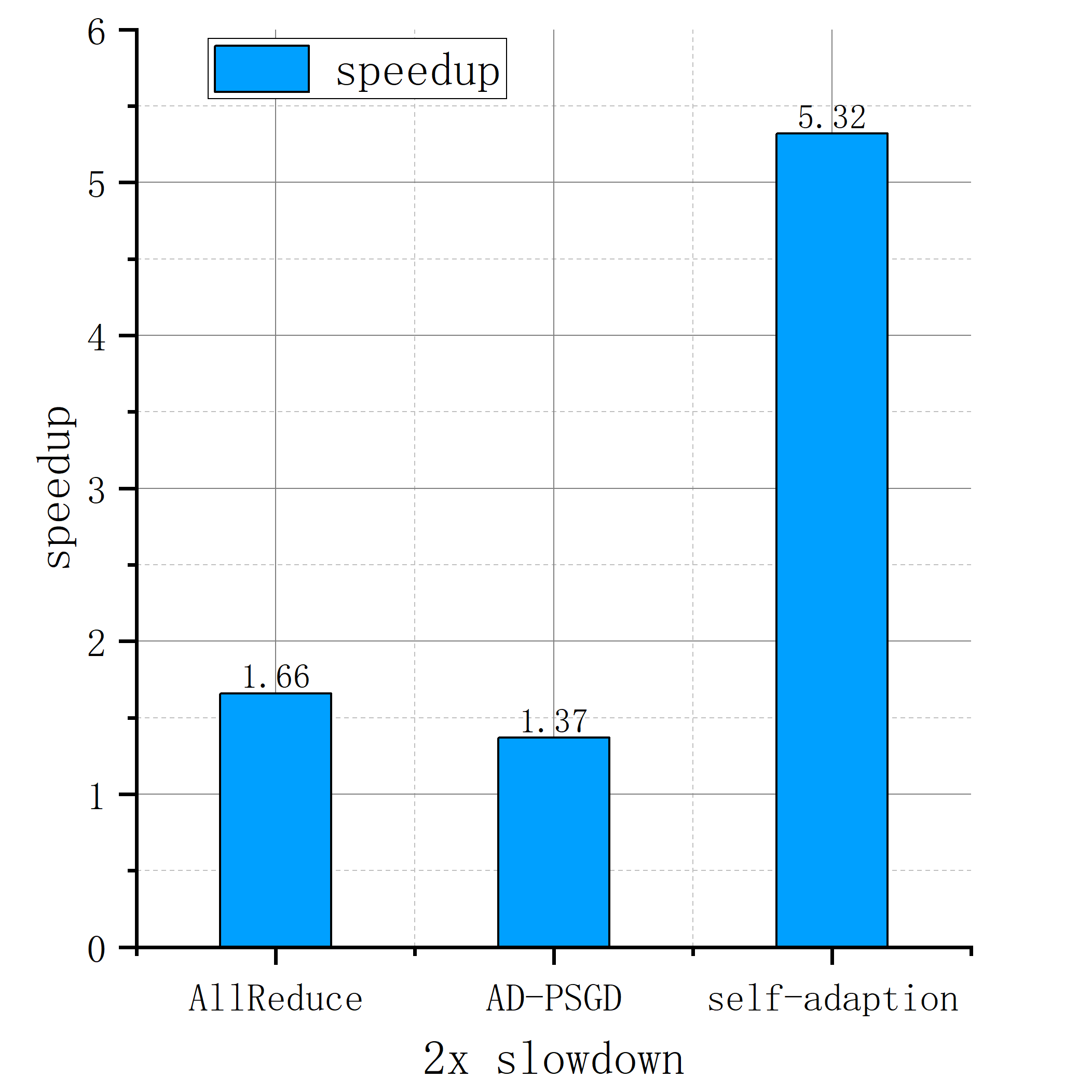}
\end{minipage}
}%
\subfigure[speedup in 5X slowdown]{
\begin{minipage}[t]{0.5\linewidth}
\centering
\includegraphics[width=1.6in]{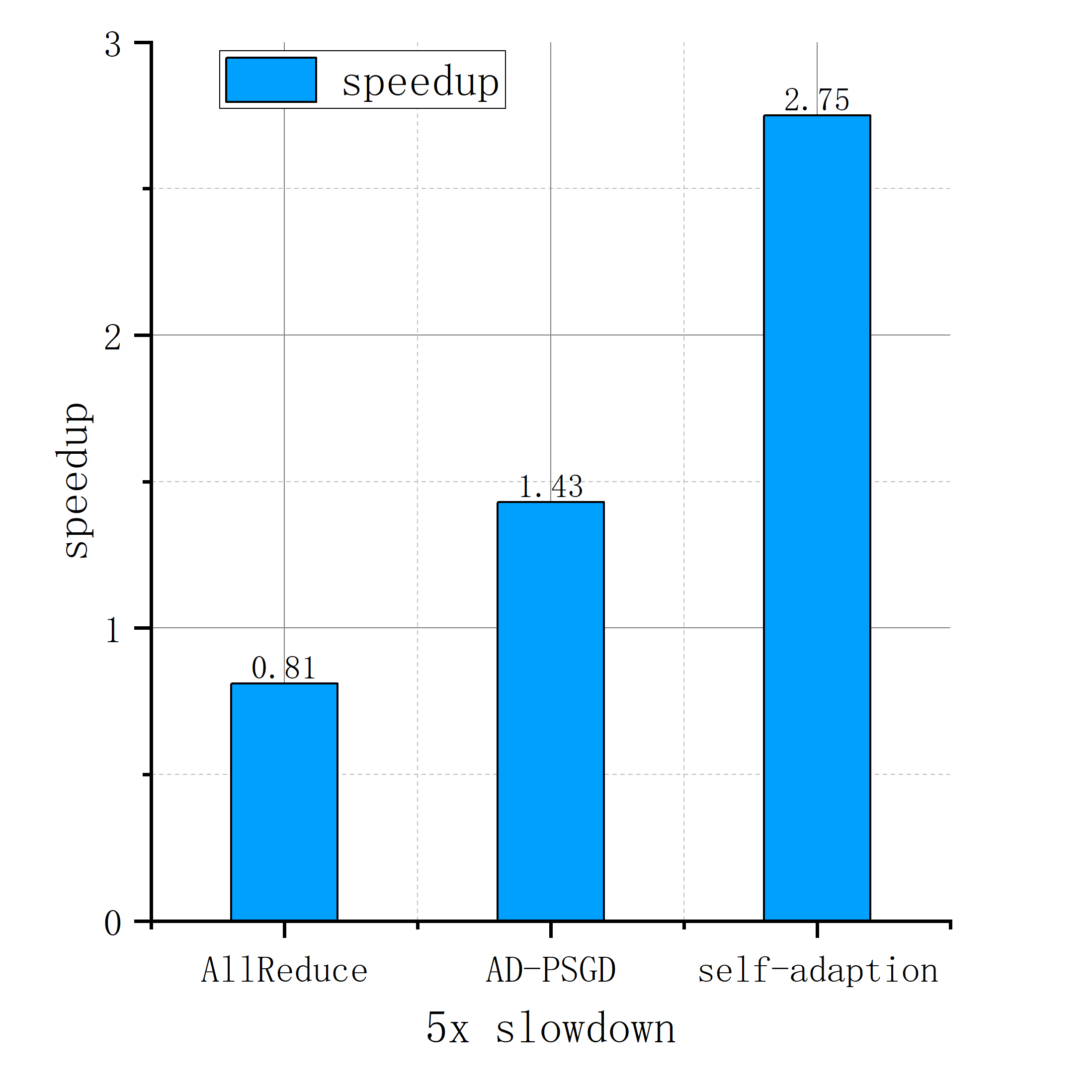}
\end{minipage}
}
\centering
\caption{the training speedup of models}
\end{figure}
\section{RELATED WORK}
Existing efforts on heterogeneous distributed deep learning algorithms can be classified into two types:   algorithms based on task allocation and algorithms based on gradient aggregation.

For algorithms based on task allocation, Yang et al. \cite{2019BOA} proposed a batch orchestration algorithm, which balances the amount of mini-batch data according to the speed of workers. FlexRR \cite{DBLP:conf/cloud/HarlapCDWGGGX16} addresses the straggler problem by integrating flexible consistency bounds with temporary peer-to-peer work reassignment. These methods have achieved certain results under the architecture of the parameter server \cite{2014Scaling}, but there is communication bottleneck existing in parameter server. They are not very practical under the architecture of Ring Allreduce \cite{2017Accurate}.

For algorithms based on gradient aggregation, DYNSGD \cite{2017Heterogeneity} can dynamically adjust the local learning rate of training node according to the delay of the node. Zhang et al. \cite{article} have a similar idea, it tracks the state of each gradient, and then adjusts the learning rate according to the state of the gradient. In addition to adjusting the learning rate during update, AD-PSGD \cite{2017Asynchronous} probabilistically reduces the effects of heterogeneity with randomized communication, by the way, this algorithm can also speed up training for some tasks in a homogeneous environment. Prague \cite{2019Heterogeneity} uses Partial All-Reduce mechanism to further accelerate training in heterogeneous environments.

Besides the work in the algorithm, it's also possible to do some work in system to mitigate the straggler problem in heterogeneous environments. Chen et al. \cite{2016Revisiting} introduce an approach of synchronous optimization with backup workers, which can avoid asynchronous noise while mitigating for the worst stragglers. Tandon et al. \cite{2017Gradient} use the method of Gradient Coding, which replicates some samples on each machine, so that each sample is repeatedly trained. In this way, the system only need to obtain the gradient of part of the training nodes to get the complete gradient. Although these methods are simple, they consume additional computing resources, so they are not a good choice.
\section{CONCLUSIONS}
To cope with task allocation, we design an implementation of a static allocation algorithm based on gradient accumulation in this paper. The dataset is artificially allocated to each worker, and each worker draws samples in proportion to train to accelerate the training speed of the network in a heterogeneous environment. The static allocation training was completed on GTX 1080ti and RTX 2080ti, and the accuracy of the experiment, epoch and training duration were recorded to verify the convergence of the network model under this method and the impact on the training speed. Furthermore, in order to adapt to various training environments and workers, we proposed an adaptive allocation algorithm. We found that the proportion of task allocation is proportional to the training speed of each worker. We use this property to calculate the change in the amount of task allocation in each round. We have verified the adaptive allocation process on multiple machines, and the training speed has changed from slow to faster. Finally, we also do experiments to verify that the speed of multi-card relative to fewer cards and the training speed of strong cards relative to weak cards are improved. 
\nocite{*}
\normalem
\printbibliography

\appendix
Assuming that the $k$th epoch task allocation of workers is $w_{1}^{(k)},w_{2}^{(k)},\cdots,w_{n}^{(k)}$, the $k+1$th epoch to be updated is $w_{1}^{(k+1)},w_{2}^{(k+1)},\cdots,w_{n}^{(k+1)}$, and the increase is $u_{1},u_{2},\cdots,u_{n}$. Therefore, the relationship of three variables can be expressed as
\begin{equation}
\left\{\begin{array}{c}
w_{1}^{(k+1)}=w_{1}^{(k)}+u_{1}\\
w_{2}^{(k+1)}=w_{2}^{(k)}+u_{2}\\
\vdots\\
\vdots\\
w_{n}^{(k+1)}=w_{n}^{(k)}+u_{n}
\end{array}\right.    
\end{equation}

According to formula (8), the waiting time of n workers must be equal, and the pairwise constraint can be expressed as $\mathrm{n}(\mathrm{n}-1)/2$ equations

\begin{equation}
\left\{\begin{array}{c}
\frac{D w_{1}^{(k+1)}}{C v_{1}}-\frac{D w_{2}^{(k+1)}}{C v_{2}}=0 \\
\frac{D w_{1}^{(k+1)}}{C v_{1}}-\frac{D w_{3}^{(k+1)}}{C v_{3}}=0 \\
\vdots \\
\vdots \\
\frac{D w_{2}^{(k+1)}}{C v_{2}}-\frac{D w_{3}^{(k+1)}}{C v_{3}}=0 \\
\frac{D w_{2}^{(k+1)}}{C v_{2}}-\frac{D w_{4}^{(k+1)}}{C v_{4}}=0 \\
\vdots \\
\vdots \\
\frac{D w_{n}^{(k+1)}}{C v_{n}}-\frac{D w_{n-1}^{(k+1)}}{C v_{n-1}}=0
\end{array}\right.
\end{equation}

However, the rank of the coefficient matrix of the linear equation system is $\mathrm{n-1}$, so the above equation system actually only has $\mathrm{n-1}$ effective equations. These n-1 equations are extracted from the above equation system to form a new homogeneous equation system:
\begin{equation}
\left\{\begin{array}{c}
\frac{D w_{1}^{(k+1)}}{C v_{1}}-\frac{D w_{2}^{(k+1)}}{C v_{2}}=0 \\
\frac{D w_{2}^{(k+1)}}{C v_{2}}-\frac{D w_{3}^{(k+1)}}{C v_{3}}=0 \\
\vdots \\
\vdots \\
\frac{D w_{n-1}^{(k+1)}}{C v_{n-1}}-\frac{D w_{n}^{(k+1)}}{C v_{n}}=0
\end{array}\right.   
\end{equation}

Simplify to get:
\begin{equation}
\left\{\begin{array}{c}
\frac{w_{1}^{(k)}+u_{1}}{v_{1}}-\frac{w_{2}^{(k)}+u_{2}}{v_{2}}=0 \\
\frac{w_{2}^{(k)}+u_{2}}{v_{2}}-\frac{w_{3}^{(k)}+u_{3}}{v_{3}}=0 \\
\vdots \\
\frac{w_{n-1}^{(k)}+u_{n-1}}{v_{n-1}}-\frac{w_{n}^{(k)}+u_{n}}{v_{n}}=0 \\
\end{array}\right.
\end{equation}

Extract the coefficient matrix to get
\begin{equation}
A^{\prime}=\left[\begin{array}{ccccccc}
\frac{1}{v_{1}} & \frac{-1}{v_{2}} & 0 & \cdots & \cdots & 0 & 0\\
0 & \frac{1}{v_{2}} & \frac{-1}{v_{3}} & \cdots & \cdots & 0 & 0 \\
0 & 0 & \frac{1}{v_{3}} & \frac{-1}{v_{4}} & \cdots & \cdots & 0\\
&&&\vdots&&\\
0 & 0 & \cdots & \cdots & 0 & \frac{1}{v_{n-1}} & \frac{-1}{v_{n}}
\end{array}\right]    
\end{equation}

It can be seen that the rank of the coefficient matrix is $\mathrm{n-1}$, so the system of equations has infinite solutions. According to the principle that the total number of batchsize remains unchanged, we set the total number of samples unchanged, which is the following equation
\begin{equation}
w_{1}+w_{2}+\cdots \cdots+w_{n}=C=Const   
\end{equation}
\begin{equation}
\begin{gathered}
u_{1}+u_{2}+\cdots \cdots+u_{n}=0 \\
u=\left[u_{1}, u_{2}, \cdots \cdots, u_{n}\right]^{T} 
\end{gathered}
\end{equation}

After the linear equations are added to the above equations, the rank of the coefficient matrix becomes n, and there is a unique solution
\begin{equation}
A=\left[\begin{array}{cccccc} 
& & A & & & \\
1 & 1 & \cdots & \cdots & 1 & 1
\end{array}\right]    
\end{equation}

\begin{equation}
A=\left[\begin{array}{ccccccc}
\frac{1}{v_{1}} & \frac{-1}{v_{2}} & 0 & \cdots & \cdots & 0 & 0 \\
0 & \frac{1}{v_{2}} & \frac{-1}{v_{3}} & \cdots & \cdots & 0 & 0 \\
0 & 0 & \frac{1}{v_{3}} & \frac{-1}{v_{4}} & \cdots & \cdots & 0 \\
&&&\vdots&&\\
0 & 0 & \cdots & \cdots & \frac{1}{v_{n-2}} & \frac{-1}{v_{n-1}} & 0 \\
0 & 0 & \cdots & \cdots & 0 & \frac{1}{v_{n-1}} & \frac{-1}{v_{n}} \\
1 & 1 & \cdots & \cdots & 1 & 1 & 1
\end{array}\right]
\end{equation}

The constant term is
\begin{equation}
b=\left[\begin{array}{c}
\frac{w_{2}^{(k)}}{v_{2}}-\frac{w_{1}^{(k)}}{v_{1}} \\
\frac{w_{3}^{(k)}}{v_{3}}-\frac{w_{2}^{(k)}}{v_{2}} \\
\vdots \\
\vdots \\
\frac{w_{n}^{(k)}}{v_{n}}-\frac{w_{n-1}^{(k)}}{v_{n-1}} \\
0
\end{array}\right]
\end{equation}

Only need to solve the subordinate system of equations
\begin{equation}
\mathrm{A} \bullet \mathrm{u}=\mathrm{b}
\end{equation}

Therefore, solutions of equations: the increase $u$ is represented:
\begin{equation}
u=\left[\begin{array}{c}
\frac{v_{1}}{\sum_{i=1}^{n} v_{i}} \sum_{i=1}^{n} w_{i}-w_{1}^{(k)} \\
\frac{v_{2}}{\sum_{i=1}^{n} v_{i}} \sum_{i=1}^{n} w_{i}-w_{2}^{(k)} \\
\vdots \\
\vdots \\
\frac{v_{n-1}}{\sum_{i=1}^{n} v_{i}} \sum_{i=1}^{n} w_{i}-w_{n-1}^{(k)} \\
\frac{v_{n}}{\sum_{i=1}^{n} v_{i}} \sum_{i=1}^{n} w_{i}-w_{n}^{(k)}
\end{array}\right]    
\end{equation}
\end{document}